\documentclass[12pt]{article}
\pdfoutput=1
\usepackage[colorlinks,linkcolor=blue,citecolor=blue,bookmarks,bookmarksnumbered]{hyperref}
\usepackage[scaled=0.85]{helvet}
\usepackage{amsmath,amssymb,accents,mathrsfs,XoohmE}
\usepackage{graphicx,color}
\usepackage{booktabs}
\usepackage{multirow}
\usepackage{placeins}
\usepackage{amsmath}
\usepackage{XoohmE}

\newcommand*{\horzbar}{\rule[.5ex]{3.5ex}{0.5pt}}

\definecolor{Green}  {rgb}{0.10,0.70,0.10} 
\definecolor{Orange} {rgb}{1.00,0.50,0.15} 
\definecolor{Red}    {rgb}{0.90,0.00,0.12} 
\definecolor{Purple} {rgb}{0.50,0.25,0.55} 
\definecolor{Turque} {rgb}{0.00,0.65,0.85} 
\definecolor{Blue}   {rgb}{0.00,0.00,1.00} 
\definecolor{Magenta}{rgb}{1.00,0.00,1.00} 
\definecolor{Gold}   {rgb}{1.00,0.75,0.25} 
\definecolor{Seaweed}{rgb}{0.01,0.24,0.09} 
\definecolor{Brown}  {rgb}{0.43,0.26,0.32} 
\definecolor{grey1}  {rgb}{0.20,0.20,0.20} 
\definecolor{grey2}  {rgb}{0.40,0.40,0.40} 
\definecolor{grey3}  {rgb}{0.60,0.60,0.60} 
\definecolor{grey4}  {rgb}{0.80,0.80,0.80} 
\definecolor{grey5}  {rgb}{0.90,0.90,0.90} 
\def\C#1#2{{\ifcase#1\or
             \color{Green}\or \color{Orange}\or \color{Red}\or
              \color{Purple}\or \color{Turque}\or \color{Blue}\or
               \color{Magenta}\or \color{Gold}\or \color{Seaweed}\or
                \color{Brown}\or\color{grey1}\or\color{grey2}\or
                 \color{grey3}\else\color{grey4}\fi#2}}

\definecolor{Slate} {rgb}{0.00,0.45,0.55}

\def\fracm#1#2{\hbox{\large{${\frac{{#1}}{{#2}}}$}}}

\def\be{\begin{equation}}
\def\ee{\end{equation}}
\newcommand{\bea}{\begin{eqnarray}}
\newcommand{\eea}{\end{eqnarray}}
\newcommand{\ena}{\end{eqnarray}}

\def\pp{{\mathchoice
          {
              \kern 1pt%
              \raise 1pt
              \vbox{\hrule width5pt height0.4pt depth0pt
                    \kern -2pt
                    \hbox{\kern 2.3pt
                          \vrule width0.4pt height6pt depth0pt
                          }
                    \kern -2pt
                    \hrule width5pt height0.4pt depth0pt}%
                    \kern 1pt
           }
            {
              \kern 1pt%
              \raise 1pt
              \vbox{\hrule width4.3pt height0.4pt depth0pt
                    \kern -1.8pt
                    \hbox{\kern 1.95pt
                          \vrule width0.4pt height5.4pt depth0pt
                          }
                    \kern -1.8pt
                    \hrule width4.3pt height0.4pt depth0pt}%
                    \kern 1pt
            }
            {
              \kern 0.5pt%
              \raise 1pt
              \vbox{\hrule width4.0pt height0.3pt depth0pt
                    \kern -1.9pt  
                    \hbox{\kern 1.85pt
                          \vrule width0.3pt height5.7pt depth0pt
                          }
                    \kern -1.9pt
                    \hrule width4.0pt height0.3pt depth0pt}%
                    \kern 0.5pt
            }
            {
              \kern 0.5pt%
              \raise 1pt
              \vbox{\hrule width3.6pt height0.3pt depth0pt
                    \kern -1.5pt
                    \hbox{\kern 1.65pt
                          \vrule width0.3pt height4.5pt depth0pt
                          }
                    \kern -1.5pt
                    \hrule width3.6pt height0.3pt depth0pt}%
                    \kern 0.5pt
            }
        }}

\def\mm{{\mathchoice
                       {
                             \kern 1pt
               \raise 1pt    \vbox{\hrule width5pt height0.4pt depth0pt
                                  \kern 2pt
                                  \hrule width5pt height0.4pt depth0pt}
                             \kern 1pt}
                       {
                            \kern 1pt
               \raise 1pt \vbox{\hrule width4.3pt height0.4pt depth0pt
                                  \kern 1.8pt
                                  \hrule width4.3pt height0.4pt depth0pt}
                             \kern 1pt}
                       {
                            \kern 0.5pt
               \raise 1pt
                            \vbox{\hrule width4.0pt height0.3pt depth0pt
                                  \kern 1.9pt
                                  \hrule width4.0pt height0.3pt depth0pt}
                            \kern 1pt}
                       {
                           \kern 0.5pt
             \raise 1pt  \vbox{\hrule width3.6pt height0.3pt depth0pt
                                  \kern 1.5pt
                                  \hrule width3.6pt height0.3pt depth0pt}
                           \kern 0.5pt}
                       }}

\def\ad{{\kern0.5pt
                   \alpha \kern-5.05pt \raise5.8pt\hbox{$\textstyle.$}\kern
0.5pt}}

\def\bd{{\kern0.5pt
                   \beta \kern-5.05pt \raise5.8pt\hbox{$\textstyle.$}\kern
0.5pt}}

\def\qd{{\kern0.5pt
                   q \kern-5.05pt \raise5.8pt\hbox{$\textstyle.$}\kern
0.5pt}}
\def\Dot#1{{\kern0.5pt
     {#1} \kern-5.05pt \raise5.8pt\hbox{$\textstyle.$}\kern
0.5pt}}

\catcode`@=11
\def\un#1{\relax\ifmmode\@@underline#1\else
        $\@@underline{\hbox{#1}}$\relax\fi}
\catcode`@=12

\def\a{\alpha}
\def\b{\beta}
\def\c{\chi}
\def\d{\delta}
\def\e{\epsilon}

\def\g{\gamma}

\def\i{\iota}
\def\j{\psi}
\def\k{\kappa}
\def\l{\lambda}

\def\o{\omega}

\def\r{\rho}

\def\t{\tau}

\def\L{\Lambda}
\def\O{\Omega}
\def\P{\Pi}

\def\S{\Sigma}

\def\ca{{\cal A}}

\def\cf{{\cal F}}

\def\dslash{\not{\hbox{\kern-2pt $\partial$}}}
\def\Dslash{\not{\hbox{\kern-4pt $D$}}}
\def\pslash{\not{\hbox{\kern-2.3pt $p$}}}
 \newtoks\slashfraction
 \slashfraction={.13}
 \def\slash#1{\setbox0\hbox{$ #1 $}
 \setbox0\hbox to \the\slashfraction\wd0{\hss \box0}/\box0 }
 
\def\kcr{{\hbox{\ro \char'170}}}                
\def\ktl{{\hbox{\ro \char'170}}}        
\def\ktr{{\hbox{\ro \char'170}}}        
\def\kbl{{\hbox{\ro \char'170}}}        
\def\kbr{{\hbox{\ro \char'170}}}        

\def\plpl{\raise-2pt\hbox{$\raise3pt\hbox{$_+$}\hskip-6.67pt\raise0.0pt
\hbox{$^+$}\hskip 0.01pt$}}
\def\mimi{\raise-2pt\hbox{$\raise3pt\hbox{$_-$}\hskip-6.67pt\raise0.0pt
\hbox{$^-$}\hskip 0.01pt$}} 

\def\bo{{\raise.15ex\hbox{\large$\Box$}}}               
\def\pa{\partial}                                       
\def\TH{{\raise.2ex\hbox{$\displaystyle \bigodot$}\mskip-4.7mu \llap H \;}}
\def\face{{\raise.2ex\hbox{$\displaystyle \bigodot$}\mskip-2.2mu \llap {$\ddot
        \smile$}}}                                      

\def\dt#1{\on{\hbox{\bf .}}{#1}}                
\def\Dot#1{\dt{#1}}

\def\Hat#1{\widehat{#1}}                        
\def\Bar#1{\overline{#1}}                       
\def\leftrightarrowfill{$\mathsurround=0pt \mathord\leftarrow \mkern-6mu
        \cleaders\hbox{$\mkern-2mu \mathord- \mkern-2mu$}\hfill
        \mkern-6mu \mathord\rightarrow$}
\def\dvec#1{\vbox{\ialign{##\crcr
        \leftrightarrowfill\crcr\noalign{\kern-1pt\nointerlineskip}
        $\hfil\displaystyle{#1}\hfil$\crcr}}}           
\def\dt#1{{\buildrel {\hbox{\LARGE .}} \over {#1}}}     

\def\fracm#1#2{\hbox{\large{${\frac{{#1}}{{#2}}}$}}}
\def\sfrac#1#2{{\vphantom1\smash{\lower.5ex\hbox{\small$#1$}}\over
        \vphantom1\smash{\raise.4ex\hbox{\small$#2$}}}} 
\def\bfrac#1#2{{\vphantom1\smash{\lower.5ex\hbox{$#1$}}\over
        \vphantom1\smash{\raise.3ex\hbox{$#2$}}}}       
\def\afrac#1#2{{\vphantom1\smash{\lower.5ex\hbox{$#1$}}\over#2}}    

\def\pa{\partial}

\def\ad{{\dot{\alpha}}}
\def\bd{{\dot{\beta}}}

 \font\rOpe=cmsy10                        
 \def\ktl{{\hbox{\rOpe\char'170}}}        
 \def\kbl{{\hbox{\rOpe\char'170}}}        
 \def\kcr{{\reflectbox{\rOpe\char'170}}}        
 \def\ktr{{\reflectbox{\rOpe\char'170}}}        
 \def\kbr{{\reflectbox{\rOpe\char'170}}}        
 \def\Border{\vbox{\hsize0pt
        \setlength{\unitlength}{1mm}
        \newcount\xco
        \newcount\yco
        \xco=-21
        \yco=12
        \begin{picture}(0,0)(-7.5,0)
        \put(\xco,\yco){$\ktl$}
        \advance\yco by-1
        {\loop
        \put(\xco,\yco){$\kcr$}
        \advance\yco by-2
        \ifnum\yco>-240
        \repeat
        \put(\xco,\yco){$\kbl$}}
        \xco=170
        \yco=12
        \put(\xco,\yco){$\ktr$}
        \advance\yco by-1
        {\loop
        \put(\xco,\yco){$\kcr$}
        \advance\yco by-2
        \ifnum\yco>-240
        \repeat
        \put(\xco,\yco){$\kbr$}}
        \put(-19.5,13){\scalebox{.6065}{%
         University of Maryland Center for String and Particle  Theory \&\ Physics Department%
        |University of Maryland Center for String and Particle  Theory \&\ Physics Department}}
        \put(-19.5,-241.5){\scalebox{.5835}{%
         ****University of Maryland * Center for String and
         Particle  Theory* Physics Department****University of Maryland *Center
        for String and Particle  Theory* Physics Department}}
        \end{picture}
        \par\vskip-8mm}}
\definecolor{UMred}{rgb}{.9,.05,.2}
\definecolor{HUblue}{rgb}{.0,.3,.7}

\definecolor{Red}    {rgb}{0.90,0.00,0.12} 
\definecolor{Blue}   {rgb}{0.00,0.00,1.00} 
\definecolor{Green}  {rgb}{0.10,0.70,0.10} 
\definecolor{Turque} {rgb}{0.00,0.65,0.85} 
\definecolor{Orange} {rgb}{1.00,0.50,0.15} 
\definecolor{Magenta}{rgb}{1.00,0.00,1.00} 
\definecolor{Gold}   {rgb}{1.00,0.75,0.25} 
\definecolor{Seaweed}{rgb}{0.01,0.24,0.09} 
\definecolor{Purple} {rgb}{0.50,0.25,0.55} 
\definecolor{Brown}  {rgb}{0.43,0.26,0.32} 
\definecolor{grey1}  {rgb}{0.20,0.20,0.20} 
\definecolor{grey2}  {rgb}{0.40,0.40,0.40} 
\definecolor{grey3}  {rgb}{0.60,0.60,0.60} 
\definecolor{grey4}  {rgb}{0.80,0.80,0.80} 
\definecolor{grey5}  {rgb}{0.90,0.90,0.90} 
\def\C#1#2{{\ifcase#1\or
             \color{Red}\or \color{Green}\or \color{Blue}\or\
              \color{Turque}\or \color{Orange}\or \color{Magenta}\or 
               \color{Gold}\or \color{Seaweed}\or \color{Purple}\or
                \color{Brown}\or\color{grey1}\or\color{grey2}\or
                 \color{grey3}\else\color{grey4}\fi#2}}

\definecolor{Slate} {rgb}{0.00,0.45,0.55}

\newdimen\parshift\parshift=\parindent
\catcode`@=11
 \long\def\@footnotetext#1{\insert\footins{\reset@font\footnotesize
           \interlinepenalty\interfootnotelinepenalty\splittopskip%
            \footnotesep\splitmaxdepth\dp\strutbox\floatingpenalty\@MM%
             \hsize\columnwidth\addtolength{\hsize}{-2\parindent}
              \@parboxrestore\protected@edef\@currentlabel%
              {\csname p@footnote\endcsname\@thefnmark}%
                \color@begingroup%
                 \@makefntext{\rule\z@\footnotesep\ignorespaces#1%
                  \@finalstrut\strutbox}%
                \color@endgroup}}
 \long\def\@makefntext#1{\hglue\parshift%
           \vbox{\noindent\baselineskip=11pt plus.5pt minus.5pt\hb@xt@0em{\hss\@makefnmark\kern1pt}#1}}
\catcode`@=12

\newskip\humongous \humongous=0pt plus 1000pt minus 1000pt
\def\caja{\mathsurround=0pt}
\def\eqalign#1{\,\vcenter{\openup2\jot \caja
        \ialign{\strut \hfil$\displaystyle{##}$&$
        \displaystyle{{}##}$\hfil\crcr#1\crcr}}\,}
\newif\ifdtup

\makeatletter
\def\section{\@startsection{section}{1}{\z@}
        {3ex plus-1ex minus-.2ex}{1pt plus1pt}{\large\sf\bfseries\boldmath}}
\def\subsection{\@startsection{subsection}{2}{\z@}
         {1.5ex plus-1ex minus-.2ex}{0.01pt plus1pt}{\sf\slshape}}
\def\subsubsection{\@startsection{subsubsection}{3}{\z@}
          {1.5ex plus-1ex minus-.2ex}{0.01pt plus0.2pt}{\sf\boldmath}}
\def\paragraph{\@startsection{paragraph}{4}{\z@}
           {.75ex \@plus.5ex \@minus.2ex}{-2mm}{\sf\bfseries\boldmath}}
\makeatother

\allowdisplaybreaks
\seceq
\numberwithin{figure}{section}

\usepackage{tikz}
\usepackage[vcentermath,enableskew]{youngtab}
\usepackage[centertableaux]{ytableau}
\let\TC=\textcolor
\definecolor{Hey}{rgb}{.9,.05,.4}
\definecolor{orange}{rgb}{1,.5,0}
\definecolor{plum}{rgb}{.4,0,.6}
\definecolor{R}{rgb}{1,0,0}
\definecolor{G}{rgb}{0,1,0}
\definecolor{B}{rgb}{0,0,1}

\long\def\CMTred#1{\leavevmode\TC{red}{\sf#1}}

\long\def\CMTR#1{\leavevmode\TC{R}{\sf#1}}

\long\def\CMTB#1{\leavevmode\TC{B}{\sf#1}}

\usepackage{lipsum}
\usepackage{listings}
\definecolor{MyDarkGreen}{rgb}{0.0,0.4,0.0} 
\newcommand{\perlscript}[2]{
\begin{itemize}
\item[]\lstinputlisting[caption=#2,label=#1]{#1.pl}
\end{itemize}
}

\begin{document}

\thispagestyle{empty}
\noindent{\small
\hfill{$~~$}  \\ 
{}
}
\begin{center}
{\large \bf
Adinkra Foundation of Component \vskip0.02in
Decomposition and the Scan for  \vskip0.02in
Superconformal Multiplets in 11D, $\mathcal{N} = 1$ Superspace
}   \\   [8mm]
{\large {
S.\ James Gates, Jr.\footnote{sylvester$_-$gates@brown.edu}${}^{,a, b}$,
Yangrui Hu\footnote{yangrui$_-$hu@brown.edu}${}^{,a,b}$, and
S.-N. Hazel Mak\footnote{sze$_-$ning$_-$mak@brown.edu}${}^{,a,b}$
}}
\\*[6mm]
\emph{
\centering
$^{a}$Brown Theoretical Physics Center,
\\[1pt]
Box S, 340 Brook Street, Barus Hall,
Providence, RI 02912, USA 
\\[10pt]
$^{b}$Department of Physics, Brown University,
\\[1pt]
Box 1843, 182 Hope Street, Barus \& Holley,
Providence, RI 02912, USA 
}
 \\*[30mm]
{ ABSTRACT}\\[5mm]
\parbox{142mm}{\parindent=2pc\indent\baselineskip=14pt plus1pt
For the first time in the physics literature, the Lorentz representations of all 2,147,483,648 bosonic degrees of freedom and 2,147,483,648 fermionic degrees of freedom in an unconstrained eleven dimensional scalar superfield are presented. Comparisons of the conceptual bases for this advance in terms of component field, superfield, and adinkra arguments, respectively, are made. These highlight the computational efficiency of the adinkra-based approach over the others. It is noted at level sixteen in the 11D, {$\cal N$} = 1 scalar superfield, the $\{ 65 \}$ representation of SO(1,10), the conformal graviton, is present.  Thus, adinkra-based arguments suggest the surprising possibility that the 11D, {$\cal N$} = 1 scalar superfield alone might describe a Poincar\' e supergravity prepotential or semi-prepotential in analogy to one of the off-shell versions of 4D, {$\cal N$} = 1 superfield supergravity.  We find the 11D, {$\cal N$} = 1 scalar superfield contains 1,494 bosonic fields, 1,186 fermionic fields, and a maximum number of 29,334 links connecting them
via orbits of the supercharges. 
} \end{center}
\vfill
\noindent PACS: 11.30.Pb, 12.60.Jv\\
Keywords: supersymmetry, superfields, supergravity, off-shell, branching rules 
\vfill
\clearpage

\newpage
{\hypersetup{linkcolor=black}
\tableofcontents
}

\newpage
\section{Introduction}

The standard ``workhorse'' of Salam-Strathdee superspace \cite{SSP} is the concept of 
the ``superfield.''  Previously, we have argued the superfield concept can be augmented 
by the newer network-centric concept of  ``adinkras''\footnote{Interested parties can also 
find the literature for these extensively cited in the works of \cite{counting10d,CoDeX}.} 
\cite{Adnk1}.  From the time of their introduction in one dimensional extended superspaces 
of the Salam-Strathdee type, $\cal {GR}$(d, $N$) algebras\footnote{The designators d and 
$N$ here for $\cal {GR}$(d, $N$) algebras refer to the number of nodes and the 
number of colors, respectively, in adinkras.} \cite{GRana1,GRana2} -  the adjacency 
matrices for the adinkra graphs - gave an explicit  solution to describing minimal irreducible 
supermultiplets in one dimensional theories for all values of degree of extension $N$.  On 
previous occasions (e.\ g.\  \cite{ENUF}), we have pointed out that $\cal {GR}$(d, $N$) matrices 
are the ($\cal {G}$)eneral ($\cal {R})$eal extensions of two-component van der 
Waerden matrices used in physics. This work also contained a description of how the $\cal 
{GR}$(d, $N$) matrices are embedded in Clifford algebras.  

As we have noted before, since $\cal {GR}$(d, $N$) matrices are the adjacency matrices
for adinkras, a direct question adinkras answer is, ``Given $N$ supercharges in a one 
dimensional system, what is the minimum number d${}_{min}$ of bosons and equal 
number of fermions required to realize the $N$ supercharges  in a linear manner?"  In the 
works of \cite{GRana1,GRana2} a function d${}_{min}(N)$ possessing Bott periodicity and 
given by
\be
{\rm d}{}_{\min}(N)=\begin{cases}
{~~}2^{\frac{N-1}{2}}\,~,&N\equiv 1,\,7 $~~~~~~~\,~~~$ \bmod{(8)}\\
{~~}2^{\frac{N}{2}}{~~~~},&N\equiv 2,\,4,\,6 $~~~~~~~$ \bmod{(8)}\\
{~~}2^{\frac{N+1}{2}}~\, ,&N\equiv 3,\,5 $~~~~\,~~~~~~$ \bmod{(8)}\\
{~~}2^{\frac{N - 2}{2}}~\, ,&N\equiv  8 $~~~~~\,~\,~~\,~~\,~~$ \bmod{(8)}\\
\end{cases}
{~~~~~~~~~~~~~~~}
\label{eqn:dmin}
\ee
(where we excluded the case of $N$ = 0, i.e. no supersymmetry) was proposed as the answer.  
Until recently, no derivation of this result that is not related to adinkra-based arguments was 
known to us. 

However,  it has been communicated to us\footnote{In a private conversation with Warran Siegel.} that another alternative 
narrative argument should lead to this same result.  W.\ Siegel has observed that as the form 
of the one dimensional $N$-extended supersymmetry algebra implies for $N$ supersymmetry 
generators $q$, satisfy
\be
\{\, q~,~q \, \} \propto~ E  ~~, 
\ee
this can be regarded as a Clifford algebra.  Thus, the quantities $q$ must be representations
of SO($N$).  Investigating the minimal such irreducible representations, he argues, must lead to the formula
above.  To this narrative, we respond the work in \cite{ENUF} (contained in its equations
(13)-(16)) precisely provides a derivation aimed at this.

The question raised in the last paragraph can be extended to the more complicated domain 
of higher dimensional ``off-shell'' theories by asking ``Given $\cal N$ supercharges in a D 
dimensional system, what is the minimum number d${}_{min}$ of bosons and equal number 
of fermions required to realize the $\cal N$ supercharges in a linear manner {\it {without the 
use of any dynamical assumptions}}?"  Thus, one is led to suspect the existence of a function 
$\Hat{\rm d}_{min}({\cal N}, \, {\rm D})$ in any dimension that gives the answer to the question 
in general but with the property
\be{
\Hat{\rm d}{}_{min}({\cal N}, \, {\rm D} \,=\, 1) ~=~ {\rm d}{}_{\min}({\cal N} {\cal F}({\rm D})) ~~~,
\label{eqn:dminND}
} 
\ee
where the function on r.h.s. of the equation is defined in (\ref{eqn:dmin}) and the function ${
\cal F}({\rm D})$ is shown in a few examples in Table \ref{tab:dD}.  The explicit form of $\Hat{
\rm d}_{min}({\cal N}, \, {\rm D})$ has remained unknown throughout the history of the subject 
of supersymmetry\footnote{The author SJG has long referred to this as one of the  ``SUSY white 
whale problems.''}, but Equation (\ref{eqn:dminND}) gives its boundary condition,  i.e. the value 
when we reduce the dimension to one.  The curious reader may question what 
is the source of the caveat regarding dynamical assumptions?  The answer is this is necessary 
to find prepotential formulations in the Salam-Strathdee superspaces for the theories under study.

When ${\cal N} \,=\, 1 ~ {\rm {and}} ~ {\rm D} \,=\, 11$, one is looking at the low-energy limit of 
M-Theory \cite{MTh}, the eleven dimensional supergravity theory \cite{crD11a,crD11b}.  

For decades, there has been little understanding created beyond these descriptions of the 
on-shell theories\footnote{Interested parties can find the literature for these theories 
extensively cited in the works of \cite{counting10d,CoDeX}.} in ordinary Salam-Strathdee 
superspace.  It is thus accurate to describe these as ``orphaned'' problems currently 
existing in an ``abandoned'' state.  Using conceptual and computational advances it 
can be argued there is a reason to expect new progress.  Analytical progress with regards 
to the M-Theory corrections to the on-shell theory has been shown in the works of 
\cite{PF1,PF2,PF3} and in \cite{Howe:2003sa,Howe:2003cy} 
group representation theory was also included in the discussions.  

Another, apparently very successful avenue, to the study of the low energy effective action (LEEA)
of M-Theory, began with the work of Green and Sethi \cite{GS} (it is useful to examine the list of 
citations to this work also).  The work in \cite{GS}, at least ``thematically" follows in an old SUSY 
tradition set in place by Mandelstam's early investigations of 4D, $\cal N$ = 4 super Yang-Mills theory.  
In the works \cite{MandL1,MandL2,MandL3} a lightcone formulation was used to establish the 
perturbative finiteness of the 4D, $\cal N$ = 4 super Yang-Mills theory.  The latest evolution 
\cite{OK} of the work which began in that of \cite{GS} is a derivation of an impressive fourteen 
orders in spacetime derivatives!

However, like the work of Mandelstam, the investigations that follow the ``GS-formalism"
\cite{GS}, have a price to pay.  The starting point of the Green-Sethi construction is clearly 
described in and around equation (2.1) of their reference.  The fundamental Grassmann 
coordinates form a complex chiral spinor in SO(1,9).  Our starting point is a Majorana 
spinor in SO(1,10).  In particular, this means that, while the Lorentz symmetry of a SO(1,9)
subgroup of SO(1,10) can be realized linearly and manifestly on the G-S model, the remaining 
symmetries of the coset SO(1,10)/SO(1,9) (expected to be present) must be realized in some 
non-linear manner.  Certainly, the existence of dualities suggests these Òcoset realizationÓ are 
possible, but we believe it would be a valuable addition to the literature to create a formalism 
demonstrating a manifest linear realization.  

We are not alone in noting these ``coset symmetries'' of 11D, superspace can occur in a
model of M-Theory.  Although the work presented in the recent posting \cite{BB} is from 
a very different perspective, it requires realization of SO(1,10)/SO(1,3) coset symmetries 
of the eleven dimensional spacetime.  These authors explicitly state the expectation of 
``higher dimensional Lorentz and supersymmetry transformations realized in a non-linear 
manner."  This shows others seeking a formalism that explicitly demonstrates of these 
symmetries.  The difference in our attempt is to realize these in a
linear and manifest manner.
    
Having encountered some unfamiliarity with the relationship between unconstrained prepotentials
and the description of higher order corrections in the LEEA, it is useful to recall the result
shown in the work of \cite{FX} which contains the first superspace description of modifications
to the open superstring effective action\footnote{We have also included this discussion in the
work of \cite{CoDeX}.}.  In this study, the direct relation between off-shell
superspace formulations and higher derivative corrections generated by superstring theory 
was shown at lowest order.  The result in this old work was derived precisely by starting from 
an off-shell superfield connection formulation. Using symmetry arguments, a unique modification 
to include the open string correction was found.  This resulting understanding of the 
relationship between off-shell superfields and higher derivative terms in the open 
superstring effective action has been verified by numerous later citations.  

A special note of attention should be directed to the works of \cite{Howe:2003sa,Howe:2003cy}
as these provide studies that are in a sense ``orthogonal'' to the direction of our works.  These 
works of Howe et.\ al.\ provide a thorough investigation of this class of problems...based on the 
study of superspace Bianchi identities (referred to as ``spinorial cohomology'') associated with 
the geometrical sector in the 11D superspace.  These are in accord with the previous analyses
of \cite{M2,2MT,2MT0} with regards to the form of deformations of the superspace torsion tensor.
However, the works in \cite{Howe:2003sa,Howe:2003cy} include analysis of the dual 6-form (and 
3-form) sector.  Indeed all these works \cite{Howe:2003sa,Howe:2003cy,M2,2MT,2MT0} may 
be considered ``orthogonal''  to our current efforts as they focus on the Bianchi identities and 
the objects that appear in them.  It is the aim of this work to provide the first exploration among 
these superfields aimed at methods of discovery of candidate prepotential superfields in this 
domain.

Computational power as well as algorithmic architecture design have advanced
tremendously since the 1980's.  Along the lines of new computational paradigms, 
there now exist breakthroughs in artificial intelligence, neural networks, 
and deep learning that emerged in the intervening period. By using conceptual 
and computational advances it can be argued, now is a propitious time to make 
new progress in many areas.

In the work of \cite{counting10d} a combination of new conceptual and computational 
tools was deployed to create progress in problems surrounding superspace geometries 
that describe supergravity in ten dimensions.  Results in the work confirmed the
analysis of the spectrum \cite{10DScLR} of the scalar superfield given by Bergshoeff 
and de Roo for the component fields in the SG limits of Type-I closed and heterotic string
theories.   Moreover, this work \cite{counting10d} also gave new similar results in the 
domains of the Type-IIA and Type-IIB SG limits.  Though these problems involve a factor 
of 65,536 more degrees of freedom than occur in Type-I closed and heterotic string
theories, the modern techniques proved to be up to the tasks of complete analysis
of these systems. As the Type-IIA SG system can be obtained from a dimensional reduction
of the eleven dimensional system, this was a signal that the supergravity limit of M-Theory
should be directly amenable to the same sort of complete analysis to {\it {provide}}
complete transparency about the SO(1,10) Lorentz representations and explicitly
{\it {demonstrate}} manifest linear realization of its spacetime symmetries.  It is the 
purpose of the present work to create an exordium for this result.

{\it {We are now able to abstract the component field content of 
superfields as well as study the existence of orbits among component fields under the 
action of the supercharges without using traditional}} $\theta$-{\it {expansions for high 
dimensions}}. 

The layout of this paper is described below.

Chapter two provides a self-contained description of the ``off-shell auxiliary field 
problem'' by beginning at the component level and discussing an {\it {ab}} {\it {initio}} 
recipe for deriving supersymmetric representations for arbitrary spacetimes.  The 
distinction between ``off-shell'' and ``on-shell'' formulations is noted.  Next a
similar high level discussion of how off-shell supermultiplets, that are equivalent to 
superfields in the context of Salam-Strathdee superspace, is presented.  The final 
portion of this chapter introduces the concept of the adinkra of a superfield or a 
supermultiplet, using the example of the 10D, $\cal N$ = 1 scalar superfield, as a 
network that encodes the Lorentz representations of the field content as well as the 
orbits of those field representations under the action of supercharges.

Chapter three follows the route of the traditional $\theta$-expansion as applied to 
the 11D, $\cal N$ = 1 scalar superfield.  A discussion involving a recursion formula 
used to move in a level-by-level manner up the $\theta$-expansion of the superfield 
is presented and the role of Duffin-Kemmer-Petiau fields is noted.  The recursive 
procedure is applied up to quartic order to show the calculational complications that 
occur in starting with a superfield and then abstracting the component field content 
from such a systematic starting point.

The fourth chapter contains the main results of this work and is dedicated to showing 
that calculational efficiency can occur in the algorithms for extracting the component 
field content from branching rules of $\mathfrak{su}(32)\supset\mathfrak{so}(11)$ and 
the concept of Plethysm instead of the superfield's $\theta$-expansion.  In order to 
implement the use of branching rules, explicit projection matrices needed are presented.

These all powerfully combine so as to make the need for any explicit calculation 
based on $\gamma$-matrices to become totally banished from these considerations.  It 
is the independence of these methods from $\gamma$-matrices that allows for substantial 
computational efficiencies which can be exploited by modern computer based algorithms.  
All these together become the ``secret sauce'' that allows unprecedented access to the 
component level structures that heretofore have been hidden within high dimensional
Salam-Strathdee superfields.  

Finally, by use of the branching rule approach to identifying the complete spectrum of
component fields within the 11D, $\cal N$ = 1 scalar superfield, we find evidence
for a new phenomenon starting at the seventh level.  Namely, consistency with the
branching rule demands that at some fixed levels $p$ of the superfield, multiple copies of the same irrep must appear.  We propose the mechanism that must be responsible for this is the fact that the expansion of the superfield at Level-$p$ must be in terms of certain linearly independent and SO(1,10) irreducible polynomials $\big[ \theta{}^{1} \, \cdots \, \theta{}^{p} \big]_{IR}$ instead of simply $ \theta{}^{1} \, \cdots \, \theta{}^{p}$.

We apply these tools to extract the SO(1,10) representations from all 1,494 bosonic 
fields and 1,186 fermionic fields contained in a 11D, $\cal N$ = 1 scalar superfield.  To 
our knowledge, these observations about the numbers of fields (both bosonic and
fermionic) as opposed to the number of degrees of freedom have not appeared previously
in the literature.  We also exploit these techniques to study the structure of orbits 
(i.\ e.\ the ``linking information'') that the supercharges generate between bosons
and fermions and vice-versa.  Our techniques permit us to count the number of such
orbits. We determine the maximum number of such orbits is 29,334 such links.  We emphasize
that to this point no Clifford algebra based calculations are utilized.

Having obtained this information, we follow the path established by 
Breitenlohner, to look for what superfield can minimally contain the conformal 11D 
graviton and gravitino.  A surprising answer is found.

The fifth chapter is devoted to describing the adinkra of the 11D, $\cal N$ = 1 superfield 
giving a level-by-level description of the number of fields contained at each level.  This 
relates back to the 1,494 bosonic fields and 1,186 fermionic fields found in the previous 
chapter.  An image of the adinkra up to Level-5 is given.  This includes a depiction 
of the orbits of the fields under the action of the supercharges.  By use of the linking
structure of the adinkra, we determine the maximum number of possible supersymmetry transformation
laws connecting bosonic and fermionic fields in the supermultiplet is 29,334.

We include our conclusions where we discuss possible implications for the superfield limit 
of M-Theory and Type-IIA superstring theory.  This is followed by six appendices. Appendix 
A contains a dictionary between Dynkin Labels and the corresponding representation 
dimensionalities.  Appendices B - D contain technical details of manipulations with 11D 
$\gamma$-matrices.  These are included for any researcher who wishes to verify independently 
the assertions we make about the properties of the $\gamma$-matrices that we are able to bypass. 
Appendix E contains an extended discussion of the role that two distinct types of Young 
Tableaux play in clarifying the manner in which $\gamma$-matrices are avoided in this approach.  
Understanding these plays a role in the final suggestion of the conclusion which is that 
calculational efficiencies are likely possible if the traditional concept of the Salam-Strathdee 
superfield is replaced by a newer concept of an ``adinkra-field'' where the fermionic Young 
Tableaux play the role of the $\theta$-coordinates and the Dynkin Labels play the role of 
the fields.

The final appendix presents the decomposition results of the 11D, $\mathcal{N} = 1$ 
scalar superfield by giving Dynkin labels.

\newpage
\section{Primers Before 11D}
\label{sec:primer}

The component formulation of supersymmetrical systems has traditionally followed
a pattern that we will review next.

\subsection{Component Primer Before 11D}
\label{sec:primer1}

At a general level, one begins with a set of bosonic representations we denote
by ${\CMTB {\{ {\cal R}^{(i)} \}}}$ and a set of fermionic representations we denote
by ${\CMTR {\{ {\cal R}^{(j)} \}}}$.  The range of the indices on the two distinct sets
need not be the same, i.e. the values taken on by ${\CMTB {(i)}}$ and ${\CMTR 
{(j)}}$ are generally different.  Next one {\it {assumes}} a set of dynamics codified
by specifying a Lagrangian, the schematic form of which is realized as (where ${\partial}$ is 
the spacetime derivative but written in an index-free notation),
\be
{\cal L} ~=~ \fracm 12 \, {\Big[} \, \sum_{ {\CMTB {(i)}}}  \,  {\CMTB {\{ {\cal R}^{(i)} \}}} \pa \, \pa  {\CMTB 
{\{ {\cal R}^{(i)} \}}} ~+~ i \,  \sum_{ {\CMTR {(j)}}} \, {\CMTR{\{ {\cal R}^{(j)}\}}} \pa \, 
{\CMTR{\{ {\cal R}^{(j)}\}}} \, {\Big]} ~~~,
\label{SUSYact0} 
\ee
which is followed by introducing a ``supercharge'' that we (once more schematically)
write as ${\CMTR{ {\rm D}}}$ together with the definitions of its realizations on the 
bosonic reps ${\CMTB {\{ {\cal R}^{(i)} \}}}$ and fermionic reps ${\CMTR {\{ {\cal 
R}^{(j)} \}}}$  according to
\be
{\CMTR{ {\rm D}}} \, {\CMTB{\{ {\cal R}^{(i)}\}}} ~=~ \sum_{ {\CMTR {(j)}}}  
 \, {\CMTB{ c }}{}_{{\CMTR{\{ {\cal R}^{(j)}\}}}}^{+{\CMTB {(i)}}  }
{\CMTR{\{ {\cal R}^{(j)}\}}} ~~~~,~~~~
{\CMTR{ {\rm D}}} \, {\CMTR{\{ {\cal R}^{(j)}\}}} ~=~ 
\sum_{ {\CMTB {(i)}}}  \,
\, {\CMTB{ c }}{}_{{\CMTB{\{ {\cal R}^{(i)}\}}}}^{-{\CMTR {(j)}}  }  \pa
{\CMTB{\{ {\cal R}^{(i)}\}}}  ~~~,
\label{SUSYact1} 
\ee
in terms of a set of constants $\, {\CMTB{ c }}{}_{{\CMTB{\{ {\cal R}^{(i)}\}}}}^{-{\CMTR 
{(j)}}  }$ and ${\CMTB{ c }}{}_{{\CMTR{\{ {\cal R}^{(j)}\}}}}^{+{\CMTB {(i)}}  }$.  From
long experience, it is known that for judicious choices of the representations
${\CMTB {\{ {\cal R}^{(i)} \}}}$ and ${\CMTR {\{ {\cal R}^{(j)} \}}}$, these constants
can be chosen so that
\be
{\CMTR{ {\rm D}}} \, {\cal L} ~=~ {\rm {purely}} ~ {\rm {surface}}~ {\rm {terms}} ~~~.
\label{SUSYact2}
\ee
For the reader interested in seeing a more explicit discussion of this in
examples, the work in \cite{C0L0R} is recommended. 

When one appropriately calculates an expression that is second
order in the ${\CMTR{ {\rm D}}}$ operator, a bifurcation occurs with two possible outcomes
\be
{~~~~~~~~~~~~}
{\CMTR{ {\rm D}}} \,\vee \, {\CMTR{ {\rm D}}} ~ \propto ~\begin{cases}
  \, &i\, 2 \,  \pa ~+~  \, \pa {\cal L}  $~~~~~~~\,~~~$ {\rm {:(a.)~on-shell}}\, \,
 {\rm {SUSY}} \\
  &i\, 2 \, \pa  $~~~~~~~\,~~~$  $~~~~~~~~~~$ {\rm {:(b.)~ off-shell}}\, \,
 {\rm {SUSY}}
\end{cases}
{~~~~~~~~~~~~~~~}
\label{eqn:dminXZ}
\ee
where the term $ \pa {\cal L}$ stands for a set of equations of motion that are 
derivable from ${\cal L}$.  To reconcile the differences between outcomes (a.)
and (b.) in (\ref{eqn:dminXZ}) above, it is most common to demand that the fields in 
the system should obey their equations of motion.

\subsection{Superfield Primer Before 11D}
\label{sec:primer3}

The idea of the superfield, or equivalently an ``off-shell supermultiplet'' is to modify
the starting point in three ways: \vskip0.01in \indent
(a.) the range of the index ${\CMTB {(i)}}$ describing the bosonic representations
${\CMTB{\{ {\cal R}^{(i)}\}}}$ is allowed
\newline $~~~~~~~~~~~$  to increase, 
\vskip0.01in \indent
(b.) the range of the index ${\CMTR {(j)}}$ describing the fermionic representations ${\CMTR{\{ {\cal R}^{(j)}\}}}$ is allowed \newline $~~~~~~~~~~~$ to increase, \vskip0.01in \indent
(c.) a ``height" or ``Level" number is introduced for all the bosonic reps ${\CMTB{\{ {\cal R}^{(i)}\}}}$
and 
\newline $~~~~~~~~~~~$
fermionic reps ${\CMTR{\{ {\cal R}^{(j)}\}}}$, each with their enhanced range of indices.

The Level numbers are non-negative integers.  It is convenient to partition the Level numbers
into even and odd integers.  In the language of superfields, this corresponds to the monomial
of Grassmann coordinates associated with the component field representation in the ``$\theta$-expansion of the superfield.''

As the ranges of the indices the ${\CMTB {(i)}}$ and ${\CMTR {(j)}}$ in this subsection
are greater than those associated with (\ref{SUSYact0}), this means new bosonic reps
and new fermionic reps are under consideration.  The new bosonic reps are called 
``auxiliary bosonic fields'' while the new fermionic reps are called ``auxiliary fermionic 
fields.''  Over the totality of the component field reps, one must now define the action of
${\CMTR{ {\rm D}}}$.  For the point covered by the following arguments, we will use
the words ``superfield'' and ``adinkra'' interchangeably.

For example, if we begin with the ``i-th'' bosonic representation ${\CMTB
{\{ {\cal R}^{(i)} \}}}{}_p$ at level $p$ in the adinkra, then the action of the spinor 
covariant (in an index-free notation) derivative ${\CMTR{ {\rm D}}}$ must take 
the form
\begin{equation}
{\CMTR{ {\rm D}}} \, {\CMTB{\{ {\cal R}^{(i)}\}}}{}_p ~=~ \sum_{ {\CMTR {(j)}}}  
 \, {\CMTB{ c }}{}_{{\CMTR{\{ {\cal R}^{(j)}\}}}}^{+(p){\CMTB {(i)}}  }
{\CMTR{\{ {\cal R}^{(j)}\}}}{}_{p + 1} ~+~ \sum_{ {\CMTR {(j)}}}  
 \,  {\CMTB{ c }}{}_{{\CMTR{\{ {\cal R}^{(j)}\}}}}^{-(p){\CMTB {(i)}}  }
{\partial} 
\, {\CMTR{\{ {\cal R}^{(j)}\}}}{}_{p - 1} ~~~, \label{adnkb}
\end{equation}
where ${\CMTR{\{ {\cal R}^{(j)}\}}}{}_{p + 1}$ and ${\CMTR{\{ {\cal R}^{(j)}\}}}{}_{
p - 1}$ correspond to the ``j-th'' fermionic representations at the $p + 1$ level 
and $p - 1$ level respectively in the adinkra.   

In a similar manner, if we begin with the ``j-th'' fermionic representation ${\CMTR{
\{ {\cal R}^{(j)}\}}}{}_p$ at level $p$ in the adinkra, then the action of the spinor 
covariant derivative ${\CMTR{ {\rm D}}}$ must take the form
\begin{equation}
{\CMTR{ {\rm D}}} \, {\CMTR{\{ {\cal R}^{(j)}\}}}{}_p ~=~ 
\sum_{ {\CMTB {(i)}}}  \,
\, {\CMTB{ c }}{}_{{\CMTB{\{ {\cal R}^{(i)}\}}}}^{+(p){\CMTR {(j)}}  }
{\CMTB{\{ {\cal R}^{(i)}\}}}{}_{p + 1} ~+~ \sum_{ {\CMTB {(i)}}} 
 \, {\CMTB{ c }}{}_{{\CMTB{\{ {\cal R}^{(i)}\}}}}^{-(p){\CMTR {(j)}}  }
 {\partial} 
\, {\CMTB{\{ {\cal R}^{(i)}\}}}{}_{p - 1} ~~~, \label{adnkf}
\end{equation}
where ${\CMTB{\{ {\cal R}^{(i)}\}}}{}_{p + 1}$ and
${\CMTB{\{ {\cal R}^{(i)}\}}}{}_{p - 1}$
correspond to the ``i-th''  bosonic representations at the $p + 1$ level and $p - 1$
level respectively in the adinkra.

The quantities ${\CMTB{ c }}{}_{{\CMTR {\{ {\cal R}^{(j)}\}}}}^{+(p){\CMTB {(i)}}  }$, 
${\CMTB{ c }}{}_{{\CMTR{\{ {\cal R}^{(j)}\}}}}^{-(p){\CMTB {(i)}}  }$,  
${\CMTB{ c }}{}_{{\CMTB{\{ {\cal R}^{(j)}\}}}}^{+(p){\CMTR {(i)}}  }$, and 
${\CMTB{ c }}{}_{{\CMTB{\{ {\cal R}^{(j)}\}}}}^{-(p){\CMTR {(i)}}  }$ (in (\ref{adnkb})
and (\ref{adnkf}) respectively) are sets of constants typically proportional to $\gamma$-matrices, 
Minkowski metric, Levi-Civita tensor, or powers of any of these when one uses
traditional Salam-Strathdee superfields.

The overarching point is by starting from the definitions in (\ref{adnkb}) and
(\ref{adnkf}) and repeating the calculation described by (\ref{eqn:dminXZ}),
the constants in two equations that define the realization of ${\CMTR{ {\rm D}}}$
are fixed by the condition that they only lead to the condition (b.), i.e. describe
an ``off-shell'' realization of supersymmetry.

\subsection{Adinkra Primer Before 11D}
\label{sec:primer4}

In the work of \cite{counting10d}, the complete descriptions of the component field 
representations required to describe an off-shell theory of scalar gravitation in 10D, $\mathcal{N} = 1$, $\mathcal{N} = 2$A, and $\mathcal{N} = 2$B superspaces were presented.  In the following, it is expedient for us to concentrate
on the 10D, $\mathcal{N} = 1$ case and focus on the adinkra.   

\begin{figure}[htp!]
\centering
\includegraphics[width=0.5\textwidth]{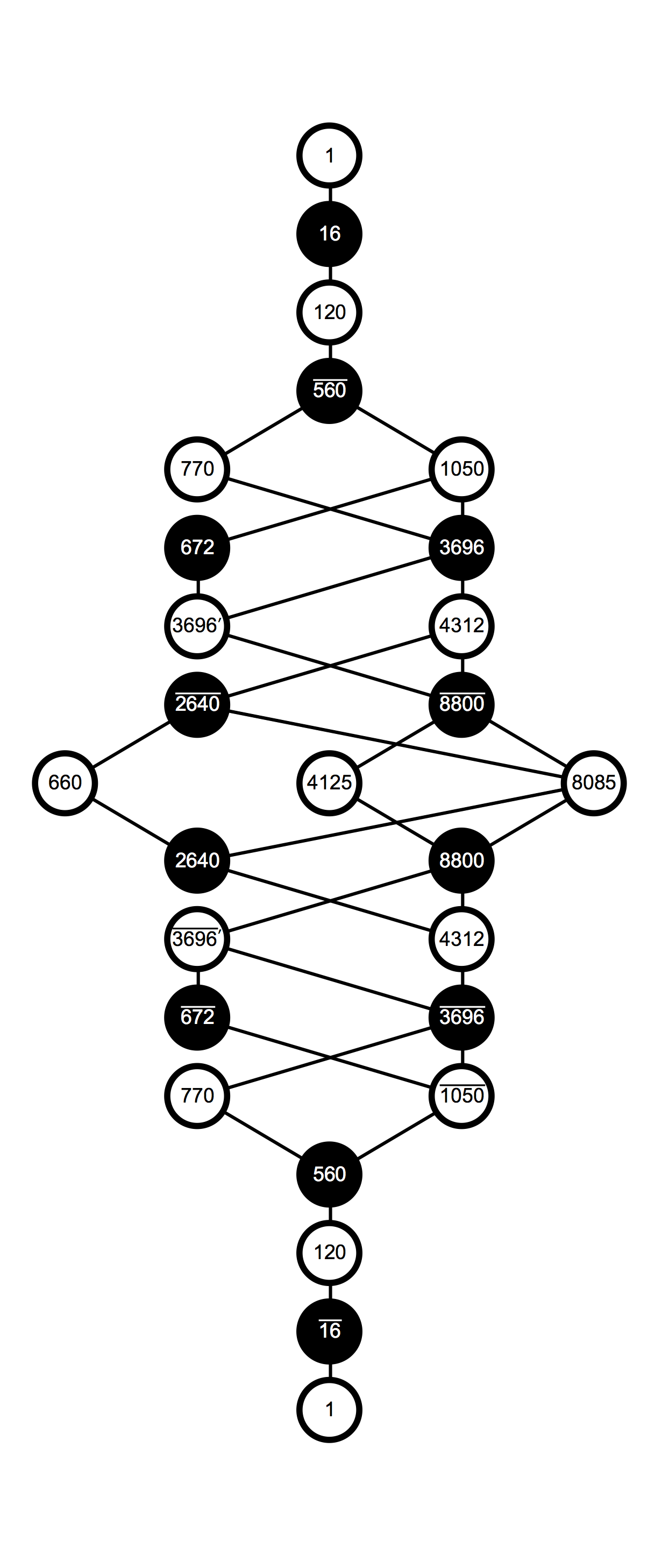}
\caption{Adinkra Diagram for 10D, $\mathcal{N} = 1$ Scalar Superfield}
\label{Fig:10D}
\end{figure}

\newpage
\noindent
This was done in the form of the adinkra shown\footnote{The image in Fig.\ \ref{Fig:10D} 
is correctly rendered using the data that follows from (\ref{L1}) - (\ref{L4}).  This corrects 
previous such rendering in the work of \cite{counting10d}.}
in Figure (\ref{Fig:10D}).

As we will discuss later in the next section, the conventional superfield 
approach while adequate for extracting the necessary information about the bosonic
reps ${\CMTB{\{ {\cal R}^{(i)}\}}}{}_p$ and fermionic reps ${\CMTR{\{ {\cal R}^{(j)}\}}}{}_p$,
becomes more unwieldy.  The adinkra approach offers a way around this.

The utility of this adinkra graph is that it provides a ``roadmap'' to the writing of 
the explicit form of the action of the supercharge on any particular component field by 
going back to equations (\ref{adnkb}) and (\ref{adnkf}). In the adinkra context, we 
note these imply $ {\CMTB{ c }}{}_{{\CMTR{\{ {\cal R}^{(j)}\}}}}^{+(p){\CMTB {(i)}}  }$,
$ {\CMTB{ c }}{}_{{\CMTR{\{ {\cal R}^{(j)}\}}}}^{-(p){\CMTB {(i)}}  }$, 
$ {\CMTB{ c }}{}_{{\CMTB{\{ {\cal R}^{(i)}\}}}}^{+(p){\CMTR {(j)}}  }$,
and ${\CMTB{ c }}{}_{{\CMTB{\{ {\cal R}^{(i)}\}}}}^{-(p){\CMTR {(j)}}  }$ are determined
by examining properties of the adinkra. In particular, there are four
calculations (implied by (\ref{adnkb}) and (\ref{adnkf}))
to be undertaken and these are respectively
\be {
{\CMTB{ c }}{}_{{\CMTR{\{ {\cal R}^{(j)}\}}}}^{+(p){\CMTB {(i)}}  } ~=~ {\cal F}{}_{1}
\left[ \,  \left(  \, \CMTred{\yng(1)}  \otimes  {\CMTB{\{ {\cal R}^{(i)}\}}}{}_p 
\, \right)   \cap   \, {\CMTR{\{ {\cal R}^{(j)}\}}}{}_{p + 1}    \, \right]  ~~~,
}  \label{L1}
\ee
\be {
 {\CMTB{ c }}{}_{{\CMTR{\{ {\cal R}^{(j)}\}}}}^{-(p){\CMTB {(i)}}  } ~=~ {\cal F}{}_{2}
 \left[ \, \left(  \, \CMTred{\yng(1)}   \otimes {\CMTR{\{ {\cal R}^{(j)}\}}}{}_{p - 1} 
 \, \right) \cap  \, {\CMTB{\{ {\cal R}^{(i)}\}}}{}_p   \, \right]  ~~~,
}  \label{L2}
\ee
\be {
{\CMTB{ c }}{}_{{\CMTB{\{ {\cal R}^{(i)}\}}}}^{+(p){\CMTR {(j)}}  } ~=~{\cal F}{}_{3}
\left[ \, \left(  \,  \CMTred{\yng(1)} \otimes  {\CMTR{\{ {\cal R}^{(j)}\}}}{}_p  
\, \right)   \cap \, {\CMTB{\{ {\cal R}^{(i)}\}}}{}_{p + 1} \, \right] ~~~,
}  \label{L3}
\ee
\be {
{\CMTB{ c }}{}_{{\CMTB{\{ {\cal R}^{(i)}\}}}}^{-(p){\CMTR {(j)}}  } ~=~ {\cal F}{}_{4}
\left[ \,  \left(   \,  \CMTred{\yng(1)} \,  \otimes  \, {\CMTB{\{ {\cal R}^{(i)}\}}}{}_{p - 1} 
 \right)   \cap  \, {\CMTR{\{ {\cal R}^{(j)}\}}}{}_p   \, \right] ~~~,
}  \label{L4}
\ee
where ${\cal F}{}_{1}$, ${\cal F}{}_{2}$, ${\cal F}{}_{3}$, and ${\cal F}{}_{4}$, are functions, 
and $\CMTR{\yng(1)}$ corresponds to the spinor representation.  All of these functions have 
the property that if the intersections indicated as their respective arguments vanish, then the 
functions output the value of zero. This is the reason why in Figure (\ref{Fig:10D}) there are 
some nodes in adjacent levels that are unconnected\footnote{One example of this can be 
seen in the adinkra shown in Fig.\ \ref{Fig:10D} where the $\CMTB{\{770\}}$ at Level-4 
is not linked to the $\CMTred{\{\Bar {672}\}}$ at Level-5.}.  The functions ${\cal F}{}_{1}$, and 
${\cal F}{}_{3}$ yield outputs of the value of one if their respective intersections are non-vanishing. The 
functions ${\cal F}{}_{2}$, and ${\cal F}{}_{4}$ yield outputs of
\be {
{\cal F}{}_{2} ~=~  K_2{}^{p \, {\CMTB {(i)}}  \,  {\CMTR {(j)}} }
~~~, ~~~ 
{\cal F}{}_{4} ~=~  K_4{}^{p \, {\CMTB {(i)}}  \,  {\CMTR {(j)}} }  ~~~,
} \label{L4}
\ee
when their respective arguments are non-vanishing.   

The quantities
$K_2{}^{p \, {\CMTB {(i)}}  \,  {\CMTR {(j)}} }$ and $ K_4{}^{p \, {\CMTB {(i)}}  \,  {\CMTR {(j)}} } $
are normalization constants determined by the conventions used to define the SUSY
algebra, i.e. enforcing the lower condition seen in (\ref{eqn:dminXZ}). The intersection 
principle can only tell us which links must be absent. However, the appearance of 
the links in the adinkra does not {\it {necessarily}} imply the corresponding 
normalization coefficients have to be non-vanishing.  Only detailed calculation
can do this for these 14,667 constants appearing in (\ref{L4}) for the 11D, $\cal N$ = 
1 scalar superfield adinkra.

\newpage
\section{Traditional Path to Superfield Component Decompositions}
\label{sec:analytical}

Before applying the same idea to eleven dimensional superspace, as we did in \cite{counting10d}, the traditional method and its problems need to be discussed. If we start from constructing the irreducible $\theta-$monomials to understand the eleven dimensional scalar superfield decomposition, two uniqueness problems will show up: (1) $\theta-$monomials have multiple expressions from the cubic level; (2) gamma matrix multiplications have multiple expressions. 
To illustrate the first problem, the quadratic level and cubic level will be discussed in detail in the following sections. 
For the second one, all gamma matrix multiplication results are listed in Appendix~\ref{appen:gamma_matrix_mul}. Moreover, constructing irreducible $\theta-$monomials requires a number of Fierz identities as shown shortly. Compared to the group representation approach
embodied by adinkras, the traditional method is much less efficient.

Each higher dimensional superspace with D bosonic dimensions, for purposes of counting 
is equivalent to some value of $d$, which is the number of real components of $\theta$.  This is shown in a few cases below (where $d$ = ${\cal F}({\rm D})$).
\begin{table}[h]
\vspace{0.4cm}
\begin{center}
\begin{tabular}{|c|c|}\hline
$d$ &  D  
\\ \hline \hline
$ ~~4 ~~$ &  $ ~~ 4 ~~$    \\ \hline
$ ~~8 ~~$ &  5   \\ \hline
$ ~~16 ~~$ &    10    \\ \hline
$ ~~32 ~~$ &  11   \\ \hline
\end{tabular}
\end{center}
\caption{Relation Between D (Number of Spacetime) Dimensions and $d$ For Some Superspaces \label{tab:dD}}
\end{table}
\begin{table}[h]
\vspace{0.4cm}
\begin{center}
\begin{tabular}{|c|c|c|c|}\hline
$d$ & $2^d$ &$n_B$ & $n_F$ 
\\ \hline \hline
$ ~~4 ~~$ &  $ ~~ 16 ~~$ &  $ ~~8~~$ &  $ ~~8 ~~$   \\ \hline
$ ~~8 ~~$ &  $ ~~ 256 ~~$ &  $ ~~128~~$ &  $ ~~128 ~~$   \\ \hline
$ ~~16 ~~$ &  $ ~~ 65,536 ~~$ &  $ ~32,768~$ &  $ ~32,768~$   \\ \hline
$ ~~32 ~~$ &  $ ~4,294,967,296~$ &  $ ~2,147,483,648~$ &  $ ~2,147,483,648~$   \\ \hline
\end{tabular}
\end{center}
\caption{Number of Independent Components in Unconstrained Scalar Superfields \label{tab:d2dnn}}
\end{table} 

So for the case of the 11D, $\cal N$ = 1 theory, the real
{\em {unconstrained}} scalar superfield $\Psi$ contains 2,147,483,648 bosonic and 2,147,483,648 fermionic degrees of freedom that are representations of supersymmetry.
While superfields easily provide a methodology for finding collections of components in principle, actually obtaining those component fields is not as
easy as it might first appear.  This is especially true in the eleven dimensional case.  

In the rest of this chapter we are going to discuss the complications
of applying the most straightforward $\theta$-expansions in the eleven 
dimensional superspace.  The discussion is meant to provide an explicit 
demonstration of the difficulties one encounters in such a program.  For 
the reader not interested in these details, it is recommended to skip to 
chapter four.

A naive expansion of a real scalar superfield $\cal V$ can be expressed  as
\be  \eqalign{
{\cal V} (\theta, \, x) ~=&~ \varphi{}^{(0)}(x) ~+~ {\theta}{}^{\a} \, \varphi^{(1)}_{\a}(x) 
~+~ \Theta {}^{(1)} \, \varphi^{(2)}(x) 
~+~  \Theta{}^{(2) \, \un{a}\un{b}\un{c}} \, \varphi^{(2)}_{\un{a}\un{b}\un{c}} (x) 
~+~  \Theta{}^{(3) \, \un{a}\un{b}\un{c}\un{d}} \, \varphi^{(2)}_{ \un{a}\un{b}\un{c}\un{d}} (x) \cr
& ~+~ \Theta {}^{(1)} \,  {\theta}{}^{\a} \,  \varphi^{(3)}_{\a}(x) ~+~  \Theta{}^{(2) \, \un{a}\un{b}\un{c}} 
\, {\theta}{}^{\a} \,\varphi{}^{(3)}_{\a \, \un{a}\un{b}\un{c} }(x) ~+~  \Theta{}^{(3) \, \un{a}\un{b}\un{c} 
\un{d}  } \, {\theta}{}^{\a} \,\varphi{}^{(3)}_{\a \, \un{a}\un{b}\un{c}\un{d} }(x)  \cr
& ~+~ \Theta {}^{(1)} \, \Theta {}^{(1)} \, {\varphi}^{(4)} (x) ~+~ \Theta{}^{(1)}  \, \Theta{}^{(2) \, \un{a}\un{b}\un{c} } \,{\varphi}^{(4)}_{ \un{a}\un{b}\un{c}} (x)  
~+~ \Theta{}^{(1)}  \, \Theta{}^{(3) \, \un{a}\un{b}\un{c}\un{d} } \, {\varphi}^{(4)}_{ \un{a}\un{b}\un{c} \un{d}} (x)   \cr
& ~+~ \Theta{}^{(2) \, \un{a}\un{b}\un{c}}  \, \Theta{}^{(2) \, \un{d}\un{e}\un{f} } \, {\varphi}^{(4)}_{ \un{a}\un{b}\un{c} \, \un{d} \un{e}\un{f} } (x) 
~+~ \Theta{}^{(2) \, \un{a}\un{b}\un{c}}  \, \Theta{
}^{(3) \, \un{d}\un{e}\un{f}\un{g} } \, {\varphi}^{(4)}_{ \un{a}\un{b}\un{c} \, \un{d} \un{e}\un{f}\un{g} 
} (x)   \cr
& ~+~ \Theta{}^{(3) \, \un{a}\un{b}\un{c}\un{d}}  \, \Theta{}^{(3) \, \un{e}\un{f}\un{g}\un{h}} \, {\varphi}^{(4)}_{ \un{a}\un{b}\un{c}\un{d} \, \un{e}\un{f}\un{g}\un{h}} (x) ~+~ \dots
} 
\label{equ:Exp1} \ee
up to the fourth order of the Grassmann coordinates. The number $n$ in the superscripts of component fields $\varphi^{(n)}_{[\text{indices}]} (x)$ indicates the $\theta$-order. In writing this expression we have introduced ``auxiliary nilpotent coordinates'' defined in (\ref{equ:quadratic-theta}) below.  

There exists a recursion formula that
can be applied at any non-trivial order $n$ in the $\theta$-expansion
to derive the form of the terms at order $(n + 1)$ in the $\theta$-expansion.
We can begin this by looking at the term linear in $\theta$,
\be{
{\cal V} (linear) ~=~ \theta{}^{\alpha} {\varphi}{}^{(1)}_{\alpha}(x)
~~~,} 
\label{equ:Exp2} \ee
and next observe the quadratic terms may be generated by a simple replacement
in this expression.
\be{
\varphi^{(1)}_{\alpha}(x) ~\to~ \left[ \, C_{\a\b} \,
{\varphi}^{(2)} (x)
~+~ (\g^{\un{a}\un{b}\un{c}})_{\a\b} \,  {\varphi}^{(2)}_{ \un{a}\un{b} \un{c}} (x)
~+~ (\g^{\un{a}\un{b}\un{c}\un{d}})_{\a\b} \,  {\varphi}^{(2)}_{ \un{a}\un{b}\un{c}\un{d}} (x)
\right] \, \theta{}^{\beta} ~\equiv ~ ({\cal B})_{\a\b} \theta{}^{\beta}
~~~.} 
\label{equ:Exp3} \ee
The quantity $({\cal B})_{\a\b}$ is a Duffin-Kemmer-Petiau \cite{DKP1,DKP2,DKP3}
field.
Therefore under the action of this replacement, we find
\be{
\varphi^{(1)}_{\alpha}(x) ~\to~  ({\cal B})_{\a\b} \theta{}^{\beta}
 \, : \, {\cal V} (linear) ~\to~ {\cal V}(quadratic)
~~~,
\label{equ:Exp4}
} \ee
with ${\cal V} (quadratic)$ given by
\be{
{\cal V} (quadratic) ~=~ \Theta {}^{(1)} \,
{\varphi}{}^{(2)}(x) ~+~  \Theta{}^{(2) \, \un{a}\un{b}\un{c}} \, {\varphi}^{(2)}_{ \un{a}\un{b}\un{c}} (x) ~+~  \Theta{}^{(3) \, \un{a}\un{b}\un{c}\un{d}} \, {\varphi}^{(2)}_{ \un{a}\un{b}\un{c}\un{d}} (x)
\label{equ:Exp4a} ~~~.
} \ee

To continue, we take the component fields
${\varphi}^{(2)} (x)$, ${\varphi}^{(2)}_{ \un{a}\un{b} \un{c}} (x)$, ${\varphi}^{(2)}_{ \un{a}\un{b}\un{c}\un{d}} (x)$ and make the
simultaneous replacements
\be{  
{\varphi}^{(2)} (x) ~\to~ \theta{}^{\alpha}
{\varphi}^{(3)}_{\alpha} (x)
~~,~~
{\varphi}^{(2)}_{ \un{a}\un{b} \un{c}} (x)
~\to~ \theta{}^{\alpha}
{\varphi}^{(3)}_{{\alpha} \, {\un a} {\un b} {\un c}} (x)
~~,~~
{\varphi}^{(2)}_{ \un{a}\un{b}\un{c}\un{d}} (x)
~\to~ \theta{}^{\alpha}
{\varphi}^{(3)}_{{\alpha} \, {\un a} {\un b} {\un c}
{\un d}} (x)
} \ee
which yield the cubic order terms 
\be{
{\cal V} (cubic) ~=~
\Theta {}^{(1)} \,  {\theta}{}^{\a} \,  \varphi{}^{(3)}_{\a}(x) ~+~  \Theta{}^{(2) \, \un{a}\un{b}\un{c}} 
\, {\theta}{}^{\a} \,\varphi{}^{(3)}_{\a \, \un{a}\un{b}\un{c} }(x) ~+~  \Theta{}^{(3) \, \un{a}\un{b}\un{c} 
\un{d}  } \, {\theta}{}^{\a} \,\varphi{}^{(3)}_{\a \, \un{a}\un{b}\un{c}\un{d} }(x)
\label{equ:Exp5}
} \ee
that appear on the second line of (\ref{equ:Exp1}).

The general rule is that if one starts with the component fields at Level-$n$, {\it {where 
$n$ is even}}, of the scalar superfield, then to obtain the component fields at Level-$(n + 1)$ 
one simply replaces the starting component fields by a $\theta$-coordinate whose index is 
contracted against a new fermionic fields in a manner that is consistent with Lorentz symmetry.

Also a general rule is that if one starts with the component fields at Level-$n$, {\it {where $n$ 
is odd}}, of the scalar superfield, then to obtain the component fields at Level-$(n + 1)$ one 
simply replaces the starting component fields by a new DKP field times a $\theta$-coordinate 
whose index is contracted against one index on new DKP fields in a manner that is consistent 
with Lorentz symmetry.

Although one can carry out this procedure to define the component fields to all orders in the 
$\theta$-expansion, it is highly inefficient and redundant.  This redundancy occurs due to the 
equivalence of many terms obtained as well as the vanishing of many terms both by the use of 
Fierz identities.  There is also the issue of irreducibility that must be enforced.  We next turn 
to the issue of irreducibility.

\subsection{Quadratic Level}
\label{sec:Q4}

We denote the 32-component Majorana Grassmann coordinate living in 11 dimensional spacetime
by $\theta^{\a}$. Since $C_{\a\b}$, $(\g^{[3]})_{\a\b}$ and $(\g^{[4]})_{\a\b}$ are the antisymmetric elements in 
the covering Clifford algebra over 11D, we can define all possible quadratic $\theta$-monomials as follows.
\begin{equation}
\begin{aligned}
    & \CMTB {\{1\} } & ~~~ \Theta^{(1)} ~=&~ C_{\a\b} \, \theta^{\a} \theta^{\b}  ~~~, \\
    & \CMTB {\{165\}}  & ~~~ \Theta^{(2) \, \un{a}\un{b}\un{c}} ~=&~ (\g^{\un{a}\un{b}\un{c}})_{\a\b} \, \theta^{\a} \theta^{\b} ~~~,  \\
    & \CMTB {\{330\}}  & ~~~ \Theta^{(3) \, \un{a}\un{b}\un{c}\un{d}} ~=&~ (\g^{\un{a}\un{b}\un{c}\un{d}})_{\a\b} \, \theta^{\a} \theta^{\b}    ~~~.
\end{aligned} \label{equ:quadratic-theta}
\end{equation}

Now if we look at the quadratic $\theta$-terms, we see the total number of 
bosonic component fields at this level can be found by simply counting the number of independent quadratic $\theta$-monomials
\begin{equation}
    \CMTred {\{ 32 \}} \, \wedge \, \CMTred {\{ 32 \}} ~=~ \frac{ \{32\} ~\times~ \{  31\} }{2}  ~=~ \{496\} ~=~ \CMTB {\{ 1\}} ~\oplus~ \CMTB {\{165\} } ~\oplus~ \CMTB {\{330\} } ~~~,
\end{equation}
so all is well as this equation gives the complete decomposition of the
product of two Grassmann coordinates into irreducible representations
of the 11D Lorentz group.

\subsection{Cubic Level}
\label{sec:cubic}

We can construct cubic $\theta$-monomials from all the possible quadratic $\theta$-monomials as listed in Equation (\ref{equ:quadratic-theta}). Since $C_{\a\b}$, $(\g^{[3]})_{\a\b}$ and $(\g^{[4]})_{\a\b}$ are the antisymmetric Clifford algebra elements in 11D, we can write all the possible cubic monomials starting with no free Lorentz vector index and going up to four free Lorentz vector indices. All of the possible {\emph {irreducible}} cubic $\theta$-monomials can be written as 
\be
\eqalign{
    \CMTred {\{5,280\}} ~~~ & ~~~ \big[ \, \Theta{}^{(3) \, \un{a}\un{b}\un{c}\un{d}}  \, \theta{}_{\a}  \, \big]_{IR} 
    ~~~,  \cr
    \CMTred {\{ 3,520\}} ~~~ & ~~~ \big[ \, \Theta{}^{(3) \, \un{a}\un{b}\un{c}\un{d}}\,  (\g{}_{\un{d}}){}_{\a\b}  \, \theta{}^{\b} \, \big]_{IR}  ~~~~,~~~ 
    \big[ \, \Theta{}^{(2) \, \un{a}\un{b}\un{c}}\,  \theta{}_{\a}  \, \big]_{IR} ~~~,  \cr
    \CMTred {\{  1,408\}} ~~~ & ~~~ \big[ \, \Theta{}^{(3) \, \un{a}\un{b}\un{c}\un{d}}\,  (\g{}_{\un{c}\un{d}}){}_{\a\b}  \, \theta{}^{\b} \, \big]_{IR}  ~~~~,~~~
    \big[ \, \Theta{}^{(2) \, \un{a}\un{b}\un{c}}\,  (\g_{\un{c}}){}_{\a\b}  \, \theta{}^{\b} \, \big]_{IR} ~~~,    \cr
    \CMTred {\{  320\}} ~~~ & ~~~ \big[ \, \Theta{}^{(3) \, \un{a}\un{b}\un{c}\un{d}}\,  (\g{}_{\un{b}\un{c}\un{d}}){}_{\a\b}  \, \theta{}^{\b} \, \big]_{IR} ~~~~,~~~ 
    \big[ \, \Theta{}^{(2) \, \un{a}\un{b}\un{c}}  \,  (\g_{\un{b}\un{c}}){}_{\a\b}  \, \theta{}^{\b} \, \big]_{IR}  ~~~, \cr
    \CMTred {\{ 32\}} ~~~ & ~~~ \Theta{}^{(3) \, \un{a}\un{b}\un{c}\un{d}}\,  (\g_{\un{a}\un{b}\un{c}\un{d}}){}_{\a\b}  \, \theta{}^{\b}  ~~~~,~~~ 
    \Theta{}^{(2) \, \un{a}\un{b}\un{c}}  \,  (\g_{\un{a}\un{b}\un{c}}){}_{\a\b}  \, \theta{}^{\b}   ~~~~,~~~ 
    \Theta {}^{(1)} \, \theta{}_{\a}   ~~~.
\label{equ:cubicirr} }
\ee
where the notation $\big[ ~~ \big]_{IR}$ simply means that a single $\g$-trace of the expression is by definition equal to zero. We will discuss each representation (except $ \CMTred {\{5,280\}}$) one by one in the following subsections. We will explain each dimension from the corresponding irreducibility condition, and prove that different versions in each representation are actually equivalent. We will show that the cubic monomials of $ \CMTred {\{320\}}$ vanish. We argue that $ \CMTred {\{5,280\}}$ cubic monomials also vanish in a similar way. Another strong reason for $ \CMTred {\{5,280\}}$ to vanish is $5,280 > 4,960$ (see equation below). We can then decompose all the cubic $\theta$-monomials by
\begin{equation}
    \CMTred { \{ 32 \}} \, \wedge \,  \CMTred {\{ 32 \}} \, \wedge \,  \CMTred {\{ 32 \}} ~=~ \frac{ \{32\} ~\times~ \{31\} ~\times~ \{30\} }{3 ~\times~ 2} ~=~ \{ 4,960\} ~=~  \CMTred {\{32\}} ~\oplus~  \CMTred {\{1,408\}}  ~\oplus~  \CMTred {\{3,520\}}  ~~~,  \label{equ:cubicdecompose}
\end{equation}
where the left hand side simply counts the number of independent cubic $\theta$-monomials one can write, and the rightmost part contains $ \CMTred {\{32\}}$, $ \CMTred {\{1,408\}}$, and $ \CMTred {\{3,520\}}$ which are irreducible representations of the 11D Lorentz group as shown in Appendix \ref{appen:so11}.

\subsubsection{$\CMTred {\{32\}}$ Cubic Monomials}

We have three versions of expressions of cubic $\theta$-monomials with no free vector index and one free spinor index as indicated in Equation (\ref{equ:cubicirr}), which are
\begin{align}
    V1 ~=&~ \Theta^{(3) \, \un{a}\un{b}\un{c}\un{d} } \,  (\g_{\un{a}\un{b}\un{c}\un{d}})_{\a\b} \, \theta^{\b} ~=~ - \frac{1}{3!}\, ( \mathcal{A}_3 )_{[\d\e\b]\a} \,  \theta^{\d}\theta^{\e}\theta^{\b}  ~~~, \\
    V2 ~=&~ \Theta^{(2) \, \un{a}\un{b}\un{c} }\,  (\g_{\un{a}\un{b}\un{c}})_{\a\b}  \, \theta^{\b} ~=~ - \frac{1}{3!}\, ( \mathcal{A}_2 )_{[\d\e\b]\a} \, \theta^{\d}\theta^{\e}\theta^{\b} ~~~,  \\
    V3 ~=&~ \Theta^{(1)}  \, \theta_{\a} ~=~ \frac{1}{3!}\, ( \mathcal{A}_1 )_{[\d\e\b]\a} \, \theta^{\d}\theta^{\e}\theta^{\b} ~~~.
\end{align}
The degrees of freedom of these monomials are thus 32, and so they are in the spinorial representation $\CMTred {\{32\}}$. Here we define these three objects
\begin{equation}
\begin{split}
    ( \mathcal{A}_1 )_{[\d\e\b]\a}  ~=&~ C_{[\d\e}\, C_{\b]\a} ~~~, \\
    ( \mathcal{A}_2 )_{[\d\e\b]\a}  ~=&~ (\g^{ \un{a}\un{b}\un{c} })_{[\d\e}\,  (\g_{\un{a}\un{b}\un{c}})_{\b]\a} 
    ~~~, \\
    ( \mathcal{A}_3 )_{[\d\e\b]\a}  ~=&~ (\g^{ \un{a}\un{b}\un{c}\un{d} })_{[\d\e} \,  (\g_{\un{a}\un{b}\un{c}\un{d}})_{\b]\a} ~~~.
    \label{equ:Aobjects}
\end{split}
\end{equation}

To find whether $V1$, $V2$, and $V3$ are related, we examine the objects $\mathcal{A}_{1}$, $\mathcal{A}_{2}$ and $\mathcal{A}_{3}$. Since $C_{\a\b}$, $(\g^{[3]})_{\a\b}$ and $(\g^{[4]})_{\a\b}$ form the complete basis for the antisymmetric elements in 11D Clifford algebra with two spinor indices, we can expand the $\mathcal{A}$ objects into this basis, i.e. find the Fierz identities. The Fierz identities in Appendix \ref{appen:cubicfierz} tell us that the objects $\mathcal{A}_{1}$, $\mathcal{A}_{2}$ and $\mathcal{A}_{3}$ are related by a system of linear equations (we suppress spinor indices here for simplicity, as all of them have the same structure),
\begin{equation}
\begin{split}
    31\mathcal{A}_1 ~=&~ \frac{1}{3!}\mathcal{A}_2 - \frac{1}{4!}\mathcal{A}_3  ~~~, \\
    37\mathcal{A}_2 ~=&~ 990\mathcal{A}_1 -\frac{11}{4}\mathcal{A}_3 ~~~,  \\
    13\mathcal{A}_3 ~=&~ -3960\mathcal{A}_1 -44\mathcal{A}_2   ~~.
\end{split}
\end{equation}
This makes sense as $\mathcal{A}_{1}$, $\mathcal{A}_{2}$ and $\mathcal{A}_{3}$ are the only three objects of this spinor index structure with no free vector indices and at least one Clifford element being antisymmetric. By solving these linear equations, we get 
\begin{equation}
\label{equ:32relations}
\left\{
    \begin{aligned}
    \mathcal{A}_2 ~=&~ 66\mathcal{A}_1 ~~~, \\
    \mathcal{A}_3 ~=&~ -528\mathcal{A}_1 ~~~, \\
    \end{aligned}
    \right.
\end{equation}
which means all three of the expressions of the cubic $\theta$-monomials are equivalent up to a multiplicative constant
\begin{equation}
    V1 ~=~ -8 \, V2 ~=~ 528 \, V3  ~~~.
\end{equation}
We can therefore take any of the cubic $\theta$-monomial constructed from linear combinations of $V1$, $V2$ and $V3$ as the fermionic irreducible representation $\CMTred{\{32\}}$ of so(11).

\subsubsection{$ \CMTred {\{320\}}$ Cubic Monomials}

For one free vector index, we have two expressions of cubic $\theta$-monomials as suggested in Equation (\ref{equ:cubicirr}). We can write all the terms from the index structures as
\begin{align}
    \big[ \, \Theta^{(3) \, \un{a}\un{b}\un{c}\un{d} }\,  (\g_{\un{b}\un{c}\un{d}})_{\a\b}  \, \theta^{\b} \, \big]_{IR}  ~=&~  \, \tilde{k}_0 \, \Big\{   \, \Theta^{(3) \, \un{a}\un{b}\un{c}\un{d}} \,  (\g_{\un{b}\un{c}\un{d}})_{\a\b}  \, \theta^{\b}  ~+~ k_{1} \, \Theta^{(3) \,  \un{b} \un{c} \un{d}\un{e}}\, ( \g^{\un{a}}{}_{\un{b}\un{c} \un{d} \un{e}})_{\a\b}  \, \theta^{\b}  \Big\}  ~~~,   \\
    \big[ \, \Theta^{(2) \, \un{a}\un{b}\un{c} }\,  (\g_{\un{b}\un{c}})_{\a\b}  \, \theta{}^{\b} \, \big]_{IR}  ~=&~  \,  \tilde{l}_0  \, \Big\{ \, \Theta^{(2) \, \un{a}\un{b}\un{c} }\,  (\g_{\un{b}\un{c}})_{\a\b}  \, \theta^{\b}  ~+~ l_{1} \, \Theta^{(2) \,   \un{b} \un{c} \un{d} } \, ( \g^{ \un{a} }_{\ \un{b}\un{c}\un{d}})_{\a\b}  \, \theta^{\b}   \Big\} ~~~.
\end{align}
The irreducibility conditions of setting the single $\gamma$-traces to zero are
\begin{align}
    (\g_{\un a})_{\a}^{\ \g}\big[ \, \Theta{}^{(3) \, \un{a}\un{b}\un{c}\un{d} }\,  (\g_{\un{b}\un{c}\un{d}}){}_{\g\b}  \, \theta{}^{\b} \, \big]_{IR}  ~=&~ 0    ~~~,
    \label{equ:320irr1} \\
    (\g_{\un a})_{\a}^{\ \g}\big[ \, \Theta{}^{(2) \, \un{a}\un{b}\un{c} }\,  (\g_{\un{b}\un{c}}){}_{\g\b}  \, \theta{}^{\b} \, \big]_{IR} ~=&~ 0  ~~~.
    \label{equ:320irr2}
\end{align}
Thus, these cubic monomials with one free vector index have $32 \times 11 - 32 = 320$ degrees of freedom and are in the $ \CMTred {\{320\}}$ representation. From the irreducibility conditions, we can fix the relative coefficients to $k_{1} = - \frac{1}{7}$ and $l_{1} = - \frac{1}{8}$. Without loss of generality, we omit the overall coefficients $\tilde{k}_{0}$ and $\tilde{l}_{0}$. Therefore, we can write
\begin{align}
    V1 ~=&~ - \frac{1}{3!} \big( \mathcal{D}_{1} ~+~ \fracm{1}{7} \, \mathcal{D}_{2} \big)^{\un{a}}_{[\d\e\b]\a} \theta^{\d}\theta^{\e}\theta^{\b}    
    \label{equ:320v1}   ~~~, \\
    V2  ~=&~  \frac{1}{3!} \big( \mathcal{C}_{1}   ~+~ \fracm{1}{8} \, \mathcal{C}_{2} \big)^{\un{a}}_{[\d\e\b]\a} \theta^{\d}\theta^{\e}\theta^{\b}  ~~~,
    \label{equ:320v2}
\end{align}
where we define five objects
\begin{equation}
\begin{split}
    ( \mathcal{B} )^{\un{a}}_{[\d\e\b]\a}  ~=&~  C_{[\d\e} (\g^{\un{a}})_{\b]\a}  ~~~, \\
    ( \mathcal{C}_1 )^{\un{a}}_{[\d\e\b]\a}  ~=&~  (\g^{\un{a}[2]})_{[\d\e}(\g_{[2]})_{\b]\a} ~~~,   \\
    ( \mathcal{C}_2 )^{\un{a}}_{[\d\e\b]\a}  ~=&~  (\g^{[3]})_{[\d\e}(\g^{\un{a}}{}_{[3]})_{\b]\a} ~~~,  \\
    ( \mathcal{D}_1 )^{\un{a}}_{[\d\e\b]]\a}  ~=&~  (\g^{\un{a}[3]})_{[\d\e}(\g_{[3]})_{\b]\a} ~~~,  \\
    ( \mathcal{D}_2 )^{\un{a}}_{[\d\e\b]\a}  ~=&~  (\g^{[4]})_{[\d\e}(\g^{\un a}{}_{[4]})_{\b]\a} ~~~.
    \label{equ:BCDobjects}
\end{split}
\end{equation}
When we consider all the objects with one free vector index constructed by two Clifford algebra basis elements with one of them being antisymmetric, we find the additional object $\mathcal{B}$ as defined above. From our past experience, we know that it has to occur in our Fierz expansions. The relevant Fierz identities are listed in Appendix \ref{appen:cubicfierz}.
They can be rewritten into a system of linear equations of objects $\mathcal{B}$, $\mathcal{C}_{1}$, $\mathcal{C}_{2}$, $\mathcal{D}_{1}$, and $\mathcal{D}_{2}$ (vector and spinor indices suppressed) as
\begin{equation}
\begin{split}
    33\mathcal{B} ~=&~ -\frac{1}{3!} \mathcal{C}_2 + \frac{1}{2}\mathcal{C}_1-\frac{1}{4!} \mathcal{D}_2 + \frac{1}{3!}\mathcal{D}_1 ~~~, \\
    90 \mathcal{C}_1 ~=&~ 180\mathcal{B} -2\mathcal{C}_2-\frac{1}{2}\mathcal{D}_2 + 2\mathcal{D}_1 ~~~, \\
    5\mathcal{C}_2 ~=&~ -90\mathcal{B} -3\mathcal{C}_1 +\frac{1}{4} \mathcal{D}_2 - \mathcal{D}_1 ~~~, \\
    5\mathcal{D}_1 ~=&~ 90\mathcal{B} - \mathcal{C}_2 + 3\mathcal{C}_1 -\frac{1}{4}\mathcal{D}_2 ~~~, \\
    17\mathcal{D}_2 ~=&~ -2520\mathcal{B} + 28 \mathcal{C}_2 -84\mathcal{C}_1 -28 \mathcal{D}_1 ~~~.
\end{split}
\end{equation}
The solutions are 
\begin{equation}
\label{equ:320relations}
\left\{
\begin{aligned}
    \mathcal{D}_1 ~=&~ - \frac{1}{7} \mathcal{D}_{2}  ~=~  48 \mathcal{B}  ~~~,  \\
    \mathcal{C}_1 ~=&~ - \frac{1}{8} \mathcal{C}_{2}  ~=~  6\mathcal{B} ~~~.
\end{aligned}
\right.
\end{equation}
Therefore, from Equations (\ref{equ:320v1}) and (\ref{equ:320v2}), it is very clear that 
\begin{equation}
    V1 ~=~ V2 ~=~ 0 ~~~.
\end{equation}
This suggests that there exists no cubic $\theta$-monomial in the $ \CMTred {\{320\}}$ irreducible representation of so(11). 

Another way of seeing that $ \CMTred {\{320\}}$ cubic $\theta$-monomials do not exist is the above constructed monomials fail to satisfy the irreducibility condition. From Equation (\ref{equ:320relations}), we see that $V1$ and $V2$ are clearly proportional to $\mathcal{B}$, for example. The irreducibility conditions in (\ref{equ:320irr1}) and (\ref{equ:320irr2}) thus read
\begin{equation}
\begin{split}
    (\g_{\un a})_{\a}^{\ \g} \, ( \mathcal{B} )^{\un{a}}_{[\d\e\b]\g} ~=~ (\g_{\un a})_{\a}^{\ \g} \, C_{[\d\e} (\g^{\un{a}})_{\b]\g} ~=~
    -11C_{[\d\e}C_{\b]\a} ~=~&  - 11 ( \mathcal{A}_{1} )_{[\d\e\b]\a} 
    ~ \neq~ 0  ~~~,
    \label{equ:320irrcheck}
\end{split}
\end{equation}
as $\mathcal{A}_{1} \neq 0$ numerically (otherwise, the $ \CMTred {\{32\}}$ cubic monomials would also vanish). Doing this with $\mathcal{C}$'s or $\mathcal{D}$'s will give us other linear combinations of $\mathcal{A}$'s, which would not vanish also as all $\mathcal{A}$'s are proportional to $\mathcal{A}_{1}$.

\subsubsection{$ \CMTred {\{1,408\}}$ Cubic Monomials}

There are two versions of expressions of cubic $\theta$-monomials with two antisymmetric vector indices as listed in Equation (\ref{equ:cubicirr}). They can be expanded in the following basis
\begin{align}
\begin{split}
    \big[ \, \Theta^{(3) \, \un{a}\un{b}\un{c}\un{d} }\,  (\g_{\un{c}\un{d}})_{\a\b}  \, \theta^{\b} \, \big]_{IR}  ~=&~ \tilde{g}_0  \, \Big\{ \Theta^{(3) \, \un{a}\un{b}\un{c}\un{d} }\,  (\g_{\un{c}\un{d}})_{\a\b}  \, \theta^{\b}  ~-~ \fracm 1 {7} \, \Theta{}^{(3) \, [  \un{a} | \un{c} \un{d}\un{e}} \, ( \g^{| \un{b} ]} {}_{\un{c} \un{d} \un{e}})_{\a\b}  \, \theta^{\b}  \\
    & \qquad ~-~  \fracm4{7} \,\fracm 1{5!4!} \,\e^{ [5] \un{a} \un{b} \un{c} \un{d} \un{e} \un{f} }  \,  \Theta{}^{(3)}_{\un{c} \un{d}\un{e} \un{f}}\,  ( \g_{ [5] } )_{\a\b} \, \theta^{\b}  \Big\}  ~~~,
\end{split}  \\
\begin{split}
    \big[ \, \Theta{}^{(2) \, \un{a}\un{b}\un{c} }\,  (\g_{\un{c}})_{\a\b}  \, \theta^{\b} \, \big]_{IR}  ~=&~ \tilde{h}_0 \, \Big\{ \Theta^{(2) \, \un{a}\un{b}\un{c} }\,  (\g_{\un{c}})_{\a\b}  \, \theta^{\b}  ~-~ \fracm 1 {8} \, \Theta^{(2) \, [  \un{a} | \un{c}\un{d} }\,  ( \g^{| \un{b} ]}{}_{\un{c} \un{d}})_{\a\b}  \, \theta^{\b}  \\
    & \qquad ~-~  \fracm1{56} \Theta^{(2) \, \un{c}\un{d}\un{e} }\, (\g^{\un{a}\un{b}}{}_{\un{c}\un{d}\un{e}}){}_{\a\b}  \, \theta^{\b}  \Big\} ~~~,
\end{split}
\end{align}
where the relative coefficients are fixed by the irreducibility conditions
\begin{align}
    (\g_{\un b})_{\a}^{\ \g}\big[ \, \Theta{}^{(3) \, \un{a}\un{b}\un{c}\un{d} }\,  (\g_{\un{c}\un{d}}){}_{\g\b}  \, \theta{}^{\b} \, \big]_{IR}  ~=&~ 0    ~~~,
    \label{equ:1408irr1} \\
    (\g_{\un b})_{\a}^{\ \g}\big[ \, \Theta{}^{(2) \, \un{a}\un{b}\un{c} }\,  (\g_{\un{c}}){}_{\g\b}  \, \theta{}^{\b} \, \big]_{IR} ~=&~ 0 ~~~.
    \label{equ:1408irr2}
\end{align}
From the conditions we know that these cubic $\theta$-monomials have $32 \times \frac{11 \times 10}{2} - 32 \times 11 = 1,408$ degrees of freedom, and thus they live in the $ \CMTred {\{1,408\}}$ representation. By omitting the overall coefficients $\tilde{g}_{0}$ and $\tilde{h}_{0}$, we can write the two versions as
\begin{align}
\begin{split}
    V1 ~=&~ \frac{1}{3!} \big( \mathcal{G}_{1} ~-~ \fracm{1}{7} \, \mathcal{G}_{2} ~-~ \fracm{4}{7} \, \mathcal{G}_{3} \big)^{\un{a}\un{b}}_{[\d\e\b]\a} \theta^{\d}\theta^{\e}\theta^{\b}
\end{split}  ~~~,  \\
\begin{split}
    V2 ~=&~ \frac{1}{3!} \big( \mathcal{F}_{1} ~+~ \fracm{1}{8} \, \mathcal{F}_{2} ~-~ \fracm1{56} \mathcal{F}_{3} \big)^{\un{a}\un{b}}_{[\d\e\b]\a} \theta^{\d}\theta^{\e}\theta^{\b} ~~~,
\end{split}
\end{align}
where we define seven objects
\begin{equation}
\begin{split}
    ( \mathcal{E} )^{\un{a}\un{b}}_{[\d\e\b]\a} ~=&~ C_{[\d\e}(\g^{\un{a}\un{b}})_{\b]\a} ~~~,  \\
    ( \mathcal{F}_{1} )^{\un{a}\un{b}}_{[\d\e\b]\a} ~=&~ (\g^{\un{a}\un{b}[1]})_{[\d\e}(\g_{[1]})_{\b]\a}  ~~~, \\
    ( \mathcal{F}_{2} )^{\un{a}\un{b}}_{[\d\e\b]\a} ~=&~ (\g^{[2][\un a})_{[\d\e}(\g^{\un b]}{}_{[2]})_{\b]\a}  ~~~, \\
    ( \mathcal{F}_{3} )^{\un{a}\un{b}}_{[\d\e\b]\a} ~=&~ (\g^{[3]})_{[\d\e}(\g^{\un{a}\un{b}}{}_{[3]})_{\b]\a}  ~~~, \\
    ( \mathcal{G}_{1} )^{\un{a}\un{b}}_{[\d\e\b]\a} ~=&~ (\g^{\un{a}\un{b}[2]})_{[\d\e}(\g_{[2]})_{\b]\a}  ~~~,\\
    ( \mathcal{G}_{2} )^{\un{a}\un{b}}_{[\d\e\b]\a} ~=&~ (\g^{[3][\un a})_{[\d\e}(\g^{\un b]}{}_{[3]})_{\b]\a} ~~~, \\
    ( \mathcal{G}_{3} )^{\un{a}\un{b}}_{[\d\e\b]\a} ~=&~ \frac{1}{4!5!} \e^{\un{a}\un{b}}{}_{[4][5]} (\g^{[4]})_{[\d\e}(\g^{[5]})_{\b]\a} ~~~.
    \label{equ:EFGobjects}
\end{split}
\end{equation}
An additional object $\mathcal{E}$ is defined to span the entire basis, which plays a similar role to $\mathcal{B}$ in the $ \CMTred {\{320\}}$ representation. The Fierz expansions of all these objects as listed in Appendix \ref{appen:cubicfierz} give us the system of linear equations
\begin{equation}
\begin{split}
    5\mathcal{G}_1 ~=&~ 9\mathcal{E} + \mathcal{F}_2 +5\mathcal{F}_1 -\mathcal{G}_3 ~~~, \\
     2\mathcal{G}_2 ~=&~ 63\mathcal{E} + \mathcal{F}_3 -21\mathcal{F}_1 - 3\mathcal{G}_3\\
    50 \mathcal{F}_1 ~=&~ -18\mathcal{E} -\mathcal{F}_3 +5\mathcal{F}_2 -\mathcal{G}_2 + 5\mathcal{G}_1 -2\mathcal{G}_3 ~~~, \\
    2 \mathcal{F}_2 ~=&~ 9\mathcal{E} +5 \mathcal{F}_1+\mathcal{G}_1 -\mathcal{G}_3 ~~~, \\
    5 \mathcal{F}_3 ~=&~ 63\mathcal{E} -21\mathcal{F}_1 + \mathcal{G}_2 -3\mathcal{G}_3 ~~~, \\
    33\mathcal{E} ~=&~ \frac{1}{3!}\mathcal{F}_3 + \frac{1}{2}\mathcal{F}_2 -\mathcal{F}_1 -\mathcal{G}_3 + \frac{1}{3!}\mathcal{G}_2 + \frac{1}{2}\mathcal{G}_1 ~~~.
\end{split}
\end{equation}
By solving these linear equations, we obtain
\begin{equation}
\label{equ:1408relations}
\left\{
    \begin{aligned}
    \mathcal{F}_2 ~=&~ 12\mathcal{E} + 4\mathcal{F}_1 ~~~,  \\
    \mathcal{F}_3 ~=&~ 30\mathcal{E} - 6\mathcal{F}_1  ~~~, \\
    \mathcal{G}_1 ~=&~ 6\mathcal{E} + 2\mathcal{F}_1  ~~~, \\
    \mathcal{G}_2 ~=&~ 60\mathcal{E} - 12\mathcal{F}_1 ~~~, \\
    \mathcal{G}_3 ~=&~ -9\mathcal{E} - \mathcal{F}_1 ~~~,
    \end{aligned}
    \right.
\end{equation}
which means
\begin{align}
    V1 ~=&~ \frac{1}{7} \big( 3\mathcal{E} + 5\mathcal{F}_1 \big)^{\un{a}\un{b}}_{[\d\e\b]\a} \theta^{\d}\theta^{\e}\theta^{\b} ~~~, \\
    V2 ~=&~ \frac{3}{56} \big( 3\mathcal{E} + 5\mathcal{F}_1 \big)^{\un{a}\un{b}}_{[\d\e\b]\a} \theta^{\d}\theta^{\e}\theta^{\b} ~~~.
\end{align}
It is of note that here we also have the freedom to choose any other two objects as our basis. For example, we can choose $\mathcal{F}_1$ and $\mathcal{F}_2$ instead. Then we will find
\begin{align}
    V1 ~=&~ \frac{1}{28} \big( 16\mathcal{F}_1 + \mathcal{F}_2 \big)^{\un{a}\un{b}}_{[\d\e\b]\a} \theta^{\d}\theta^{\e}\theta^{\b} ~~~,  \\
    V2 ~=&~ \frac{3}{224} \big( 16\mathcal{F}_1 + \mathcal{F}_2 \big)^{\un{a}\un{b}}_{[\d\e\b]\a} \theta^{\d}\theta^{\e}\theta^{\b} ~~~,
\end{align}
$V1$ and $V2$ are always proportional to each other,
\begin{equation}
    V1 = \frac{8}{3} V2 ~~~,
\end{equation}
thus they are equivalent and there's only one $ \CMTred {\{1,408\}}$ irreducible representation sitting in the cubic sector.

Let's check the irreducibility condition. If we choose $\mathcal{E}$ and $\mathcal{F}_1$ as the basis, the conditions in (\ref{equ:1408irr1}) and (\ref{equ:1408irr2}) translate to
\begin{equation}
    (\g_{\un b})_{\a}{}^{\g} \big( 3\mathcal{E} + 5\mathcal{F}_1 \big)^{\un{a}\un{b}}_{[\d\e\b]\a} \theta^{\d}\theta^{\e}\theta^{\b} = 0 ~~~.
\end{equation}
After some quick calculations, we can simplify this condition as
\begin{equation}
    -30 C_{[\d\e}(\g^{\un{a}})_{\b]\a} + 5 (\g^{\un{a}[2]})_{[\d\e}(\g_{[2]})_{\b]\a} = \big( -30 \mathcal{B} + 5 \mathcal{C}_1 \big)^{\un{a}}_{[\d\e\b]\a} = 0 ~~~,
    \label{equ:1408irrcheck}
\end{equation}
which is exactly satisfied by Equation (\ref{equ:320relations}). If we choose another basis, like $\mathcal{F}_1$ and $\mathcal{F}_2$, and simplify the irreducible condition, we will get the relation between objects $\mathcal{B}$ and $\mathcal{D}_2$ in Equation (\ref{equ:320relations}). 

With the experiences of checking the irreducibility conditions in $ \CMTred {\{320\}}$ and $ \CMTred {\{1,408\}}$, we observe a nested structure. In Equations (\ref{equ:Aobjects}), (\ref{equ:BCDobjects}) and (\ref{equ:EFGobjects}), we considered all the objects constructed by two Clifford elements with one of them being antisymmetric on spinor indices, and with zero, one and two free vector indices for $ \CMTred {\{32\}}$, $ \CMTred {\{320\}}$ and $ \CMTred {\{1,408\}}$ respectively. Applying an irreducibility condition involves doing a single $\gamma$-trace, which contracts one vector index out. Therefore, the irreducibility condition of $ \CMTred {\{320\}}$ in Equation (\ref{equ:320irrcheck}) can be written in terms of objects in $ \CMTred {\{32\}}$, and different versions of irreducibility conditions of $ \CMTred {\{1,408\}}$ (such as Equation (\ref{equ:1408irrcheck})) can be written in terms of objects in $ \CMTred {\{320\}}$, which in turn give us the relations between the $ \CMTred {\{320\}}$ objects in Equation (\ref{equ:320relations}). 

We can then comment further on why $ \CMTred {\{320\}}$ cubic monomials must vanish. We observe that there is one non-vanishing independent object in $ \CMTred {\{32\}}$, therefore the irreducibility condition in $ \CMTred {\{320\}}$ implies that there are more than one independent objects in $ \CMTred {\{320\}}$. Meanwhile, there are two independent objects in $ \CMTred {\{1,408\}}$, and the irreducibility condition in $ \CMTred {\{1,408\}}$ implies that there is only one independent object in $ \CMTred {\{320\}}$. Thus, we reach a contradiction. Therefore, sandwiching from $ \CMTred {\{32\}}$ and $ \CMTred {\{1,408\}}$ would force the $ \CMTred {\{320\}}$ objects to vanish.

Following this line of logical arguments, the next representation with three free vector indices, $ \CMTred {\{3,520\}}$, would have to have three independent objects if not vanishing, and its irreducibility conditions should reduce to the relations between objects in $ \CMTred {\{1,408\}}$ in Equation (\ref{equ:1408relations}), as we will see.

\subsubsection{$ \CMTred {\{3,520\}}$ Cubic Monomials}

The two versions of cubic $\theta$-monomials with three totally antisymmetric vector indices in Equation (\ref{equ:cubicirr}) can be expressed as
\begin{align}
\begin{split}
    \big[ \, \Theta^{(3) \, \un{a}\un{b}\un{c}\un{d} }\,  (\g_{\un{d}})_{\a\b}  \, \theta^{\b} \, \big]_{IR}  ~=&~ \tilde{m}_0 \, \Big\{  \Theta^{(3) \, \un{a}\un{b}\un{c}\un{d} }\,  (\g_{\un{d}})_{\a\b}  \, \theta^{\b}  ~-~ \fracm 1 {14} \, \Theta^{(3) \, [  \un{a}\un{b}| \un{d}\un{e}}\, ( \g^{| \un{c} ]} {}_{\un{d}\un{e}})_{\a\b}  \, \theta^{\b}  \\
    & \qquad ~-~  \fracm 1{84} \,  \Theta^{(3) \, [\un{a}| \un{d}\un{e}\un{f} } \, ( \g^{|\un{b}\un{c}]}{}_{\un{d}\un{e}\un{f}} )_{\a\b} \, \theta^{\b}   \\
    & \qquad
    ~+~  \fracm{4}{35} \,\fracm{1}{4!4!} \,\e^{ [4] \un{a}\un{b}\un{c}\un{d}\un{e}\un{f}\un{g} }  \,  \Theta^{(3)}{}_{ \un{d}\un{e}\un{f}\un{g}}\, ( \g_{ [4] } )_{\a\b} \, \theta^{\b}    \Big\} ~~~,
\end{split}  \\
\begin{split}
    \big[ \, \Theta^{(2) \, \un{a}\un{b}\un{c} }\,  \theta_{\a} \, \big]_{IR} ~=&~   \tilde{n}_0  \, \Big\{  \Theta^{(2) \, \un{a}\un{b}\un{c} }\,   \theta_{\a} ~+~ \fracm 1 {16} \, \Theta^{(2) \,  [\un{a}\un{b} | \un{d} }  \,  ( \g^{|\un{c}] } {}_{\un{d} })_{\a\b} \, \theta^{\b}   \\
    & \qquad ~+~ \fracm 1{112}  \, \Theta^{(2) \, [\un{a}| \un{d}\un{e}} \, ( \g^{|\un{b}\un{c}]}{}_{\un{d}\un{e}} )_{\a\b} \, \theta^{\b}   \\
    & \qquad
    ~-~ \fracm{1}{56} \, \fracm 1{5!3!} \, \e^{ [5] \un{a}\un{b}\un{c}\un{d}\un{e}\un{f}  }  \Theta^{(2)}{}_{ \un{d}\un{e}\un{f} }\,  ( \g_{ [5] } )_{\a\b} \, \theta^{\b}    \Big\}  ~~~,
\end{split}
\end{align}
where the relative coefficients are fixed by the irreducibility conditions
\begin{align}
    (\g_{\un{c}})_{\a}^{\ \g}\big[ \, \Theta{}^{(3) \, \un{a}\un{b}\un{c}\un{d} }\,  (\g_{\un{d}}){}_{\g\b}  \, \theta^{\b} \, \big]_{IR}  ~=&~ 0   ~~~,
    \label{equ:3520irr1} \\
    (\g_{\un{c}})_{\a}^{\ \g}\big[ \, \Theta{}^{(2) \, \un{a}\un{b}\un{c} } \, \theta_{\g} \, \big]_{IR} ~=&~ 0
    ~~~.
    \label{equ:3520irr2}
\end{align}
Let us do the counting. By subtracting the degrees of freedom from the irreducibility conditions, these cubic $\theta$-monomials have $32 \times \frac{11 \times 10 \times 9}{3 \times 2} - 32 \times \frac{11 \times 10}{2} = 3,520$ degrees of freedom. Hence they sit in the $ \CMTred {\{3,520\}}$ representation. Again, we can omit the overall coefficients $\tilde{m}_{0}$ and $\tilde{n}_{0}$ and write
\begin{align}
\begin{split}
    V1 ~=&~ \frac{1}{3!}\, \big( \mathcal{H}_1 ~+~ \fracm{1}{14}\, \mathcal{H}_2 ~+~ \fracm{1}{84}\, \mathcal{H}_3 ~-~ \fracm{4}{35}\,\mathcal{H}_4 \big)^{\un{a}\un{b}\un{c}}_{[\d\e\b]\a} \,  \theta^{\d}\theta^{\e}\theta^{\b}
\end{split} ~~~,   \\
\begin{split}
    V2 ~=&~ \frac{1}{3!}\, \big( \mathcal{I}_0 ~+~ \fracm{1}{16}\, \mathcal{I}_1 ~-~ \fracm{1}{112}\,\mathcal{I}_2 ~-~ \fracm{1}{56}\, \mathcal{I}_3 \big)^{\un{a}\un{b}\un{c}}_{[\d\e\b]\a} \, \theta^{\d}\theta^{\e}\theta^{\b}
    ~~~,
\end{split}
\end{align}
where we define nine objects
\begin{equation}
\begin{split}
    ( \mathcal{J} )^{\un{a}\un{b}\un{c}}_{[\d\e\b]\a} ~=&~ C_{[\d\e}(\g^{\un{a}\un{b}\un{c}})_{\b]\a}  ~~~,  \\
    ( \mathcal{H}_{1} )^{\un{a}\un{b}\un{c}}_{[\d\e\b]\a} ~=&~ (\g^{\un{a}\un{b}\un{c}[1]})_{[\d\e}(\g_{[1]})_{\b]\a} 
    ~~~, \\
    ( \mathcal{H}_{2} )^{\un{a}\un{b}\un{c}}_{[\d\e\b]\a} ~=&~ (\g^{[2][\un{a}\un{b}})_{[\d\e}(\g^{\un{c}]}{}_{[2]})_{\b]\a} ~~~,  \\
    ( \mathcal{H}_{3} )^{\un{a}\un{b}\un{c}}_{[\d\e\b]\a} ~=&~ (\g^{[3][\un{a}})_{[\d\e}(\g^{\un{b}\un{c}]}{}_{[3]})_{\b]\a} ~~~,   \\
    ( \mathcal{H}_{4} )^{\un{a}\un{b}\un{c}}_{[\d\e\b]\a} ~=&~ \frac{1}{4!4!} \, \e^{\un{a}\un{b}\un{c}[4][\bar{4}]}\, (\g_{[4]})_{[\d\e}(\g_{[\bar{4}]})_{\b]\a}  \\
    ( \mathcal{I}_{0} )^{\un{a}\un{b}\un{c}}_{[\d\e\b]\a} ~=&~ (\g^{\un{a}\un{b}\un{c}})_{[\d\e}C_{\b]\a}
    ~~~,   \\
    ( \mathcal{I}_{1} )^{\un{a}\un{b}\un{c}}_{[\d\e\b]\a} ~=&~ (\g^{[1][\un{a}\un{b}})_{[\d\e}(\g^{\un{c}]}{}_{[1]})_{\b]\a}  ~~~,  \\
    ( \mathcal{I}_{2} )^{\un{a}\un{b}\un{c}}_{[\d\e\b]\a} ~=&~ (\g^{[2][\un{a}})_{[\d\e}(\g^{\un{b}\un{c}]}{}_{[2]})_{\b]\a}  ~~~,  \\
    ( \mathcal{I}_{3} )^{\un{a}\un{b}\un{c}}_{[\d\e\b]\a} ~=&~ \frac{1}{3!5!} \, \e^{\un{a}\un{b}\un{c}[3][5]}\, (\g_{[3]})_{[\d\e}(\g_{[5]})_{\b]\a} ~~~,
\end{split}
\end{equation}
with an additional $\mathcal{J}$ playing the role of $\mathcal{B}$ and $\mathcal{E}$ in representations $ \CMTred {\{320\}}$ and $ \CMTred {\{1,408\}}$. Again, relevant Fierz identities are listed in Appendix \ref{appen:cubicfierz}. We can rewrite them as
\begin{equation}
\begin{split}
    33\mathcal{I}_0 ~=&~ -\mathcal{J} + \mathcal{I}_3 + \frac{1}{4}\mathcal{I}_2-\frac{1}{2}\mathcal{I}_1-\mathcal{H}_4+ \frac{1}{12}\mathcal{H}_3 + \frac{1}{4}\mathcal{H}_2 + \mathcal{H}_1 ~~~, \\
    33\mathcal{J} ~=&~ - \mathcal{I}_3 + \frac{1}{4}\mathcal{I}_2+\frac{1}{2}\mathcal{I}_1 -\mathcal{I}_0 - \mathcal{H}_4 - \frac{1}{12}\mathcal{H}_3 + \frac{1}{4}\mathcal{H}_2 - \mathcal{H}_1 ~~~, \\
    26\mathcal{H}_1 ~=&~ -8\mathcal{J} - 2\mathcal{I}_3 - \mathcal{I}_2+3\mathcal{I}_1 +8\mathcal{I}_0  + \frac{1}{6}\mathcal{H}_3 + \mathcal{H}_2  ~~~, \\
    38\mathcal{I}_1 ~=&~ 48\mathcal{J} - 12\mathcal{I}_3 -2 \mathcal{I}_2 -48\mathcal{I}_0  - \frac{1}{3}\mathcal{H}_3 + 2\mathcal{H}_1  ~~~, \\
    28\mathcal{H}_2 ~=&~ 336\mathcal{J} - 24\mathcal{I}_3 +4 \mathcal{I}_2 +28\mathcal{I}_1 +336\mathcal{I}_0 -48\mathcal{H}_4  + \frac{2}{3}\mathcal{H}_3 + 168\mathcal{H}_1 ~~~,  \\
    28\mathcal{I}_2 ~=&~ 336\mathcal{J} + 24\mathcal{I}_3  -28\mathcal{I}_1 +336\mathcal{I}_0 -48\mathcal{H}_4  - \frac{2}{3}\mathcal{H}_3 +4 \mathcal{H}_2 - 168\mathcal{H}_1  ~~~.
\end{split}
\end{equation}
Solving these linear equations gives us
\begin{equation}
\left\{
    \begin{aligned}
    \mathcal{I}_1~=&~ 4\mathcal{J} - 4\mathcal{I}_0 + 2\mathcal{H}_1  ~~~, \\
    \mathcal{I}_2 ~=&~ 20\mathcal{J} +28 \mathcal{I}_0 - 8\mathcal{H}_1 ~~~, \\
    \mathcal{I}_3~=&~ -4\mathcal{J} + 4\mathcal{I}_0 - \mathcal{H}_1 ~~~, \\
    \mathcal{H}_2 ~=&~ 28\mathcal{J} + 20\mathcal{I}_0 + 8\mathcal{H}_1 ~~~, \\
    \mathcal{H}_3 ~=&~ -120 \mathcal{J} + 120\mathcal{I}_0 + 12\mathcal{H}_1 ~~~,\\
    \mathcal{H}_4 ~=&~ -5\mathcal{J} - 5\mathcal{I}_0  ~~~.
    \end{aligned}
    \right.
\end{equation}
Here we choose $\mathcal{J}$, $\mathcal{I}_0$ and $\mathcal{H}_1$ as our basis, since they are the three simplest objects. Then
\begin{align}
    V1 ~=&~ \frac{4}{7} \big( 2\mathcal{J} + 6\mathcal{I}_0 + 3\mathcal{H}_1 \big)^{\un{a}\un{b}\un{c}}_{[\d\e\b]\a} \theta^{\d}\theta^{\e}\theta^{\b} ~~~,  \\
    V2 ~=&~ \frac{1}{14} \big( 2\mathcal{J} + 6\mathcal{I}_0 + 3\mathcal{H}_1 \big)^{\un{a}\un{b}\un{c}}_{[\d\e\b]\a} \theta^{\d}\theta^{\e}\theta^{\b} ~~~, 
\end{align}
which means
\begin{equation}
    V1 ~=~ 8 \, V2  ~~~,
\end{equation}
and that there is precisely one independent $\CMTred{\{3,520\}}$ representation sitting in the space of cubic monomials.

Now rewrite the irreducibility conditions in (\ref{equ:3520irr1}) and (\ref{equ:3520irr2}) to  
\begin{equation}
    (\g_{\un c})_{\a}{}^{\g} \, \big( 2\mathcal{J} + 6\mathcal{I}_0 + 3\mathcal{H}_1 \big)^{\un{a}\un{b}\un{c}}_{[\d\e\b]\a} \, \theta^{\d}\theta^{\e}\theta^{\b} = 0 ~~~.
\end{equation}
This condition can be simplified as
\begin{equation}
    - 6\, C_{[\d\e}(\g^{\un{a}\un{b}})_{\b]\a} 
    + \,(\g^{\un{a}\un{b}[2]})_{[\d\e}(\g_{[2]})_{\b]\a} 
    - 2\, (\g^{\un{a}\un{b}[1]})_{[\d\e}(\g_{[1]})_{\b]\a}
    = \big( -6 \mathcal{E} + \mathcal{G}_{1} - 2 \mathcal{F}_{1} \big)^{\un{a}\un{b}}_{[\d\e\b]\a} = 0 ~~~,
\end{equation}
which exactly satisfies Equation (\ref{equ:1408relations}), as predicted in the last subsection.

The importance of the observations so far in this chapter is that an expansion to 
third order of  a real scalar superfield $\cal V$ must be written in terms of complete
sets of irreducible monomials.  So to this order we have
\be  \eqalign{
{\cal V} (\theta, \, x) ~=&~ \varphi{}^{(0)}(x) ~+~ {\theta}{}^{\a} \, \varphi^{(1)}_{\a}(x) 
~+~ \Theta {}^{(1)} \, \varphi^{(2)}(x) 
~+~  \Theta{}^{(2) \, \un{a}\un{b}\un{c}} \, \varphi^{(2)}_{\un{a}\un{b}\un{c}} (x) 
~+~  \Theta{}^{(3) \, \un{a}\un{b}\un{c}\un{d}} \, \varphi^{(2)}_{ \un{a}\un{b}\un{c}\un{d}} (x) \cr
& ~+~ \Theta {}^{(1)} \,  {\theta}{}^{\a} \,  \varphi^{(3)}_{\a}(x) ~+~  \big[ \, \Theta{}^{(2) \, \un{a}\un{b}\un{c}} 
\, {\theta}{}^{\a}  \, \big]_{IR}  \,\varphi{}^{(3)}_{\a \, \un{a}\un{b}\un{c} }(x) ~+~  \big[ \Theta{}^{(3) \, \un{a}\un{b}\un{c} 
\un{d}  } \, {\theta}{}^{\a}  \, \big]_{IR} \,\varphi{}^{(3)}_{\a \, \un{a}\un{b}\un{c}\un{d} }(x) 
 ~+~ \dots ~~~.
} 
\label{equ:Exp3rd} \ee

\subsection{Quartic Level}
\label{sec:Q2}

At fourth order, a new problem appears in the expression in (\ref{equ:Exp1}). 
Let us first pick out the relevant terms,
\be 
\begin{aligned}
{\cal V} (quartic) ~=&~ \Theta {}^{(1)} \, \Theta {}^{(1)} \, {\varphi}^{(4)} (x) ~+~ \Theta{}^{(1)}  \, \Theta{}^{(2) \, \un{a}\un{b}\un{c} } \,{\varphi}^{(4)}_{ \un{a}\un{b}\un{c}} (x)
~+~ \Theta{}^{(1)}  \, \Theta{}^{(3) \, \un{a}\un{b}\un{c}\un{d} } \, {\varphi}^{(4)}_{ \un{a}\un{b}\un{c} \un{d}} (x)   \\
& ~+~ \Theta{}^{(2) \, \un{a}\un{b}\un{c}}  \, \Theta{}^{(2) \, \un{d}\un{e}\un{f} } \, {\varphi}^{(4)}_{ \un{a}\un{b}\un{c} \, \un{d} \un{e}\un{f} } (x) 
~+~ \Theta{}^{(2) \, \un{a}\un{b}\un{c}}  \, \Theta{
}^{(3) \, \un{d}\un{e}\un{f}\un{g} } \, {\varphi}^{(4)}_{ \un{a}\un{b}\un{c} \, \un{d} \un{e}\un{f}\un{g} 
} (x)   \\
& ~+~ \Theta{}^{(3) \, \un{a}\un{b}\un{c}\un{d}}  \, \Theta{}^{(3) \, \un{e}\un{f}\un{g}\un{h}} \, {\varphi}^{(4)}_{ \un{a}\un{b}\un{c}\un{d} \, \un{e}\un{f}\un{g}\un{h}} (x) ~+~ \dots ~~~~~.
\end{aligned} \label{eqn:quarticexp}
\ee
The problem is seen by the
following argument.  From the antisymmetry of the $\theta$-coordinates, we know the number of 
degrees of freedom at this order is given by
\begin{equation}
\CMTred { \{ 32 \}} \, \wedge \,  \CMTred {\{ 32 \}} \, \wedge \,  \CMTred {\{ 32 \}} 
\, \wedge \,  \CMTred {\{ 32 \}}
    ~=~ \frac{ \{32\} ~\times~ \{31\} ~\times~ \{30\}  ~\times~ \{29\}  }{4 ~\times~ 3 ~\times~2} 
    ~=~ \{  35,960  \} 
    ~~~.  \label{equ:Qticdecompose}
\end{equation}

Next, one can count the degrees of freedom of the bosonic component fields at the fourth order superfield expansion in (\ref{eqn:quarticexp}). We have
\be
\begin{aligned}
    &\CMTB{\{1\}} ~~~ & ~~~ & \Theta {}^{(1)} \, \Theta {}^{(1)} ~~~,  \cr
    &\CMTB{\{165\}} ~~~ & ~~~ & \Theta{}^{(1)}  \, \Theta{}^{(2) \, \un{a}\un{b}\un{c} } ~~~,  \cr
    &\CMTB{\{330\}} ~~~ & ~~~ & \Theta{}^{(1)}  \, \Theta{}^{(3) \, \un{a}\un{b}\un{c}\un{d} } ~~~,  \cr
    &\{ 13,695 \} ~~~ & ~~~ & \Theta{}^{(2) \, \un{a}\un{b}\un{c}}  \, \Theta{}^{(2) \, \un{d}\un{e}\un{f} }  ~~~, \cr
    &\{ 54,450 \} ~~~ & ~~~ & \Theta{}^{(2) \, \un{a}\un{b}\un{c}}  \, \Theta{}^{(3) \, \un{d}\un{e}\un{f}\un{g} }  ~~~, \cr
    &\{ 54,615 \} ~~~ & ~~~ & \Theta{}^{(3) \, \un{a}\un{b}\un{c}\un{d}}  \, \Theta{}^{(3) \, \un{e}\un{f}\un{g}\un{h}}  ~~~.
\end{aligned}
\ee
The last three numbers comes from
\begin{equation}
\begin{split}
    \{ 13,695 \} ~=&~ [\, \CMTB{\{165\}} ~\otimes~ \CMTB{\{165\}} \,]_{S} ~=~ \frac{\{165\} ~\times~ \{166\}}{2} ~~~, \\
    \{ 54,450 \} ~=&~ \CMTB{\{165\}} ~\otimes~ \CMTB{\{330\}} ~~~, \\
    \{ 54,615 \} ~=&~ [\, \CMTB{\{330\}} ~\otimes~ \CMTB{\{330\}} \,]_{S} ~=~ \frac{\{330\} ~\times~ \{331\}}{2} ~~~,
\end{split} \label{eqn:165330products}
\end{equation}
where $[~~]_{S}$ means the symmetric part of the product, as $\Theta^{(2)}$ and $\Theta^{(3)}$ carry two spinor indices and they commute with themselves.
Note that these three numbers are not the dimensions of any $\mathfrak{so}(11)$ irrep. They are reducible representations that carry more than the necessary irreducible components contained in the superfield. 
This is clear as symmetry properties are not fully utilized to constrain the degrees of freedom. 
For example in the second line, the spinor indices on $\Theta^{(2)}$ and $\Theta^{(3)}$ could also be swapped, but the tensor product among them clearly include the maximal degrees of freedom.
Also, in the first and third lines, one only considered exchanging the positions of the entire $\Theta^{(2)}$'s or $\Theta^{(3)}$'s, but not the possibilities of exchanging individual Lorentz indices.
Moreover, 
\be
{\CMTB {\{ 1 \} }} ~\oplus~ {\CMTB { \{ 165 \} }} ~\oplus~ {\CMTB { \{ 330 \} }} ~\oplus~ \{ 13,695 \}
~\oplus~ \{ 54,450 \} ~\oplus~ \{ 54,615 \} ~=~ 
 \{ 123,256 \}  ~~~.
\label{equ:SF4cntng}
\ee
Clearly, the dimension of $\{ 123,256 \}$ far exceeds that of $\{  35,960  \}$.  

Therefore, by following the path in the cubic sector, one immediate thought is to write down all possible index structures, 
i.e. first constructing the irreducible quartic monomials just as was done for the cubic order.
That would essentially be breaking down the reducible representations $\{13,695\}$, $\{54,450\}$, and $\{54,615\}$ into sums of $\mathfrak{so}(11)$ irreducible representations. 
In other words, one would allow the component fields 
${\varphi}^{(4)}_{ \un{a}\un{b}\un{c} \, \un{d} \un{e}\un{f} } (x)$, 
${\varphi}^{(4)}_{ \un{a}\un{b} \un{c} \, \un{d} \un{e}\un{f}\un{g} } (x)$, and ${\varphi}^{(4)}_{ \un{a}\un{b}\un{c}\un{d} \, \un{e} \un{f}\un{g}\un{h}} (x)$ 
to admit linear constraints that identify various irreducible component 
fields with each other. 
Due to the overwhelming excess of irreducible components, 
one could hope to ameliorate this situation by finding that some of the
bosonic fields are actually zero. 
As with the cubic sector, there exist a sufficient number of Fierz identities 
such that when these fourth order terms are expanded over an irreducible basis, 
one finds relations between the seemingly independent ones in such a way that 
some terms vanish and the counting of the remaining field components adds to 35,960. 
That would be an excessive amount of tedious calculations. 

In later sections, one would see how these terrible decompositions could be turned into neat group-theory problems that could be solved by efficient algorithms. 

The main message of this chapter of our work is that explicit $\theta$-expansion of the eleven
dimensional scalar superfield is considerably more complicated than in lower dimensions.  
One must contend with four separate problems: 
\newline\indent
(a.) there are multiple equivalent ways to express
the required $\theta$-monomials,
\newline\indent
(b.) some apparently reasonable monomial combinations actually vanish, 
\newline\indent
(c.) the requirement of irreducibility of the $\theta$-monomial expansion requires
\newline\indent $~~~~~$
carefully
constructed combinations, and
\newline\indent
(d.) the over abundance of bosonic fields encountered from the most obvious $\theta$-expansion.
 \newline  \noindent
The resolution of the first two of these problems relies of the derivation of Fierz identities.  With 
regard to the third problem, the only methodology known to use is brute force establishment of 
their existences.  The final problem requires a careful choice of constraints.

At all higher orders, up to the sixteenth, in the $\theta$-expansion this problem of deriving
explicit irreducible $\theta$-monomials occurs. Above this order, the form of the required higher
order irreducible $\theta$-monomials
can be deduced from the lower order irreducible $\theta$-monomials.
For terms in odd orders, an actual derivation
of the irreducible $\theta$-monomials involves derivations of Fierz identities as we explicitly
demonstrated at cubic order.  For terms in even orders, an actual derivation of the irreducible 
$\theta$-monomials involves derivations along the lines we implicitly discussed at quartic order.  
To our knowledge, these impediments have not been recognized previously in the literature.  A
most disappointing realization would be that all of these need to be sorted out before any
discussion of dynamics using off-shell 11D, $\cal N$ = 1 superfields.

All of these point to the fact that superfields and their accompanying $\theta$-expansions become 
increasing unwieldy as a ``technological platform" for the study of supermultiplets in higher 
dimensions.  This is due to the fact that in order to calculate efficiently, polynomials
$\big[ \theta{}^1 \cdots \theta{}^p \big]{}_{IR}$ at Level-$p$ are required to be explicitly
constructed.
We will return to this inefficiency in our conclusions.

For all the reasons enunciated in an illustrative manner throughout this chapter, we 
are motivated to search for an approach that
is free of these burdensome complications.

\newpage
\section{11D $\mathcal{N} = 1$ Scalar Superfield Decomposition}

In 11D, $\cal N$ = 1 superspace, the number of independent Grassmann coordinates is $2^{(11-1)/2} = 32$ with the Majorana condition since we use the Minkowski signature $(-,+,+,\dots,+)$. Then the superspace has coordinates $(x^{\un{a}},\theta^{\a})$, where $\un{a} = 0, 1, \ldots, 10$ and $\a = 1, \ldots, 32$. Hence, the $\theta$-expansion of the eleven dimensional scalar superfield begins at Level-0 and continues to Level-32, where Level-$n$ corresponds to the order $\mathcal{O}(\theta^{n})$. The unconstrained real scalar superfield ${\cal V}$ contains $2^{32-1} = 2,147,483,648$ bosonic and $2^{32-1} = 2,147,483,648$ fermionic components. Expressed in terms of a $\theta$-expansion of ${\cal V}$ we have
\begin{equation}
    {\cal V} (x^{\un{a}}, \, \theta^{\a}) ~=~ \varphi^{(0)} (x^{\un{a}}) ~+~ \theta^{\a} \, \varphi^{(1)}_{\a} (x^{\un{a}}) ~+~ \theta^{\a} \theta^{\b} \, \varphi^{(2)}_{\a\b}  (x^{\un{a}}) ~+~ \dots  ~~~.  \label{equ:10DIsuperfield}
\end{equation}
We can decompose $\theta$-monomials $\theta^{\a_{1}} \cdots \theta^{\a_{n}}$ into a direct sum of irreducible representations of Lorentz group SO(1,10). With the antisymmetric property of Grassmann coordinates, we have
\begin{equation}
{\cal V} ~=~ \begin{cases}
{~~}{\rm {Level}}-0 \,~~~~~~~~~~ \{ 1 \} ~~~,  \\
{~~}{\rm {Level}}-1 \,~~~~~~~~~~ \{ 32 \} ~~~,  \\
{~~}{\rm {Level}}-2 \,~~~~~~~~~~ \{ 32 \} \, \wedge \, \{ 32 \} ~~~,  \\
{~~}{\rm {Level}}-3 \,~~~~~~~~~~ \{ 32 \} \, \wedge \, \{ 32 \} \, \wedge \, \{ 32 \} ~~~,  \\
{~~~~~~}  {~~~~} \vdots  {~~~~~~~~~\,~~~~~~} \vdots \\
{~~}{\rm {Level}}-n \,~~~~~~~~~ \underbrace{\{ 32 \} \, \wedge \, ~\dots~ \, \wedge \, \{ 32 \}}_\text{$n$ times}  ~~~,  \\
{~~~~~~}  {~~~~} \vdots  {~~~~~~~~~\,~~~~~~} \vdots \\
{~~}{\rm {Level}}-32 ~~~~~~~~~ \{ 1 \} ~~~.
\end{cases}
{~~~~~~~~~~~~~~~}
\end{equation}
All even levels are bosonic representations, while all odd levels are fermionic representations. Note that in a 32-dimensional Grassmann space, the Hodge dual of a $p$-form is a $(32-p)$-form. Therefore, Level-$(32-n)$ is the dual of Level-$n$ for $n = 0, \ldots, 16$, and they have the same dimensions. By simple use of the values of the function ``32 choose n,'' these dimensions are found to be the ones that follow
\begin{equation}
{\cal V} ~=~ \begin{cases}
{~~}{\rm {Level}}-0 \,~~~~~~~~~~ 1 ~~~,  \\
{~~}{\rm {Level}}-1 \,~~~~~~~~~~ 32 ~~~,  \\
{~~}{\rm {Level}}-2 \,~~~~~~~~~~   496~~~,  \\
{~~}{\rm {Level}}-3 \,~~~~~~~~~~ 4960  ~~~,  \\
{~~~~~~}  {~~~~} \vdots  {~~~~~~~~~\,~~~~~~} \vdots \\
{~~}{\rm {Level}}-n \,~~~~~~~~~ \frac {32!}{n! (32 - n)!}  ~~~,  \\
{~~~~~~}  {~~~~} \vdots  {~~~~~~~~~\,~~~~~~} \vdots \\
{~~}{\rm {Level}}-32 ~~~~~~~~~  1 ~~~.
\end{cases}
{~~~~~~~~~~~~~~~}
\end{equation}

Since $\cal V$ is a scalar superfield, the conjecture given in the
conclusion of the work of \cite{counting10d} implies the following statements
\begin{equation} 
\frac {32!}{n! (32 - n)!} ~=~
\sum_{ \CMTB {\cal R}} \, b_{ \CMTB {\{  {\cal R}  \}  } }    \,  d_{ \CMTB {\{  {\cal R}  \}  } }  
~~~,
\label{equ:BosBIN}
\end{equation}
for even values of $n$ and 
\begin{equation} 
\frac {32!}{n! (32 - n)!} ~=~
\sum_{ \CMTR {\cal R}} \, b_{ \CMTR {\{  {\cal R}  \}  } }    \,  d_{ \CMTR {\{  {\cal R}  \}  } }  
~~~,
\label{equ:FerBIN}
\end{equation}
for odd values of $n$ apply to it. In these equations, $b_{ \CMTB {\{ {\cal R}\}}}$ and $b_{ \CMTR {\{ {\cal R}\}}}$ are non negative integers, $n$ refers to the Level-$n$ in the $\theta$-expansion of the superfield, while finally $d_{ \CMTB {\{ {\cal R}\}}}$ and $d_{ \CMTR {\{ {\cal R}\}}}$ refer to the dimensions of bosonic representations $\CMTB {\{{\cal R} \} }$ and fermionic representations $\CMTR {\{ {\cal R} \} }$ of the SO(1,10) Lorentz algebra. In Appendix \ref{appen:so11}, $\mathfrak{so}(11)$ irreps with small dimensions are listed.  The data shown in Section \ref{sec:decomp_result} provides the explicit information about the quantities appearing in (\ref{equ:BosBIN}) and (\ref{equ:FerBIN}).

\subsection{Methodology 1: Branching Rules for $\mathfrak{su}(32)\supset\mathfrak{so}(11)$}
\label{sec:M1}

In \cite{counting10d} while we have applied branching rules to find component decompositions of scalar superfields in ten dimensions, we didn't explain the details of branching rule calculations. In this section, we will present the explicit algorithmic~\cite{yamatsu2015} calculations needed
for finding branching rules, in particular for the case of $\mathfrak{su}(32)\supset\mathfrak{so}(11)$. 

First, a branching rule is a relation between a representation of Lie algebra $\mathfrak{g}$ and representations of one of its Lie subalgebras $\mathfrak{h}$. For a simple Lie algebra $\mathfrak{g}$, its Lie subalgebras can be classified as regular subalgebras and special subalgebras. Regular subalgebras can be obtained by deleting dots from extended Dynkin diagrams, while special subalgebras cannot. Moreover, subalgebras can be classified as maximal subalgebras and non-maximal subalgebras. The definition of a maximal subalgebra $\mathfrak{h}$ of $\mathfrak{g}$ is that there is no any subalgebra $\mathfrak{l}$ satisifies $\mathfrak{h}\subset \mathfrak{l}\subset \mathfrak{g}$. 

Branching rules between $\mathfrak{g}$ and $\mathfrak{h}$ are completely determined by the projection matrix $P_{\mathfrak{g}\supset\mathfrak{h}}$. Suppose the rank of Lie algebra $\mathfrak{g}$ is $n$ and the rank of Lie algebra $\mathfrak{h}$ is $m$, then the projection matrix $P_{\mathfrak{g}\supset\mathfrak{h}}$ is a $m\times n$ matrix. Given a weight vector $w_{\mathfrak{g}}$ in $\mathfrak{g}$, the projected weight vector $v_{\mathfrak{h}}$ in $\mathfrak{h}$ is 
\begin{equation}
\label{equ:def_prjmat}
    v_{\mathfrak{h}}^T ~=~ P_{\mathfrak{g}\supset\mathfrak{h}}w_{\mathfrak{g}}^T ~~~,
\end{equation}
where weight vectors are row vectors. Thus, the algorithm for calculating a branching rule of an irrep $R$ of $\mathfrak{g}$ given the projection matrix can be summarized as follows. 
\begin{enumerate}
    \item Write the weight diagram of $R$;
    \item Calculate the projected weight vector in $\mathfrak{h}$ by Equation~(\ref{equ:def_prjmat}) for every weight vector in the weight diagram of $R$ and get the projected weight diagram;
    \item Find irrep(s) in $\mathfrak{h}$ corresponding to the projected weight diagram.
\end{enumerate}

As for how to obtain the projection matrix $P_{\mathfrak{g}\supset\mathfrak{h}}$, the recipe depends on which class of subalgebra of $\mathfrak{g}$ does $\mathfrak{h}$ belong to. If $\mathfrak{h}$ is a maximal subalgebra of $\mathfrak{g}$, one can calculate the projection matrix by a ``reverse'' process.
\begin{enumerate}
    \item Find one branching rule of $\mathfrak{g}\supset\mathfrak{h}$ and write down the weight diagrams of the reps of $\mathfrak{g}$ and $\mathfrak{h}$;
    \item Find an appropriate correspondence between weight vectors;
    \item Calculate the matrix elements by Equation~(\ref{equ:def_prjmat}).
\end{enumerate}
If $\mathfrak{h}$ is a non-maximal subalgebra of $\mathfrak{g}$, you can calculate the projection matrix by matrix multiplications:
\begin{enumerate}
    \item Find Lie subalgebras $\mathfrak{l}^{(a)}\ (a = 1,...,l)$ satisfying $\mathfrak{h}\subset \mathfrak{l}^{(1)}\subset\cdots \subset\mathfrak{l}^{(l)}\subset \mathfrak{g}$ where each $\mathfrak{g}\supset \mathfrak{l}^{(l)}$, $\mathfrak{l}^{(l)}\supset \mathfrak{l}^{(l-1)}$, ..., $\mathfrak{l}^{(1)}\supset \mathfrak{h}$ is a pair of one Lie algebra and its maximal subalgebra;
    \item The projection matrix is $P_{\mathfrak{g}\supset\mathfrak{h}}~=~ P_{\mathfrak{l}^{(1)}\supset \mathfrak{h}}P_{\mathfrak{l}^{(2)}\supset \mathfrak{l}^{(1)}}\cdots P_{\mathfrak{l}^{(l)}\supset \mathfrak{l}^{(l-1)}}P_{\mathfrak{g}\supset \mathfrak{l}^{(l)}}$. 
\end{enumerate}


Since $\mathfrak{so}(11)$ is a maximal special subalgebra of $\mathfrak{su}(32)$, 
we can find the projection matrix from the weight systems of the irreducible representation $\{32\}$ in $\mathfrak{su}(32)$ and $\mathfrak{so}(11)$, as we know one $\mathfrak{su}(32) \supset \mathfrak{so}(11)$ branching rule
\begin{equation}
    \{32\} \rightarrow \{32\}   ~~~,
\end{equation}
or more clearly in Dynkin labels,
\begin{equation}
    (1000000000000000000000000000000) \rightarrow (00001)   ~~~.
\end{equation}
The definition of the projection matrix for the branching rules of $\mathfrak{su}(32)\supset\mathfrak{so}(11)$ is
\begin{equation}
    (v_{\mathfrak{so}(11)})^{T} ~=~ P_{\mathfrak{su}(32)\supset\mathfrak{so}(11)} \, (w_{\mathfrak{su}(32)})^{T}  ~~~.
\end{equation}
The matrix $P_{\mathfrak{su}(32)\supset\mathfrak{so}(11)}$ is a $5\times 31$ matrix, as the ranks of $\mathfrak{so}(11)$ and $\mathfrak{su}(32)$ are 5 and 31 respectively. 

The weight system of $\{32\} = (1000000000000000000000000000000)$ in $\mathfrak{su}(32)$ has 32 weights given by
\begin{equation}
\begin{gathered}
\overbrace{\boxed{ ~ 1 ~~~~ 0 ~~~~ 0 ~~~~ 0 ~~~~ \cdots ~~~~ 0 ~ }}^\text{31 digits}   \\
\boxed{ -1 ~~~~ 1 ~~~~ 0 ~~~~ 0 ~~~~ \cdots ~~~~ 0 ~ }   \\
\boxed{ ~ 0 ~~ -1 ~~~~ 1 ~~~~ 0 ~~~~ \cdots ~~~~ 0 ~ }   \\
\vdots   \\
\boxed{ ~ 0 ~~~~ \cdots ~~~~ 0 ~~ -1 ~~~~ 1 ~~~~ 0 ~ }   \\
\boxed{ ~ 0 ~~~~ \cdots ~~~~ 0 ~~~~ 0 ~~ -1 ~~~~ 1 ~ }   \\
~~~~~~ \boxed{ ~ 0 ~~~~ \cdots ~~~~ 0 ~~~~ 0 ~~~~ 0 ~~ -1 ~ }   ~~~~~,
\end{gathered} \vspace{1em}
\end{equation}
while the weight system of $\{32\} = (00001)$ in $\mathfrak{so}(11)$ also has 32 weights given by
$$
\begin{gathered}
\boxed{ ~ 0 ~~~~ 0 ~~~~ 0 ~~~~ 0 ~~~~ 1 ~ }   \\
\boxed{ ~ 0 ~~~~ 0 ~~~~ 0 ~~~~ 1 ~~ -1 ~ }   \\
\boxed{ ~ 0 ~~~~ 0 ~~~~ 1 ~~ -1 ~~~~ 1 ~ }   \\
\boxed{ ~ 0 ~~~~ 0 ~~~~ 1 ~~~~ 0 ~~ -1 ~ }   \\
\boxed{ ~ 0 ~~~~ 1 ~~ -1 ~~~~ 0 ~~~~ 1 ~ }   \\
\boxed{ ~ 0 ~~~~ 1 ~~ -1 ~~~~ 1 ~~ -1 ~ }   \\
\boxed{ ~ 1 ~~ -1 ~~~~ 0 ~~~~ 0 ~~~~ 1 ~ }   
\end{gathered}
$$
\begin{equation}
\begin{gathered}
\boxed{ -1 ~~~~ 0 ~~~~ 0 ~~~~ 0 ~~~~ 1 ~ }   \\
\boxed{ ~ 0 ~~~~ 1 ~~~~ 0 ~~ -1 ~~~~ 1 ~ }   \\
\boxed{ ~ 1 ~~ -1 ~~~~ 0 ~~~~ 1 ~~ -1 ~ }   \\
\boxed{ -1 ~~~~ 0 ~~~~ 0 ~~~~ 1 ~~ -1 ~ }   \\
\boxed{ ~ 0 ~~~~ 1 ~~~~ 0 ~~~~ 0 ~~ -1 ~ }   \\
\boxed{ ~ 1 ~~ -1 ~~~~ 1 ~~ -1 ~~~~ 1 ~ }   \\
\boxed{ -1 ~~~~ 0 ~~~~ 1 ~~ -1 ~~~~ 1 ~ }   \\
\boxed{ ~ 1 ~~ -1 ~~~~ 1 ~~~~ 0 ~~ -1 ~ }   \\
\boxed{ ~ 1 ~~~~ 0 ~~ -1 ~~~~ 0 ~~~~ 1 ~ }   \\
\boxed{ -1 ~~~~ 0 ~~~~ 1 ~~~~ 0 ~~ -1 ~ }   \\
\boxed{ -1 ~~~~ 1 ~~ -1 ~~~~ 0 ~~~~ 1 ~ }   \\
\boxed{ ~ 1 ~~~~ 0 ~~ -1 ~~~~ 1 ~~ -1 ~ }   \\
\boxed{ -1 ~~~~ 1 ~~ -1 ~~~~ 1 ~~ -1 ~ }   \\
\boxed{ ~ 0 ~~ -1 ~~~~ 0 ~~~~ 0 ~~~~ 1 ~ }   \\
\boxed{ ~ 1 ~~~~ 0 ~~~~ 0 ~~ -1 ~~~~ 1 ~ }   \\
\boxed{ -1 ~~~~ 1 ~~~~ 0 ~~ -1 ~~~~ 1 ~ }   \\
\boxed{ ~ 0 ~~ -1 ~~~~ 0 ~~~~ 1 ~~ -1 ~ }   \\
\boxed{ ~ 1 ~~~~ 0 ~~~~ 0 ~~~~ 0 ~~ -1 ~ }   \\
\boxed{ -1 ~~~~ 1 ~~~~ 0 ~~~~ 0 ~~ -1 ~ }   \\
\boxed{ ~ 0 ~~ -1 ~~~~ 1 ~~ -1 ~~~~ 1 ~ }   \\
\boxed{ ~ 0 ~~ -1 ~~~~ 1 ~~~~ 0 ~~ -1 ~ }   \\
\boxed{ ~ 0 ~~~~ 0 ~~ -1 ~~~~ 0 ~~~~ 1 ~ }   \\
\boxed{ ~ 0 ~~~~ 0 ~~ -1 ~~~~ 1 ~~ -1 ~ }   \\
\boxed{ ~ 0 ~~~~ 0 ~~~~ 0 ~~ -1 ~~~~ 1 ~ }   \\
~~~~~~ \boxed{ ~ 0 ~~~~ 0 ~~~~ 0 ~~~~ 0 ~~ -1 ~ }  ~~~~~.
\end{gathered} \vspace{1em}
\end{equation}
If we put together all the weights in the weight system of $\{32\}$ in $\mathfrak{su}(32)$ and $\mathfrak{so}(11)$ into two matrices, 
\begin{align}
W_{\mathfrak{su}(32)} ~=&~
\left( \begin{array}{ccc}
    \horzbar & w_{\mathfrak{su}(32)}{}^{1}  & \horzbar \\
    \horzbar & w_{\mathfrak{su}(32)}{}^{2}  & \horzbar \\
             & \vdots                       &          \\
    \horzbar & w_{\mathfrak{su}(32)}{}^{32} & \horzbar
\end{array} \right)_{32\times 31}  ~~~, {~~~~} \text{and} {~~~~}
V_{\mathfrak{so}(11)} ~=~
\left( \begin{array}{ccc}
    \horzbar & v_{\mathfrak{so}(11)}{}^{1}  & \horzbar \\
    \horzbar & v_{\mathfrak{so}(11)}{}^{2}  & \horzbar \\
             & \vdots                       &          \\
    \horzbar & v_{\mathfrak{so}(11)}{}^{32} & \horzbar
\end{array} \right)_{32\times 5}   ~~~,
\end{align}
where the superscript indices $i = 1, \dots, 32$ of $w_{\mathfrak{su}(32)}{}^{i}$ and $v_{\mathfrak{so}(11)}{}^{i}$ label the weights in the $\{32\}$ weight system, then
\begin{equation}
    V_{\mathfrak{so}(11)}{}^{T} ~=~ P_{\mathfrak{su}(32)\supset\mathfrak{so}(11)} \, W_{\mathfrak{su}(32)}{}^{T}    ~~~.
\end{equation}
If a matrix $A_{m \times n}$ has rank $m$ ($m \leq n$), then it has right inverse $B_{n \times m}$ such that $AB ~=~ \mathbb{I}_{m\times m}$. Now the matrix $(W_{\mathfrak{su}(32)}{}^{T})_{31\times 32}$ has rank 31. Hence, there exists a right inverse $\big((W_{\mathfrak{su}(32)}{}^{T})^{-1}\big)_{32\times 31}$, so we can invert the formula
\begin{equation}
    P_{\mathfrak{su}(32)\supset\mathfrak{so}(11)} ~=~ V_{\mathfrak{so}(11)}{}^{T} \, \big( W_{\mathfrak{su}(32)}{}^{T} \big)^{-1}     ~~~,
\end{equation}
and find the explicit projection matrix to be
\begingroup
\footnotesize
\begin{equation}
\begin{split}
    & P_{\mathfrak{su}(32)\supset\mathfrak{so}(11)}  \\
    &=
\left(\begin{array}{*{31}c}
 0 & 0 & 0 & 0 & 0 & 0 & 1 & 0 & 0 & 1 & 0 & 0 & 1 & 0 & 1 & 2 & 1 & 0 & 1 & 0 & 0 & 1 & 0 & 0 & 1 & 0 & 0 & 0 & 0 & 0 & 0 \\
 0 & 0 & 0 & 0 & 1 & 2 & 1 & 1 & 2 & 1 & 1 & 2 & 1 & 1 & 0 & 0 & 0 & 1 & 1 & 2 & 1 & 1 & 2 & 1 & 1 & 2 & 1 & 0 & 0 & 0 & 0 \\
 0 & 0 & 1 & 2 & 1 & 0 & 0 & 0 & 0 & 0 & 0 & 0 & 1 & 2 & 3 & 2 & 3 & 2 & 1 & 0 & 0 & 0 & 0 & 0 & 0 & 0 & 1 & 2 & 1 & 0 & 0 \\
 0 & 1 & 0 & 0 & 0 & 1 & 1 & 1 & 0 & 1 & 2 & 2 & 1 & 0 & 0 & 0 & 0 & 0 & 1 & 2 & 2 & 1 & 0 & 1 & 1 & 1 & 0 & 0 & 0 & 1 & 0 \\
 1 & 0 & 1 & 0 & 1 & 0 & 1 & 2 & 3 & 2 & 1 & 0 & 1 & 2 & 1 & 2 & 1 & 2 & 1 & 0 & 1 & 2 & 3 & 2 & 1 & 0 & 1 & 0 & 1 & 0 & 1 
\end{array}\right)    ~.
\end{split}
\end{equation}
\endgroup

\subsection{Methodology 2: Plethysms}
\label{subsec:plethysm}

Generally, branching rules of $\mathfrak{su}(m)\supset\mathfrak{so}(n)$ where $\mathfrak{so}(n)$ is the maximal special subalgebra of $\mathfrak{su}(m)$ are equivalent to symmetrized tensor powers of the generating irrep with respect to the partition, which is called Plethysm~\cite{Plethysm,Susyno,LiE}.
For example, for $\mathfrak{su}(10)\supset\mathfrak{so}(10)$, the generating irrep is the defining representation of $\mathfrak{so}(10)$ $(10000)$, and for $\mathfrak{su}(16)\supset\mathfrak{so}(10)$, the generating irrep is the spinor representation of $\mathfrak{so}(10)$ $(00001)$. This equivalence is based on an important fact that there is a bijection between irreducible representations of $\mathfrak{su}(m)$ and partitions which are vectors with length $m$. The bijection is realized by the following algorithm: the highest weight of an irreducible representation of $\mathfrak{su}(m)$ $w$ is a row vector with length $(m-1)$ and the $i-$th entry of its corresponding partition is the summation of $j-$th entry of $w$ for $j\ge i$. Conversely, the $i-$th entry of $w$ is $i-$th entry minus $(i+1)-$th entry of the partition. Here are several examples:

\begin{table}[htp!]

\centering
\begin{tabular}{|c|c|}
\hline
    The Highest Weight  & Partition   \\ \hline
    $(d0\dots0)$  & $[d,0,\dots,0]$  \\ \hline
    $(0\dots010\dots0)$ ($d$-th entry is 1) & $[1,1,\dots,1,0,\dots,0]$ (\# of 1 = $d$) \\ \hline
    $(111\dots1)$ (length = $d$) & $[d,d-1,\dots,1,0]$  \\ \hline
\end{tabular}
\end{table}

Therefore, the component decomposition at Level-$n$ for an unconstrained scalar superfield in 11D, $\mathcal{N}=1$ superspace is the $n$-th completely antisymmetric tensor power of the spinor representation of $\mathfrak{so}(11)$ $\{32\}$ with Dynkin label $(00001)$. The partition corresponding to $n$-th completely antisymmetric tensor power is $[1,1,\dots,1]$ where the number of 1 is $n$. 

Plethysm is basically based on the manipulation of the character polynomial. Consider a Lie group $G$ with rank $n$ and a representation of it $R$. For any group element $g\in G$, the representation $D_{R}(g)$ is a square matrix and the character $\chi_R(g)$ is defined as the trace of the matrix $D_R(g)$. Suppose the Lie group $G$ has $m$ generators $T_1$, $T_2$, $\dots$, $T_m$ and the first $n$ generators form the Cartan subalgebra. 
Then the group element $g$ can be expressed as
\begin{equation}
g~=~ {\rm exp}\{ i\alpha_1T_1+ \dots + i\alpha_mT_m\}  ~~~,
\end{equation}
and diagonalized as 
\begin{equation}
{\rm exp}\{ ia_1T_1+ \dots + ia_nT_n\}  ~~~.
\end{equation}
Since the character is invariant under the conjugacy class, 
\begin{equation}
\chi_R(g) ~=~ {\rm Tr}\Big[ {\rm exp}\{ ia_1D_R(T_1)+ \dots + ia_nD_R(T_n)\} \Big] ~~~,
\end{equation} 
where $D_R(T_1)$ to $D_R(T_n)$ are diagonal matrices. Moreover their diagnoal entries form the weight vectors $w_i \, (i=1,2,\dots,{\rm dim}(R))$ of representation $R$: the $k$-th entry of $w_i$ is the $i$-th diagonal entry of $D_R(T_k)$. 
Therefore 
\begin{equation}
\chi_R(g) ~=~ \sum_{i=1}^{{\rm dim}(R)}e^{iv_g\cdot w_i}   ~~~~~~,
\end{equation} 
where $v_g=(a_1,\dots,a_n)$. Then actually we can treat $e^{iv_g}$ as $X$ and rewrite the character as the character polynomial
\begin{equation}
\chi_R(g) ~=~ \sum_{i=1}^{{\rm dim}(R)}X^{w_i} ~~~~~~.
\end{equation}

The character polynomials of symmetrized tensor power with respect to the partition $\lambda$ of $R$ can be obtained by the character polynomials of $R$. General algorithm can be found in \cite{Plethysm}. Here we only list the algorithm for completely antisymmetric and completely symmetric cases~\cite{Susyno,LiE}. 

\begin{itemize}
\item Completely antisymmetric: the character polynomials of $k$-th completely antisymmetric tensor power of $R$ is the summation of all products of $k$ distinct monomials;
\item Completely symmetric: the character polynomials of $k$-th completely antisymmetric tensor power of $R$ is the summation of all products of $k$ monomials.
\end{itemize}
For example, the character polynomial of antisymmetric square of R is 
\begin{equation}
\begin{split}
X^{w_1+w_2}+X^{w_1+w_3}+\dots+X^{w_1+w_{{\rm dim}(R)}}+X^{w_2+w_3}+X^{w_2+w_4}\\
+\dots
+X^{w_2+w_{{\rm dim}(R)}}+\dots+X^{w_{{\rm dim}(R)-1}+w_{{\rm dim}(R)}}   ~~~~.
\end{split}
\end{equation}  
The character polynomial of symmetric square of R is 
\begin{equation}
\begin{split}
X^{2w_1}+X^{w_1+w_2}+\dots+X^{w_1+w_{{\rm dim}(R)}}+X^{2w_2}+X^{w_2+w_3}+\dots\\
+X^{w_2+w_{{\rm dim}(R)}}+\dots+X^{2w_{{\rm dim}(R)}} ~~~~.
\end{split}
\end{equation}  

One can quickly check the dimension. The dimension of $k$-th completely antisymmetric tensor power of $R$ is $\frac{{\rm dim}(R)!}{k![{\rm dim}(R)-k]!}$, which is also the number of monomials occur in its character polynomial. The dimension of $k$-th completely symmetric tensor power of $R$ is $\frac{{\rm dim}(R)\times [{\rm dim}(R)+1]\times \dots \times [{\rm dim}(R)+k-1] }{k!}$, which is also the number of monomials occur in its character polynomial.

Thus, based on the character polynomial approach, one can obtain the whole weight system of $k$-th completely (anti)symmetric tensor power of $R$ directly from the weight system of $R$, which is much more efficient than the projection matrix approach. 

This chapter describes available software \cite{Plethysm,Susyno,LiE,LieART}
and principles for designing algorithms to use for exploring the component field 
content of Salam-Strathdee superfields.  This is the {\bf {main}} result presented in 
this work.  The discussion has been presented at a high enough level that we believe 
these apply to superfields in a space of arbitrary dimensionality, either lower or 
higher than eleven.  One interesting example, to which this might be applied, is the 
a space where the Lorentz group is SO(2,10) as this has long been conjectured to 
relate to F-Theory \cite{FT}.

\subsection{Component Decomposition Results\label{sec:decomp_result}}
\label{sec:CDR}

By using the projection matrix and the Plethysm function with the Susyno Mathematica application \cite{Susyno}, we obtain the explicit Lorentz decomposition results of the 11D, $\mathcal{N}=1$ scalar superfield as follows. The decomposition results can also be expressed in terms of Dynkin Labels, which are listed in Appendix \ref{appen:dynkin}. 

\begin{itemize}
\sloppy
\item Level-0: $ \CMTB {\{1\}}$
\item Level-1: $ \CMTred {\{32\}}$
\item Level-2: $ \CMTB {\{1\}} \oplus \CMTB {\{165\}} \oplus \CMTB {\{330\}}$
\item Level-3: $\CMTred {\{32\}}  \oplus  \CMTred {\{1,408\}}  \oplus  \CMTred {\{3,520\}}$
\item Level-4: $ \CMTB {\{1\}} \oplus  \CMTB {\{165\}} \oplus  \CMTB {\{330\}} \oplus  \CMTB {\{1,144\}} \oplus  \CMTB {\{4,290\}} \oplus  \CMTB {\{5,005\}} \oplus  \CMTB {\{7,865\}} \oplus  \CMTB {\{17,160\}}$ 
\item Level-5: $\CMTred {\{32\}} \oplus \CMTred {\{1,408\}} \oplus \CMTred {\{3,520\}} \oplus \CMTred {\{4,224\}} \oplus \CMTred {\{10,240\}} \oplus \CMTred {\{24,960\}} \oplus \CMTred {\{28,512\}} \oplus \CMTred {\{36,960\}} \oplus \CMTred {\{91,520\}}$
\item Level-6: $\CMTB{\{1\}} \oplus \CMTB{\{165\}} \oplus \CMTB{\{330\}} \oplus \CMTB{\{1,144\}} \oplus \CMTB{\{4,290\}} \oplus \CMTB{\{5,005\}} \oplus \CMTB{\{7,128\}} \oplus \CMTB{\{7,865\}} \oplus \CMTB{\{15,400\}} \oplus (2) \CMTB{\{17,160\}} \oplus \CMTB{\{28,314\}} \oplus \CMTB{\{33,033\}} \oplus \CMTB{\{37,752\}} \oplus \CMTB{\{70,070\}} \oplus \CMTB{\{78,650\}} \oplus \CMTB{\{117,975\}} \oplus \CMTB{\{175,175\}} \oplus \CMTB{\{289,575\}}$
\item Level-7: $\CMTred{\{32\}} \oplus \CMTred{\{1,408\}} \oplus \CMTred{\{3,520\}} \oplus \CMTred{\{4,224\}} \oplus \CMTred{\{7,040\}} \oplus \CMTred{\{10,240\}} \oplus \CMTred{\{24,960\}} \oplus (2) \CMTred{\{28,512\}} \oplus \CMTred{\{36,960\}} \oplus \CMTred{\{45,056\}} \oplus \CMTred{\{45,760\}} \oplus (2) \CMTred{\{91,520\}} \oplus \CMTred{\{134,784\}} \oplus \CMTred{\{137,280\}} \oplus \CMTred{\{147,840\}} \oplus \CMTred{\{160,160\}} \oplus \CMTred{\{219,648\}} \oplus \CMTred{\{264,000\}} \oplus \CMTred{\{274,560\}} \oplus \CMTred{\{573,440\}} \oplus \CMTred{\{1,034,880\}}$
\item Level-8: $\CMTB{\{1\}} \oplus \CMTB{\{165\}} \oplus \CMTB{\{330\}} \oplus \CMTB{\{935\}} \oplus \CMTB{\{1,144\}} \oplus \CMTB{\{4,290\}} \oplus \CMTB{\{5,005\}} \oplus \CMTB{\{7,128\}} \oplus \CMTB{\{7,293\}} \oplus (2) \CMTB{\{7,865\}} \oplus (2) \CMTB{\{15,400\}} \oplus (2) \CMTB{\{17,160\}} \oplus \CMTB{\{22,275\}} \oplus \CMTB{\{23,595\}} \oplus \CMTB{\{23,595'\}} \oplus \CMTB{\{28,314\}} \oplus \CMTB{\{28,798\}} \oplus \CMTB{\{33,033\}} \oplus \CMTB{\{37,752\}} \oplus \CMTB{\{57,915\}} \oplus \CMTB{\{58,344\}} \oplus \CMTB{\{70,070\}} \oplus \CMTB{\{72,930\}} \oplus (2) \CMTB{\{78,650\}} \oplus \CMTB{\{85,085\}} \oplus \CMTB{\{112,200\}} \oplus (2) \CMTB{\{117,975\}} \oplus (2) \CMTB{\{175,175\}} \oplus \CMTB{\{178,750\}} \oplus \CMTB{\{188,760\}} \oplus \CMTB{\{255,255\}} \oplus \CMTB{\{268,125\}} \oplus (2) \CMTB{\{289,575\}} \oplus \CMTB{\{333,234\}} \oplus \CMTB{\{382,239\}} \oplus \CMTB{\{503,965\}} \oplus \CMTB{\{802,230\}} \oplus \CMTB{\{868,725\}} \oplus \CMTB{\{875,160\}} \oplus \CMTB{\{984,555\}} \oplus \CMTB{\{1,274,130\}} \oplus \CMTB{\{1,519,375\}}$
\item Level-9: $\CMTred{\{32\}} \oplus \CMTred{\{1,408\}} \oplus \CMTred{\{3,520\}} \oplus \CMTred{\{4,224\}} \oplus (2) \CMTred{\{7,040\}} \oplus \CMTred{\{10,240\}} \oplus \CMTred{\{22,880\}} \oplus \CMTred{\{24,960\}} \oplus (3) \CMTred{\{28,512\}} \oplus \CMTred{\{36,960\}} \oplus (2) \CMTred{\{45,056\}} \oplus (2) \CMTred{\{45,760\}} \oplus (3) \CMTred{\{91,520\}} \oplus \CMTred{\{128,128\}} \oplus (2) \CMTred{\{134,784\}} \oplus \CMTred{\{137,280\}} \oplus (2) \CMTred{\{147,840\}} \oplus \CMTred{\{157,696\}} \oplus (2) \CMTred{\{160,160\}} \oplus \CMTred{\{183,040\}} \oplus (3) \CMTred{\{219,648\}} \oplus \CMTred{\{251,680\}} \oplus (2) \CMTred{\{264,000\}} \oplus \CMTred{\{274,560\}} \oplus \CMTred{\{292,864\}} \oplus \CMTred{\{480,480\}} \oplus \CMTred{\{570,240\}} \oplus (2) \CMTred{\{573,440\}} \oplus \CMTred{\{798,720\}} \oplus \CMTred{\{896,896\}} \oplus \CMTred{\{901,120\}} \oplus (3) \CMTred{\{1,034,880\}} \oplus \CMTred{\{1,351,680\}} \oplus \CMTred{\{1,921,920\}} \oplus \CMTred{\{1,936,000\}} \oplus \CMTred{\{2,114,112\}} \oplus \CMTred{\{2,168,320\}} \oplus \CMTred{\{2,288,000\}} \oplus \CMTred{\{4,212,000\}}$
\item Level-10: $\CMTB{\{1\}} \oplus \CMTB{\{165\}} \oplus \CMTB{\{330\}} \oplus \CMTB{\{935\}} \oplus \CMTB{\{1,144\}} \oplus \CMTB{\{4,290\}} \oplus \CMTB{\{5,005\}} \oplus (2) \CMTB{\{7,128\}} \oplus \CMTB{\{7,293\}} \oplus (2) \CMTB{\{7,865\}} \oplus (3) \CMTB{\{15,400\}} \oplus (3) \CMTB{\{17,160\}} \oplus \CMTB{\{22,275\}} \oplus \CMTB{\{23,595\}} \oplus \CMTB{\{23,595'\}} \oplus \CMTB{\{26,520\}} \oplus \CMTB{\{28,314\}} \oplus \CMTB{\{28,798\}} \oplus (2) \CMTB{\{33,033\}} \oplus (2) \CMTB{\{37,752\}} \oplus \CMTB{\{57,915\}} \oplus (2) \CMTB{\{58,344\}} \oplus (2) \CMTB{\{70,070\}} \oplus \CMTB{\{72,930\}} \oplus (3) \CMTB{\{78,650\}} \oplus \CMTB{\{81,510\}} \oplus (2) \CMTB{\{85,085\}} \oplus \CMTB{\{112,200\}} \oplus (3) \CMTB{\{117,975\}} \oplus \CMTB{\{137,445\}} \oplus (3) \CMTB{\{175,175\}} \oplus (2) \CMTB{\{178,750\}} \oplus \CMTB{\{181,545\}} \oplus \CMTB{\{182,182\}} \oplus (2) \CMTB{\{188,760\}} \oplus \CMTB{\{255,255\}} \oplus \CMTB{\{268,125\}} 
\newline
\oplus (4) \CMTB{\{289,575\}} \oplus (2) \CMTB{\{333,234\}} \oplus (2) \CMTB{\{382,239\}} \oplus \CMTB{\{386,750\}} \oplus \CMTB{\{448,305\}} \oplus (3) \CMTB{\{503,965\}} \oplus \CMTB{\{525,525\}} \oplus \CMTB{\{616,616\}} \oplus \CMTB{\{650,650\}} \oplus \CMTB{\{715,715\}} \oplus (2) \CMTB{\{802,230\}} \oplus (2) \CMTB{\{868,725\}} \oplus (2) \CMTB{\{875,160\}} \oplus (2) \CMTB{\{984,555\}} \oplus \CMTB{\{1,002,001\}} \oplus \CMTB{\{1,100,385\}} \oplus (2) \CMTB{\{1,274,130\}} \oplus (2) \CMTB{\{1,310,309\}} \oplus \CMTB{\{1,412,840\}} \oplus (2) \CMTB{\{1,519,375\}} \oplus \CMTB{\{1,673,672\}} \oplus \CMTB{\{1,786,785\}} \oplus \CMTB{\{2,571,250\}} \oplus \CMTB{\{3,128,697\}} \oplus \CMTB{\{3,641,274\}} \oplus \CMTB{\{3,792,360\}} \oplus \CMTB{\{4,506,040\}} \oplus \CMTB{\{5,214,495\}} \oplus \CMTB{\{7,900,750\}}$
\item Level-11: $\CMTred {\{32\}} \oplus \CMTred {\{1,408\}} \oplus \CMTred {\{3,520\}} \oplus (2) \CMTred {\{4,224\}} \oplus (2) \CMTred {\{7,040\}} \oplus (2) \CMTred {\{10,240\}} \oplus \CMTred {\{22,880\}} \oplus (2) \CMTred {\{24,960\}} \oplus (4) \CMTred {\{28,512\}} \oplus (2) \CMTred {\{36,960\}} \oplus (2) \CMTred {\{45,056\}} \oplus (3) \CMTred {\{45,760\}} \oplus (4) \CMTred {\{91,520\}} \oplus \CMTred {\{128,128\}}  \oplus (4) \CMTred {\{134,784\}} \oplus (2) \CMTred {\{137,280\}} \oplus (3) \CMTred {\{147,840\}}{~} \oplus {~} (2) \CMTred {\{157,696\}} {~} \oplus {~} (3)  \CMTred {\{160,160\}} {~~~} \oplus 
\newline
(2) \CMTred {\{183,040\}} \oplus (4) \CMTred {\{219,648\}} \oplus \CMTred {\{251,680\}} \oplus (2) \CMTred {\{264,000\}} \oplus (2) \CMTred {\{274,560\}} \oplus (2) \CMTred {\{292,864\}} \oplus \CMTred {\{457,600\}} \oplus (3) \CMTred {\{480,480\}} \oplus (2) \CMTred {\{570,240\}} \oplus (4) \CMTred {\{573,440\}} \oplus \CMTred {\{672,672\}} \oplus (2) \CMTred {\{798,720\}} \oplus (2) \CMTred {\{896,896\}} \oplus (2) \CMTred {\{901,120\}} \oplus (5) \CMTred {\{1,034,880\}} \oplus \CMTred {\{1,140,480\}} \oplus \CMTred {\{1,351,680\}} \oplus \CMTred {\{1,425,600\}} \oplus \CMTred {\{1,757,184\}} \oplus (2) \CMTred {\{1,921,920\}} \oplus \CMTred {\{1,936,000\}} \oplus \CMTred {\{2,013,440\}} \oplus \CMTred {\{2,038,400\}} \oplus (3) \CMTred {\{2,114,112\}} \oplus (2) \CMTred {\{2,168,320\}} 
 \oplus (3) \CMTred {\{2,288,000\}} \oplus \CMTred {\{2,358,720\}} \oplus \CMTred {\{3,706,560\}} \oplus (3) \CMTred {\{4,212,000\}} \oplus \CMTred {\{5,857,280\}} \oplus \CMTred {\{5,930,496\}} \oplus \CMTred {\{6,040,320\}} \oplus \CMTred {\{7,208,960\}} \oplus \CMTred {\{8,781,696\}} \oplus \CMTred {\{9,123,840\}} \oplus \CMTred {\{11,714,560\}}$
\item Level-12: $\CMTB{\{1\}} \oplus \CMTB{\{165\}} \oplus \CMTB{\{330\}} \oplus \CMTB{\{935\}} \oplus (2) \CMTB{\{1,144\}} \oplus (2) \CMTB{\{4,290\}} \oplus (2) \CMTB{\{5,005\}} \oplus (2) \CMTB{\{7,128\}} \oplus \CMTB{\{7,150\}} \oplus \CMTB{\{7,293\}} \oplus (3) \CMTB{\{7,865\}} \oplus (3) \CMTB{\{15,400\}} \oplus (4) \CMTB{\{17,160\}} \oplus (2) \CMTB{\{22,275\}} \oplus \CMTB{\{23,595\}} \oplus \CMTB{\{23,595'\}} \oplus \CMTB{\{26,520\}} \oplus (2) \CMTB{\{28,314\}} \oplus (2) \CMTB{\{28,798\}} \oplus (3) \CMTB{\{33,033\}} \oplus (2) \CMTB{\{37,752\}} \oplus \CMTB{\{47,190\}} \oplus (2) \CMTB{\{57,915\}} \oplus (2) \CMTB{\{58,344\}} \oplus (2) \CMTB{\{70,070\}} \oplus \CMTB{\{72,930\}} \oplus (4) \CMTB{\{78,650\}} \oplus \CMTB{\{81,510\}} \oplus (3) \CMTB{\{85,085\}} \oplus \CMTB{\{91,960\}} \oplus \CMTB{\{112,200\}} \oplus (5) \CMTB{\{117,975\}} \oplus (2) \CMTB{\{137,445\}} \oplus (4) \CMTB{\{175,175\}} \oplus (3) \CMTB{\{178,750\}} \oplus \CMTB{\{181,545\}} 
\oplus \CMTB{\{182,182\}} \oplus (2) \CMTB{\{188,760\}} \oplus \CMTB{\{235,950\}} \oplus \CMTB{\{251,680'\}} \oplus (3) \CMTB{\{255,255\}} \oplus \CMTB{\{266,266\}} \oplus (2) \CMTB{\{268,125\}} 
\oplus (5) \CMTB{\{289,575\}} \oplus (3) \CMTB{\{333,234\}} \oplus (3) \CMTB{\{382,239\}} \oplus \CMTB{\{386,750\}} \oplus (2) \CMTB{\{448,305\}} \oplus (5) \CMTB{\{503,965\}} \oplus (2) \CMTB{\{525,525\}} \oplus \CMTB{\{616,616\}} \oplus \CMTB{\{650,650\}} \oplus \CMTB{\{715,715\}} \oplus \CMTB{\{722,358\}} \oplus (4) \CMTB{\{802,230\}} \oplus \CMTB{\{862,125\}} \oplus (4) \CMTB{\{868,725\}} \oplus (3) \CMTB{\{875,160\}} \oplus \CMTB{\{948,090\}} \oplus (3) \CMTB{\{984,555\}} \oplus \CMTB{\{1,002,001\}} \oplus (2) \CMTB{\{1,100,385\}} \oplus \CMTB{\{1,115,400\}} \oplus \CMTB{\{1,123,122\}} \oplus \CMTB{\{1,245,090\}} \oplus (3) \CMTB{\{1,274,130\}} \oplus (3) \CMTB{\{1,310,309\}} \oplus \CMTB{\{1,412,840\}} \oplus (3) \CMTB{\{1,519,375\}} \oplus (3) \CMTB{\{1,673,672\}} \oplus \CMTB{\{1,718,496\}} \oplus (2) \CMTB{\{1,786,785\}} \oplus \CMTB{\{2,147,145\}} \oplus \CMTB{\{2,450,250\}} \oplus \CMTB{\{2,571,250\}} \oplus \CMTB{\{2,743,125\}} \oplus (3) \CMTB{\{3,128,697\}} \oplus (2) \CMTB{\{3,641,274\}} \oplus (2) \CMTB{\{3,792,360\}} \oplus \CMTB{\{3,993,990\}} \oplus (2) \CMTB{\{4,506,040\}} \oplus \CMTB{\{4,708,704\}} \oplus (3) \CMTB{\{5,214,495\}} \oplus \CMTB{\{5,651,360\}} \oplus \CMTB{\{5,834,400\}} \oplus \CMTB{\{6,276,270\}} \oplus \CMTB{\{7,468,032\}} \oplus \CMTB{\{7,487,480\}} \oplus (2) \CMTB{\{7,900,750\}} \oplus \CMTB{\{11,981,970\}} \oplus \CMTB{\{14,889,875\}} \oplus \CMTB{\{20,084,064\}}$
\item Level-13: $\CMTred{\{32\}} \oplus (2) \CMTred{\{1,408\}} \oplus (2) \CMTred{\{3,520\}} \oplus (2) \CMTred{\{4,224\}} \oplus (2) \CMTred{\{7,040\}} \oplus (3) \CMTred{\{10,240\}} \oplus \CMTred{\{22,880\}} \oplus (3) \CMTred{\{24,960\}} \oplus (5) \CMTred{\{28,512\}} \oplus (3) \CMTred{\{36,960\}} \oplus (3) \CMTred{\{45,056\}} \oplus (4) \CMTred{\{45,760\}} \oplus (5) \CMTred{\{91,520\}} \oplus (2) \CMTred{\{128,128\}} 
\oplus (5) \CMTred{\{134,784\}} \oplus (3) \CMTred{\{137,280\}} \oplus (4) \CMTred{\{147,840\}} \oplus (3) \CMTred{\{157,696\}} \oplus (4) \CMTred{\{160,160\}} \oplus (3) \CMTred{\{183,040\}} \oplus (5) \CMTred{\{219,648\}} \oplus \CMTred{\{251,680\}} \oplus (2) \CMTred{\{264,000\}} \oplus (3) \CMTred{\{274,560\}} \oplus (2) \CMTred{\{292,864\}} \oplus \CMTred{\{302,016\}} \oplus (2) \CMTred{\{457,600\}} \oplus (4) \CMTred{\{480,480\}} \oplus (3) \CMTred{\{570,240\}} \oplus (6) \CMTred{\{573,440\}} \oplus (2) \CMTred{\{672,672\}} \oplus (3) \CMTred{\{798,720\}} \oplus (4) \CMTred{\{896,896\}} \oplus (3) \CMTred{\{901,120\}} \oplus (7) \CMTred{\{1,034,880\}} \oplus (2) \CMTred{\{1,140,480\}} \oplus \CMTred{\{1,171,456\}} \oplus \CMTred{\{1,351,680\}} \oplus (2) \CMTred{\{1,425,600\}} \oplus (2) \CMTred{\{1,757,184\}} \oplus (2) \CMTred{\{1,921,920\}} \oplus (2) \CMTred{\{1,936,000\}} \oplus (2) \CMTred{\{2,013,440\}} \oplus (2) \CMTred{\{2,038,400\}} \oplus (4) \CMTred{\{2,114,112\}} \oplus (3) \CMTred{\{2,168,320\}} \oplus (5) \CMTred{\{2,288,000\}} \oplus \CMTred{\{2,342,912\}} \oplus (2) \CMTred{\{2,358,720\}} \oplus \CMTred{\{2,402,400\}} \oplus (2) \CMTred{\{3,706,560\}} \oplus \CMTred{\{3,706,560'\}} \oplus (2) \CMTred{\{3,794,560\}} \oplus (5) \CMTred{\{4,212,000\}} \oplus \CMTred{\{5,720,000\}} \oplus (2) \CMTred{\{5,857,280\}} \oplus \CMTred{\{5,930,496\}} \oplus (2) \CMTred{\{6,040,320\}} \oplus \CMTred{\{6,864,000\}} \oplus (2) \CMTred{\{7,208,960\}} \oplus (2) \CMTred{\{8,781,696\}} \oplus (2) \CMTred{\{9,123,840\}} \oplus \CMTred{\{10,570,560'\}} \oplus (2) \CMTred{\{11,714,560\}} \oplus \CMTred{\{11,927,552\}} \oplus \CMTred{\{12,390,400\}} \oplus \CMTred{\{13,246,464\}} \oplus \CMTred{\{13,453,440\}} \oplus \CMTred{\{33,554,432\}}$
\item Level-14: $\CMTB{ \{1\}} \oplus (2) \CMTB{ \{165\}} \oplus (2) \CMTB{ \{330\}} \oplus \CMTB{ \{935\}} \oplus (2) \CMTB{ \{1,144\}} \oplus \CMTB{ \{1,430\}} \oplus \CMTB{ \{3,003\}} \oplus (2) \CMTB{ \{4,290\}} \oplus (2) \CMTB{ \{5,005\}} \oplus (3) \CMTB{ \{7,128\}} \oplus \CMTB{ \{7,150\}} \oplus \CMTB{ \{7,293\}} \oplus (3) \CMTB{ \{7,865\}} \oplus \CMTB{ \{11,583\}} \oplus (4) \CMTB{ \{15,400\}} \oplus (5) \CMTB{ \{17,160\}} \oplus (2) \CMTB{ \{22,275\}} \oplus (2) \CMTB{ \{23,595\}} \oplus \CMTB{ \{235,95'\}} \oplus (2) \CMTB{ \{26,520\}} \oplus (2) \CMTB{ \{28,314\}} \oplus (2) \CMTB{ \{28,798\}} \oplus (3) \CMTB{ \{33,033\}} \oplus (3) \CMTB{ \{37,752\}} \oplus \CMTB{ \{47,190\}} \oplus (2) \CMTB{ \{57,915\}} \oplus (3) \CMTB{ \{58,344\}} \oplus (3) \CMTB{ \{70,070\}} \oplus \CMTB{ \{72,930\}} \oplus (5) \CMTB{ \{78,650\}} \oplus (2) \CMTB{ \{81,510\}} \oplus (3) \CMTB{ \{85,085\}} \oplus \CMTB{ \{91,960\}} \oplus \CMTB{ \{112,200\}} \oplus (6) \CMTB{ \{117,975\}} \oplus (2) \CMTB{ \{137,445\}} \oplus \CMTB{ \{162,162\}} \oplus (5) \CMTB{ \{175,175\}} \oplus (4) \CMTB{ \{178,750\}} \oplus (2) \CMTB{ \{181,545\}} \oplus (2) \CMTB{ \{182,182\}} \oplus (2) \CMTB{ \{188,760\}} \oplus \CMTB{ \{218,295\}} \oplus \CMTB{ \{235,950\}} \oplus \CMTB{ \{2516,80'\}} \oplus (3) \CMTB{ \{255,255\}} \oplus \CMTB{ \{266,266\}} \oplus (2) \CMTB{ \{268,125\}} \oplus (7) \CMTB{ \{289,575\}} \oplus (4) \CMTB{ \{333,234\}} \oplus (4) \CMTB{ \{382,239\}} \oplus (2) \CMTB{ \{386,750\}} \oplus (2) \CMTB{ \{448,305\}} \oplus \CMTB{ \{490,490\}} \oplus (6) \CMTB{ \{503,965\}} \oplus (3) \CMTB{ \{525,525\}} \oplus \CMTB{ \{526,240\}} \oplus \CMTB{ \{616,616\}} \oplus (2) \CMTB{ \{650,650\}} \oplus \CMTB{ \{715,715\}} \oplus \CMTB{ \{722,358\}} \oplus (5) \CMTB{ \{802,230\}} \oplus \CMTB{ \{825,825\}} \oplus \CMTB{ \{862,125\}} \oplus (5) \CMTB{ \{868,725\}} \oplus (3) \CMTB{ \{875,160\}} \oplus (2) \CMTB{ \{948,090\}} \oplus (4) \CMTB{ \{984,555\}} \oplus \CMTB{ \{1,002,001\}} \oplus (3) \CMTB{ \{1,100,385\}} \oplus \CMTB{ \{1,115,400\}} \oplus \CMTB{ \{1,123,122\}} \oplus \CMTB{ \{1,190,112\}} \oplus \CMTB{ \{1,191,190\}} \oplus \CMTB{ \{1,245,090\}} \oplus (4) \CMTB{ \{1,274,130\}} \oplus (5) \CMTB{ \{1,310,309\}} \oplus (2) \CMTB{ \{1,412,840\}} \oplus (4) \CMTB{ \{1,519,375\}} \oplus \CMTB{ \{1,533,675\}} \oplus (4) \CMTB{ \{1,673,672\}} \oplus \CMTB{ \{1,718,496\}} \oplus (3) \CMTB{ \{1,786,785\}} \oplus \CMTB{ \{2,147,145\}} \oplus \CMTB{ \{2,450,250\}} \oplus (2) \CMTB{ \{2,571,250\}} \oplus (2) \CMTB{ \{2,743,125\}} \oplus \CMTB{ \{3,083,080\}} \oplus (4) \CMTB{ \{3,128,697\}} \oplus \CMTB{ \{3,586,440\}} \oplus (3) \CMTB{ \{3,641,274\}} \oplus (2) \CMTB{ \{3,792,360\}} \oplus \CMTB{ \{3,993,990\}} \oplus \CMTB{ \{4,332,042\}} \oplus (4) \CMTB{ \{4,506,040\}} \oplus (2) \CMTB{ \{4,708,704\}} \oplus \CMTB{ \{4,781,920\}} \oplus (5) \CMTB{ \{5,214,495\}} \oplus \CMTB{ \{52,144,95'\}} \oplus \CMTB{ \{5,651,360\}} \oplus \CMTB{ \{5,834,400\}} \oplus (2) \CMTB{ \{6,276,270\}} \oplus \CMTB{ \{7,468,032\}} \oplus (2) \CMTB{ \{7,487,480\}} \oplus \CMTB{ \{7,865,000\}} \oplus (3) \CMTB{ \{7,900,750\}} \oplus \CMTB{ \{9,845,550\}} \oplus \CMTB{ \{10,830,105\}} \oplus (2) \CMTB{ \{11,981,970\}} \oplus \CMTB{ \{12,972,960\}} \oplus \CMTB{ \{14,889,875\}} \oplus \CMTB{ \{17,606,160\}} \oplus \CMTB{ \{18,718,700\}} \oplus (2) \CMTB{ \{20,084,064\}} \oplus \CMTB{ \{31,082,480\}}$
\item Level-15: $(2) \CMTred{\{32\}} \oplus \CMTred{\{320\}} \oplus (2) \CMTred{\{1,408\}} \oplus \CMTred{\{1,760\}} \oplus (3) \CMTred{\{3,520\}} \oplus (2) \CMTred{\{4,224\}} \oplus \CMTred{\{5,280\}} \oplus (3) \CMTred{\{7,040\}} \oplus (3) \CMTred{\{10,240\}} \oplus (2) \CMTred{\{22,880\}} \oplus (3) \CMTred{\{24,960\}} \oplus (6) \CMTred{\{28,512\}} \oplus (3) \CMTred{\{36,960\}} \oplus (4) \CMTred{\{45,056\}} \oplus (4) \CMTred{\{45,760\}} \oplus \CMTred{\{64,064\}} \oplus (6) \CMTred{\{91,520\}} \oplus (3) \CMTred{\{128,128\}} \oplus (6) \CMTred{\{134,784\}} \oplus (3) \CMTred{\{137,280\}} \oplus (4) \CMTred{\{147,840\}} \oplus (3) \CMTred{\{157,696\}} \oplus (5) \CMTred{\{160,160\}} \oplus \CMTred{\{160,160'\}} \oplus (3) \CMTred{\{183,040\}} \oplus (6) \CMTred{\{219,648\}} \oplus \CMTred{\{251,680\}} \oplus (3) \CMTred{\{264,000\}} \oplus (3) \CMTred{\{274,560\}} \oplus (3) \CMTred{\{292,864\}} \oplus \CMTred{\{302,016\}} \oplus \CMTred{\{366,080\}} \oplus (2) \CMTred{\{457,600\}} \oplus (5) \CMTred{\{480,480\}} \oplus (3) \CMTred{\{570,240\}} \oplus (7) \CMTred{\{573,440\}} \oplus (2) \CMTred{\{672,672\}} \oplus (4) \CMTred{\{798,720\}} \oplus (5) \CMTred{\{896,896\}} \oplus (4) \CMTred{\{901,120\}} \oplus (8) \CMTred{\{1,034,880\}} \oplus (3) \CMTred{\{1,140,480\}} \oplus \CMTred{\{1,171,456\}} \oplus \CMTred{\{1,208,064\}} \oplus (2) \CMTred{\{1,351,680\}} \oplus (3) \CMTred{\{1,425,600\}} \oplus (2) \CMTred{\{1,757,184\}} \oplus (2) \CMTred{\{1,921,920\}} \oplus (3) \CMTred{\{1,936,000\}} \oplus (3) \CMTred{\{2,013,440\}} \oplus (2) \CMTred{\{2,038,400\}} \oplus (5) \CMTred{\{2,114,112\}} \oplus (3) \CMTred{\{2,168,320\}} \oplus (6) \CMTred{\{2,288,000\}} \oplus \CMTred{\{2,342,912\}} \oplus (3) \CMTred{\{2,358,720\}} \oplus (2) \CMTred{\{2,402,400\}} \oplus \CMTred{\{2,446,080\}} \oplus (3) \CMTred{\{3,706,560\}} \oplus (2) \CMTred{\{3,706,560'\}} \oplus (3) \CMTred{\{3,794,560\}} \oplus \CMTred{\{4,026,880\}} \oplus (6) \CMTred{\{4,212,000\}} \oplus (2) \CMTred{\{5,720,000\}} \oplus (2) \CMTred{\{5,857,280\}} \oplus \CMTred{\{5,930,496\}} \oplus (3) \CMTred{\{6,040,320\}} \oplus \CMTred{\{6,307,840\}} \oplus \CMTred{\{6,864,000\}} \oplus (3) \CMTred{\{7,208,960\}} \oplus (3) \CMTred{\{8,781,696\}} \oplus (3) \CMTred{\{9,123,840\}} \oplus \CMTred{\{10,570,560\}} \oplus \CMTred{\{10,570,560'\}} \oplus (2) \CMTred{\{11,714,560\}} \oplus \CMTred{\{11,927,552\}} \oplus (2) \CMTred{\{12,390,400\}} \oplus (2) \CMTred{\{13,246,464\}} \oplus (2) \CMTred{\{13,453,440\}} \oplus \CMTred{\{15,375,360\}} \oplus \CMTred{\{30,201,600\}} \oplus \CMTred{\{33,116,160\}} \oplus \CMTred{\{33,554,432\}}$
\item Level-16: $(2) \CMTB{\{1\}} \oplus \CMTB{\{11\}} \oplus \CMTB{ \{65\}} \oplus (2) \CMTB{\{165\}} \oplus \CMTB{\{275\}} \oplus (2) \CMTB{\{330\}} \oplus \CMTB{\{462\}} \oplus (2) \CMTB{\{935\}} \oplus (2) \CMTB{\{1,144\}} \oplus \CMTB{\{1,430\}} \oplus \CMTB{\{2,717\}} \oplus \CMTB{\{3,003\}} \oplus (3) \CMTB{\{4,290\}} \oplus (2) \CMTB{\{5,005\}} \oplus \CMTB{\{7,007\}} \oplus (3) \CMTB{\{7,128\}} \oplus \CMTB{\{7,150\}} \oplus \CMTB{\{7,293\}} \oplus (4) \CMTB{\{7,865\}} \oplus \CMTB{\{11,583\}} \oplus (4) \CMTB{\{15,400\}} \oplus \CMTB{\{16,445\}} \oplus (5) \CMTB{\{17,160\}} \oplus (3) \CMTB{\{22,275\}} \oplus (3) \CMTB{\{23,595\}} \oplus (2) \CMTB{\{23,595'\}} \oplus (2) \CMTB{\{26,520\}} \oplus (2) \CMTB{\{28,314\}} \oplus (2) \CMTB{\{28,798\}} \oplus (3) \CMTB{\{33,033\}} \oplus \CMTB{\{35,750\}} \oplus (3) \CMTB{\{37,752\}} \oplus \CMTB{\{47,190\}} \oplus (3) \CMTB{\{57,915\}} \oplus (3) \CMTB{\{58,344\}} \oplus (3) \CMTB{\{70,070\}} \oplus \CMTB{\{72,930\}} \oplus (5) \CMTB{\{78,650\}} \oplus (2) \CMTB{\{81,510\}} \oplus (4) \CMTB{\{85,085\}} \oplus \CMTB{\{91,960\}} \oplus (2) \CMTB{\{112,200\}} \oplus (6) \CMTB{\{117,975\}} \oplus (2) \CMTB{\{137,445\}} \oplus \CMTB{\{162,162\}} \oplus (5) \CMTB{\{175,175\}} \oplus (5) \CMTB{\{178,750\}} \oplus (2) \CMTB{\{181,545\}} \oplus (2) \CMTB{\{182,182\}} \oplus (3) \CMTB{\{188,760\}} \oplus \CMTB{\{218,295\}} \oplus \CMTB{\{235,950\}} \oplus \CMTB{\{251,680'\}} \oplus (4) \CMTB{\{255,255\}} \oplus (2) \CMTB{\{266,266\}} \oplus (3) \CMTB{\{268,125\}} \oplus (7) \CMTB{\{289,575\}} \oplus (4) \CMTB{\{333,234\}} \oplus (4) \CMTB{\{382,239\}} \oplus (2) \CMTB{\{386,750\}} \oplus (2) \CMTB{\{448,305\}} \oplus \CMTB{\{490,490\}} \oplus (6) \CMTB{\{503,965\}} \oplus (3) \CMTB{\{525,525\}} \oplus \CMTB{\{526,240\}} \oplus \CMTB{\{616,616\}} \oplus \CMTB{\{628,320\}} \oplus (2) \CMTB{\{650,650\}} \oplus \CMTB{\{674,817\}} \oplus \CMTB{\{715,715\}} \oplus (2) \CMTB{\{722,358\}} \oplus (6) \CMTB{\{802,230\}} \oplus \CMTB{\{825,825\}} \oplus (2) \CMTB{\{862,125\}} \oplus (6) \CMTB{\{868,725\}} \oplus (4) \CMTB{\{875,160\}} \oplus (2) \CMTB{\{948,090\}} \oplus (4) \CMTB{\{984,555\}} \oplus \CMTB{\{1,002,001\}} \oplus (3) \CMTB{\{1,100,385\}} \oplus (2) \CMTB{\{1,115,400\}} \oplus (2) \CMTB{\{1,123,122\}} \oplus \CMTB{\{1,190,112\}} \oplus \CMTB{\{1,191,190\}} \oplus \CMTB{\{1,245,090\}} \oplus (4) \CMTB{\{1,274,130\}} \oplus (5) \CMTB{\{1,310,309\}} \oplus (2) \CMTB{\{1,412,840\}} \oplus (5) \CMTB{\{1,519,375\}} \oplus \CMTB{\{1,533,675\}} \oplus (4) \CMTB{\{1,673,672\}} \oplus (2) \CMTB{\{1,718,496\}} \oplus \CMTB{\{1,758,120\}} \oplus (3) \CMTB{\{1,786,785\}} \oplus \CMTB{\{2,147,145\}} \oplus (2) \CMTB{\{2,450,250\}} \oplus (2) \CMTB{\{2,571,250\}} \oplus \CMTB{\{2,598,960\}} \oplus (3) \CMTB{\{2,743,125\}} \oplus \CMTB{\{2,858,856\}} \oplus \CMTB{\{3,056,625\}} \oplus \CMTB{\{3,083,080\}} \oplus (4) \CMTB{\{3,128,697\}} \oplus \CMTB{\{3,586,440\}} \oplus (3) \CMTB{\{3,641,274\}} \oplus (2) \CMTB{\{3,792,360\}} \oplus \CMTB{\{3,993,990\}} \oplus \CMTB{\{4,332,042\}} \oplus (4) \CMTB{\{4,506,040\}} \oplus (2) \CMTB{\{4,708,704\}} \oplus \CMTB{\{4,781,920\}} \oplus (6) \CMTB{\{5,214,495\}} \oplus (2) \CMTB{\{5,214,495'\}} \oplus (2) \CMTB{\{5,651,360\}} \oplus \CMTB{\{5,834,400\}} \oplus (2) \CMTB{\{6,276,270\}} \oplus \CMTB{\{7,468,032\}} \oplus (3) \CMTB{\{7,487,480\}} \oplus (2) \CMTB{\{7,865,000\}} \oplus (3) \CMTB{\{7,900,750\}} \oplus \CMTB{\{8,893,500\}} \oplus \CMTB{\{9,845,550\}} \oplus \CMTB{\{10,696,400'\}} \oplus \CMTB{\{10,830,105\}} \oplus (2) \CMTB{\{11,981,970\}} \oplus \CMTB{\{12,972,960\}} \oplus \CMTB{\{14,889,875\}} \oplus \CMTB{\{17,606,160\}} \oplus \CMTB{\{18,718,700\}} \oplus (3) \CMTB{\{20,084,064\}} \oplus \CMTB{\{30,604,288\}} \oplus \CMTB{\{31,082,480\}}$
\end{itemize}
Level-17 to 32 are the same as Level-15 to 0 respectively since all irreducible representations in SO(11) are self-conjugate. Moreover, the irreps corresponding to component fields are the same as the $\theta-$monomials.

This is also consistent with the existence of the spinor metric $C_{\a\b}$ and $C^{\a\b}$. Consider a field with a upstairs spinor index $\chi^{\a}$ and assign it with the irrep $\CMTred{\{32\}}$. We can lower the spinor index by
\begin{equation}
    \chi_{\b} ~=~ \chi^{\a}C_{\a\b} ~~,
\end{equation}
and the irrep corresponding to $\chi_{\b}$ is also $\CMTred{\{32\}}$. That means in 11D, $\mathcal{N} = 1$ case, the position of the spinor index doesn't matter in the context of representation theory. 

There is an aspect of the results shown over pages 31-34 that a sophisticated reader may find puzzling.
At numbers of levels, there are multiple occurrence of the same representation.  Thus, one is led to wonder
how this can occur?  Given the necessity (for the sake of efficiency) of expanding the superfield over the irreducible polynomials $\big[ \theta{}^1 \cdots \theta{}^p \big]{}_{IR}$ at Level-$p$, and not just 
$ \theta{}^1 \cdots \theta{}^p$, is a requirement, a mechanism by which this phenomenon occurs can 
easily be identified.

At low order it was shown (see (\ref{equ:cubicirr})) that starting from the most general Lorentz covariant possibilities, multiple
expressions could be written.  However at those orders, all such irreducible polynomials $\big[ \theta{}^1 \cdots \theta{}^p \big]{}_{IR}$ of the same dimensionality were found to be proportional to one another among such sets.  This was proven by investigations of Fierz identities.

Starting at Level-7, one sees that two independent $\CMTR { \{ 91,520 \} }$ representations are {\it {required}} 
by the branching rule, etc. to appear.  This can be accommodated {\it {only}} if there are two {\it {linearly}} {\it {independent}}
irreducible polynomials at Level-$7$, i. e. $\big[ \theta{}^1 \cdots \theta{}^7 \big]{}_{IR}^{1, \, {\CMTR { \{ 91,520 \} }}} $ and $\big[ \theta{}^1 \cdots \theta{}^7 \big]{}_{IR}^{2, \, {\CMTR { \{ 91,520 \} }}} $.

\subsection{11D, $\mathcal{N} = 1$ Theory to 10D, $\mathcal{N} = 2$A Theory: $\mathfrak{so}(11)\supset\mathfrak{so}(10)$}
\label{sec:11DN2A}

Since Type IIA theory can be obtained by the projection from 11D, $\mathcal{N} = 1$ theory, we can reproduce the scalar superfield decomposition results in 10D, $\mathcal{N} = 2$A superspace, which was listed in Chapter six of \cite{counting10d} by projecting 11D, $\mathcal{N} = 1$ component decomposition results into 10D. In one specified level, we restrict each irreps of $\mathfrak{so}(11)$ into $\mathfrak{so}(10)$ and consequently obtain a direct sum of several irreps of $\mathfrak{so}(10)$. The projection matrix of $\mathfrak{so}(11)\supset\mathfrak{so}(10)$ is 
\begin{equation}
\begin{split}
     P_{\mathfrak{so}(11)\supset\mathfrak{so}(10)}  
    ~=~
\begin{pmatrix}
0 & 0 & 0& 1& 0\\
0& 0& 1 &0 &0 \\
0& 1& 0 &0& 0\\
1& 0& 0 &0& 0\\
-1 &-2 &-2 &-2 &-1 \\
\end{pmatrix}    ~~~.
\end{split}
\end{equation}

\subsection{11D, $\mathcal{N} = 1$ Breitenlohner Approach}
\label{sec:11DB}

In 11D, $\mathcal{N} = 1$ superspace, the graviton has $(11\times 12)/2 = 66$ degrees of freedom and can be split into the conformal part and non-conformal part
\begin{equation}
\begin{split}
    \tilde{h}_{\un a\un b} ~=&~ h_{\un a\un b} ~+~ \eta_{\un a\un b} h  ~~~, \\
     \{66\} ~=&~ \CMTB{\{65\}} ~\oplus~ \CMTB{ \{1\}} ~~~.
\end{split}
\end{equation}
Similarly, the gravitino has $11\times32=352$ degrees of freedom and can be split as 
\begin{equation}
\begin{split}
    \tilde{\psi}_{\un a}{}^{\b} ~=&~ \psi_{\un a}{}^{\b} ~-~ \frac{1}{11}(\g_{\un a})^{\b\g}\psi_{\g}  ~~~, \\
    \CMTB{\{11\}} ~\otimes~ \CMTred{\{32\}}~=&~ \CMTred{\{320\}} ~\oplus~ \CMTred{\{32\}} ~~~,
\end{split}
\end{equation}
where the non-conformal ``spin-$\frac{1}{2}$ part'' of the gravitino is defined as $\psi_{\b} ~\equiv~ (\g^{\un{a}})_{\b\g} \tilde{\psi}_{\un{a}}{}^{\g}$.  A final interesting
note to make concerns the three-form gauge field $b_{\un a\un b \un c}$ which is known to occur in the on-shell eleven dimensional supergravity theory.  Since this bosonic gauge field is a form, it is already obvious that it is irreducible and it follows as far as representation goes ${b}_{\un a\un b \un c} = \CMTB{\{165\}}$.

It can be seen that at Level-16 there occurs a boson in the $\CMTB {\{ 65 \}} $ representation.  Also at this same level, there occur two bosons in the $\CMTB {\{ 165 \}} $ representation.
Finally, at Level-15, there occurs one fermion in the $\CMTR {\{ 320 \}} $ representation which implies at Level-17 there occurs one fermion in the $\CMTR {\{ 320 \}} $ representation. This is consistent with SUSY transformation laws of the graviton $h_{\un a\un b}$ in 11D, $\mathcal{N} = 1$ theory. Acting the  D-operator on the graviton gives a term proportional to the gravitino in the on-shell case, 
\begin{equation}
    D_{\alpha}h_{\un a\un b} ~\propto~ (\gamma_{(\un a})_{\a\b}\psi_{\un b)}{}^{\b} ~~~,
\end{equation}
while in the off-shell case, there are several auxiliary fields showing up in the r.h.s. besides the gravitino.

This tells us the scalar superfield gives one possible embedding for the graviton, two possible embeddings for the gravitino, along with a number of auxiliary fields. It's possible that the scalar superfield itself plays the roles both as prepotential and conformal compensator.
This is not a {\rm {new}} phenomenon.  In 4D, $\cal N$ = 1 supergravity
among a number of off-shell distinct formulations, there exists one
where the vector superfield $H{}^{\un a}$ provides the superconformal supermultiplet as well as the compensator supermultiplet.  Momentarily in our discussion, let us depart
the domain of the 11D theory to discuss this particular 
4D, $\cal N$ = 1 supergravity theory.

Among the many forms of {\it {irreducible}} off-shell 4D, $\cal N$ = 1 supergravity, there is the one first described in the work \cite{VaR}.  This form of the off-shell theory 
possesses one prepotential: the conformal prepotential ${H}{}^{\un a}$ and all the component fields of the theory reside in it.  The component fields associated with this form of supergravity include the 4D graviton, gravitino, the axial vector auxiliary field, and two auxiliary 3-forms,
\begin{equation}
    h{}_{\un a\un b}(x), ~\psi_{\un a}{}^{\b}(x),~
    A{}_{\un a}(x), ~ b{}_{\un a\un b  \un c}(x), ~ {\Bar b}{}_{\un a\un b
    \un c}(x) ~~~.
\end{equation}
Although often overlooked, this is one of the original off-shell formulations (the Stelle-West formulation) known in the
literature  \cite{0VRT}.  In this limit the linearized frame superfields take the form as shown in\footnote{These equations first appeared in the works {\it {Superspace}}
\cite{SpRSp8c} and are written in the conventions of that work.} \cite{CoDeX}
\begin{equation}
\begin{split}
{\rm E}_{\alpha} ~=&~  {\rm D}_{\alpha} + \Bar{X} {\rm D}_{\alpha} + i \frac 12 \,
({\rm D}_{\alpha} H^{\un{b}}) \pa_{\un{b}} ~~~,   \\
{\rm E}_{\Dot{\alpha}} ~=&~  \Bar{{\rm D}}_{\Dot{\alpha}} + X \Bar{{\rm D}}_{\Dot{\alpha}}
- i \frac 12 \, ( \Bar{{\rm D}}_{\Dot{\alpha}} H^{\un{b}}) \pa_{\un{b}} ~~~,
\end{split}
\end{equation}
\begin{equation}
\label{equ:LiN1}
\begin{split}
{\rm E}_{\un{a}} ~=&~ \pa_{\un{a}} + i \Big[ \frac{1}{2} \Bar{{\rm D}}^{2} {\rm D}_{(\alpha} 
H^{\gamma)}{}_{\Dot{\alpha}} - (\Bar{{\rm D}}_{\Dot{\alpha}} \Bar{X}) \delta_{\alpha}{}^{
\gamma} \Big] {\rm D}_{\gamma} + i \Big[ - \frac{1}{2} {\rm D}^{2} \Bar{{\rm D}}_{(\Dot{
\alpha}} H_{\alpha}{}^{\Dot{\gamma})} - ({\rm D}_{\alpha} X)  \delta_{\Dot{\alpha}}{}^{
\Dot{\gamma}} \Big] \Bar{{\rm D}}_{\Dot{\gamma}}  \\
&  + \Big[ -\, \frac 12 ( \, [ {\rm D}_{\alpha} ~,~ \Bar{{\rm D}}_{\Dot{\alpha}} ] H^{\un{b}}) + 
(X + \Bar{X}) \delta_{\un{a}}{}^{\un{b}} \Big] \pa_{\un{b}} ~~~,  \\
& {~}  \\
& X ~=~ - \, \frac 16 \left( \, 2 \, \Bar{{\rm D}}_{\Dot{\alpha}}
{{\rm D}}_{\alpha} ~+~{{\rm D}}_{\alpha}  \Bar{{\rm D}}_{\Dot{\alpha}} \, \right) \, H^{\un{a}}
~~~.
\end{split}
\end{equation}
As promised all of the SG supermultiplet component fields in this formulation and in this WZ gauge arise from the $\theta$-expansion of $H{}^{\un a}$.  From (\ref{equ:LiN1}) we extract the conformal
part of the linearized supergraviton and find
\begin{equation}
\label{equ:LiN2}
\eqalign{ {~~~~~}
h{}_{\un{a} \un{b}} ~=&~ -\, \frac 12 \, \Big[  \big( \, [ {\rm D}_{\alpha} ~,~ \Bar{{\rm D}}_{\Dot{\alpha}} ] H{}_{\un{b}} \big)  ~+~ 
[ {\rm D}_{\beta} ~,~ \Bar{{\rm D}}_{\Dot{\beta}} ] H{}_{\un{a}} \big)  \, \Big]  ~+~ \frac{1}{4}
\, C{}_{\a \b} \, C{}_{{\Dot \a} {\Dot \b}} \,
\big( [ {\rm D}_{\g} ~,~ \Bar{{\rm D}}_{\Dot{\g}} ] H{}^{\un{c}} \big) 
  \cr
~=&~ -\, \frac 12 \, \Big[ \d{}_{\a}{}^{\g} \,  
\d{}_{\Dot \a}{}^{\Dot \g} \, \d{}_{\un b}{}^{\un d} ~+~ \d{}_{\b}{}^{\g} \,  
\d{}_{\Dot \b}{}^{\Dot \g} \, \d{}_{\un a}{}^{\un d} \, 
~-~ \fracm{1}{2}
\, C{}_{\a \b} \, C{}_{{\Dot \a} {\Dot \b}} \, C{}^{\g \d} \, C{}^{{\Dot \g} {\Dot \d}} 
\Big] \, \big( \, [ {\rm D}_{\g} ~,~ \Bar{{\rm D}}_{\Dot{\g}} ] H{}_{\un{d}} \big)  
   \cr
~=&~ -\, \frac 12 \, \Big[ \d{}_{\un a}{}^{\un c} \, \d{}_{\un b}{}^{\un d} ~+~ \d{}_{\un b}{}^{\un c} \, \d{}_{\un a}{}^{\un d} \, ~-~ \fracm{1}{2}\, \eta{}_{{\un a}{\un b}} \eta{}^{{\un c}{\un d}} \, 
\Big] \, \big( \, [ {\rm D}_{\g} ~,~ \Bar{{\rm D}}_{\Dot{\g}} ] H{}_{\un{d}} \big) \cr  
 ~\equiv&~
 {\cal T}{}^{\,\, {\g} \, {\Dot \g} \, {\un d}}
 {}_{{\un a}\,  {\un b}  } \, \big( \, [ {\rm D}_{\g} ~,~ \Bar{{\rm D}}_{\Dot{\g}} ] H{}_{\un{d}} \big) 
~~~,}
\end{equation}
where the quantity ${\cal T}{}^{\,\, {\g} \, {\Dot \g} \, {\un d}} {}_{{\un a}\,  {\un b}  } $ on the
final line of (\ref{equ:LiN2}) is defined by the first factor on the preceding line.
Let us rewrite this final equation by first introducing the 4D, $\cal N$ = 1 scalar superfield $v$
\begin{equation}
\label{equ:4D_superfield_expansion}
\begin{split}
    {v}(x^{\un{a}},\theta^{\a}) ~=&~ f(x^{\un{a}}) ~+~ \theta^{\a} \, \psi_{\a}(x^{
    \un{a}}) ~+~ \theta^{\a}\theta^{\b} [ \, C_{\a\b} \, g(x^{\un{a}})~+~ i(\g^{5})_{\a\b} \, h(x^{\un{a}}) ~+~
     i(\g^{5}\g^{\un b})_{\a\b} \, ] v_{\un b}(x^{\un{a}}) \\
    &+~ \theta^{\a}\theta^{\b}\theta^{\g} C_{\a\b}C_{\g\d} \, \chi^{\d}(x^{\un{a}}) 
    ~+~ \theta^{\a}\theta^{\b} \theta^{\g}\theta^{\d} C_{\a\b}C_{\g\d} \, N(x^{\un{a}})
\end{split}
\end{equation}
and this representation can be ``tensored'' with the $\CMTB{ \{10\} }$ and $\CMTB{\{4\}}$ 
representations of the SO(1,3) algebra.  Doing this we find
\be
\left[ \, {v} \otimes {\CMTB{\{10\} } } \, \right] ~=~ h{}_{\un{a} \un{b}}
~~,~~ 
\left[ \, {v} \otimes {\CMTB{\{4 \} }} \, \right] ~=~ H{}_{\un{a}} ~~~,
\ee
so that the final line of (\ref{equ:LiN2}) yields
\begin{equation}
\label{equ:LiN2xx}
\eqalign{ 
\left[ \, 
v \, \otimes \, {\CMTB{\{10\} } }
\,\right] 
~=~
{\cal T}{}^{\,\, {\g} \, {\Dot \g} \, {\un d}}
 {}_{{\un a}\,  {\un b}  } \, \left( \, [ {\rm D}_{\g} ~,~ \Bar{{\rm D}}_{\Dot{\g}} ] \,
[ v \, \otimes \, {\CMTB{\{4\} } } ] \right)
~~~.}
\end{equation}

Returning to the case of 11D, one may impose the condition
\begin{equation}
\left[ \,{\cal V} \otimes \CMTB{\{65\}} \, \right]~=~  {\cal T}{}^{{\a}_1 \cdots{\a}_{16}}{}_{\CMTB{\{65\}} \,  } \big( \, {\rm D}_{ [ {\a}_1} \cdots {\rm D}_{{\a}_{16} ] } \, {\cal V}\, \big) ~~~~,
\label{equ:CnSTRNT}
\end{equation}
which has the effect of implying that the graviton candidate at Level-0 in ${\cal V} \otimes \CMTB{\{65\}} $ and the graviton candidate at Level-16 in the ${\cal V} $ representation are one and the same field.  In this
equation, the term ${\cal T}{}^{{\g}_1 \cdots{\g}_{16}}{}_{\CMTB{\{65\}} \, }$ are a set of quantities chosen so that the equation is consistent
with SO(1,10) Lorentz symmetry.  In other words,
${\cal T}{}^{{\g}_1 \cdots{\g}_{16}}{}_{\CMTB{\{65\}} \, }$ in (\ref{equ:CnSTRNT}) is the analog of
${\cal T}{}^{\,\, {\g} \, {\Dot \g} \, {\un d}} {}_{{\un a}\,  {\un b}  } $ in (\ref{equ:LiN2}).
As in the 4D theory, the graviton occurs at second
order in the $\theta$-expansion of $H{}^{\un a}$,
two spinorial derivatives occur in (\ref{equ:LiN2}).
For the proposed 11D theory, the graviton occurs at sixteenth
order in the $\theta$-expansion of ${\cal V}$, hence
sixteen spinorial derivatives occur in (\ref{equ:CnSTRNT}).

\subsection{Using The $\cal V$ Gateway }
\label{sec:USV}

In case the value in the results listed across pages 31 - 34 or equivalently in Appendix \ref{appen:dynkin} 
are not apparent, let us here use the symbol $\cal V$ for this listing, also we can use $\CMTB {\{ \cal B 
\}}$, and $\CMTR { \{ \cal F \}}$ for any bosonic or fermionic representation (respectively) of 11D spacetime 
symmetry. The {\it {explicit}} spectrum of component fields in {\it {any}} representation of the 11D Lorentz 
symmetry is found from the multiplications of representations ${\cal V}\otimes \CMTB {\{ \cal B \}}$ or ${\cal V}\otimes\CMTR { \{ 
\cal F \}}$.  Thus, $\cal V$ is a gateway to the explicit component spectrum of {\it {all}} 11D superfields.  
Below, we apply this technology to execute other scans for the conformal graviton, etc..

With the explicit knowledge of $\cal V$ in hand, it is possible to construct scans in many different ways.
The use of this $\cal V$-gateway allows for rather flexible scans in addition a simple enumeration of the
component field spectrum of the 11D, $\cal N$ = 1 scalar superfield.  Let us work through some examples.  

We know that the conformal graviton occurs in the scalar superfield. That led us to wonder how frequently it appears in other superfields. We have performed a computer-based search involving tensoring the scalar superfield up to the irrep dimension 260,338.

However, if we demand that the graviton $\CMTB{\{65\}}$ must occur at the middle level (Level-16) {\emph{only}}, the gravitino $\CMTred{\{320\}}$ must appear at the next level (Level-17), and the 3-form $\CMTB{\{165\}}$ must appear at the same level as the graviton (Level-16), the number of superfields that satisfy all these conditions drops drastically, from 91 to 4. They are listed in Table \ref{Tab:graviton16}, where the numbers of graviton(s), gravitino(s) and 3-form(s) that occur at Level-16, Level-17 and Level-16 (let them be $b_{\CMTB{\{65\}}}$, $b_{\CMTred{\{320\}}}$ and $b_{\CMTB{\{165\}}}$ respectively) are shown. 
\begingroup
\def\arraystretch{1.8}
\begin{table}[h!]
\centering
\begin{tabular}{|c|c|c|} \hline
    Dynkin Label & Irrep & $(b_{\CMTB{\{65\}}}, b_{\CMTred{\{320\}}}, b_{\CMTB{\{165\}}})$ \\ \hline \hline
    $\CMTB{(00000)}$ & $\CMTB{\{1\}}$ & $(1,1,2)$  \\ \hline
    $\CMTB{(10000)}$ & $\CMTB{\{11\}}$ & $(2,6,3)$  \\ \hline 
    $\CMTB{(70000)}$ & $\CMTB{\{16,445\}}$ & $(2,4,2)$  \\ \hline 
    $\CMTB{(80000)}$ & $\CMTB{\{35,750\}}$ & $(2,2,1)$  \\ \hline 
\end{tabular}
\caption{Summary of bosonic superfields that contain graviton(s) only at Level-16, gravitino(s) at Level-17 and 3-form(s) at Level-16 \label{Tab:graviton16}}
\end{table}
\endgroup
\newline \noindent

The appearance of the conformal graviton representation in the case of the 11D, $\cal N$ = 1 theory is very different than the behavior seen in the case of the 10D, $\cal N$ = 1 theory. In the latter case, the conformal graviton representation only occurs in some particular cases of tensoring between either bosonic or spinorial irreps of SO(1,9) and the scalar superfield. In the former case, the conformal graviton represenation appears in every case where either a bosonic or a spinorial irrep of SO(1,10) is tensored with the 11D, $\cal N$ = 1 scalar superfield up to the case of the $\CMTB{\{255,255\}}$ irrep. The next irrep, $\CMTB{\{260,338\}}$, does not contain the conformal graviton at any level.

Below we describe one final scan, though it will not be undertaken.
The work of \cite{M2} noted 11D, $\cal N$ = 1 superspace geometry is consistent\footnote{These
constraints also appear consistent with the analysis given in the works of \cite{Howe:2003sa,Howe:2003cy}.} 
with superspace scale symmetry if the constraints
\be {
\eqalign{
 i \,  \fracm 1{32} \,  (\g_{\un a})^{\a \b} \, T_{\a \b}{}^{\un b} &=~   \d_{\un a} \, {}^{\un b} 
~~\,~~~~,{~~}~~~~~~~~~~~~~~~~~~~~ (\g_{\un a})^{\a \b} \, T_{\a \b}{}^{\g} ~=~ 0 ~~~~~~~,
~~~~~
 \cr 
{~~~~~~~~~~}
T_{\a \, [ {\un d} {\un e}]} ~-~ \fracm 2{55} \, (\g_{{\un d} {\un e}})_{\a}{}^{\g}  
\, T_{\g {\un b}}{}^{\un b} ~&=~0 ~~~~\,~~~\,~, {~~}~~~~~~~~~~~~~~~~~~~
(\g_{\un a})^{\a \b} \, R_{\a \b}{}^{{\un d}{\un e}} ~=~ 0 ~~~~~~~,~~~ ~~
\cr 
(\g_{{\un a}{\un b}})^{\a \b} \,T_{\a \b}{}^{\un b} ~&=~ 0 ~\,~
~~\,~~~~,{~~}~~~~~~~~~~~~~\,~ ~~~(\g_{[{\un a} {\un b}|})^{\a \b} \, T_{\a \b}{}_{ | {\un c}]} ~=~ 0 
~~~~~~~, ~~\cr
(\g_{{\un a}  {\un b}  {\un c}  {\un d} {\un e}})^{\a \b} \, T_{\a \b}{ }^{\un e} &=~ 0 ~~~\,~~~\,~~, 
{~~}~~~~~~~
(\g_{[  {\un a}{}_{1}    {\un a}{}_{2}    {\un a}{}_{3}    {\un a}{}_{4}    {\un a}{}_{5} |    })^{\a \b} \, 
T_{\a \b}{}_{  | {\un a}{}_{6} ]  } ~=~ 0  ~~~~~~~,
} 
\label{eq:constrts} }
\ee
are imposed on the torsion and curvature superfields.  These were derived \cite{M2} from demanding  
the frame superfield in 11D should depend solely  on a conformal compensating superfield and a 
superconformal semi-prepotential.  They imply only three independent conformal tensors appear in torsion and curvature superfields: the Weyl tensor ${\cal W}{}_{{\un a}{\un b}{\un c}{\un d}}$ 
in the Riemann tensor, and two other tensors $X_{ [{\un a} {\un b} ]}{}^{\un c} ~\equiv~ \fracm 1{32} \, (\g_{{\un a} {\un b}})^{\a \b} \,  T_{\a \b}{}^{\un c}$ ,
and $X_{[ {\un a}  {\un b}  {\un c}  {\un d} {\un e}]}{}^{\un f} ~ \equiv~ i 
\, \fracm 1{32} \, (\g_{{\un a}  {\un b}  {\un c}  {\un d} {\un e}})^{\a \b} \, T_{\a \b}{}^{\un f}
~$. The constraints in (\ref{eq:constrts}) imply these two tensors respectively correspond to the
$\CMTB { \{429 \}}$ and $\CMTB{ \{4290\}}$ irreps.  Interestingly enough, the scalar superfield
has the $\CMTB{ \{4290 \}}$ irrep but not the $\CMTB { \{429 \}}$ irrep at Level-16.
This means either $X_{ [{\un a} {\un b} ]}{}^{\un c}$ can be set to zero, or another prepotential
must be sought by an additional scan.

This discussion illustrates how the full supergeometry is constrained by the off-shell prepotential
superfields.  Since the full geometry controls all interactions, and the prepotential controls
the supergeometry, the prepotential also controls the interactions.

\newpage
\section{11D, $\mathcal{N} = 1$ Adinkra Diagram}\label{sec:11DAdinkra}

In \cite{counting10d}, we have developed ten dimensional adinkra diagrams for the first time. In this chapter, 
we will apply the same technique to define the 11D, $\mathcal{N} = 1$ adinkra diagram.

Let us first list the number of {\it {independent}} component fields at each level up to Level-16.
\begin{table}[h]
\begin{center}
\begin{tabular}{|c|c|}\hline
${\rm {Level}}~ \#$ &  {\rm {Component ~Field~ Count}}
\\ \hline \hline
$ ~~0 ~~$ &  1   \\ \hline
$ ~~1 ~~$ &  1   \\ \hline
$ ~~2 ~~$ &  3   \\ \hline
$ ~~3 ~~$ &  3   \\ \hline
$ ~~4 ~~$ &  8   \\ \hline
$ ~~5 ~~$ &  9   \\ \hline
$ ~~6 ~~$ &  19   \\ \hline
$ ~~7 ~~$ &  23   \\ \hline
$ ~~8 ~~$ &  49   \\ \hline
$ ~~9 ~~$ &  55   \\ \hline
$ ~~10 ~~$ &  99   \\ \hline
$ ~~11 ~~$ &  106   \\ \hline
$ ~~12 ~~$ &  173   \\ \hline
$ ~~13 ~~$ &  171   \\ \hline
$ ~~14 ~~$ &  247   \\ \hline
$ ~~15 ~~$ &  225   \\ \hline
$ ~~16 ~~$ &  296   \\ \hline
\end{tabular}
\end{center}
\caption{Number of Independent Fields at Each Level  \label{tabx}}
\end{table}

As usual, beyond the middle level in a superfield
(and thus its adinkra), the number of fields at
Level-$n$ when 17 $\le$ $n$ $\le 32$ is equal to the number Level-($32 - n$) since 32 is the top level of the expansion.  We thus find 1,198 bosonic fields in the even levels 0-14 together with the even levels 18-32, and 296 at
the middle level.  So the total number of bosonic fields in the 11D, $\cal N$ = 1 scalar
superfield is 1,494 fields. There are
1,186 fermionic fields in the odd levels 1-15 together with the odd levels 17-31.  The equality in the number of degrees of freedom
is accomplished by, on average, having fermions
appear in representations that are larger
than that of the bosons.  So the total number
of fields in the 11D $\cal N$ = 1 scalar 
superfield is 1,494 bosonic fields and 1,186
fermionic fields.

Now we come to the adinkra itself. 

Based on the component decomposition results shown in Sec.~\ref{sec:decomp_result}, we can explicitly demonstrate the 11D, $\mathcal{N} = 1$ adinkra by the same process as we described in~\cite{counting10d}: use open nodes to denote bosonic component fields and put their corresponding irreps in the center. For fermionic component fields, use closed nodes. The number of level represents the height assignment. Black edges connect nodes in the adjacent levels, meaning SUSY transformations.  In order to determine the linkages between the nodes in adjacent levels, we apply the process described in Section \ref{sec:primer4}.  While in principle 
it is possible to draw the adinkra exactly showing all 1,494 bosonic nodes, all 1,186 fermionic nodes, and a maximum of 29,334 links\footnote{This number might be smaller 
depending on the number of $K$-parameters described in the final equations in the subsection (2.3) that vanish.} connecting bosons to fermions and vice-versa, for reasons of practicality we will only draw it up to the quintic level.

\begin{figure}[htp!]
\centering
\includegraphics[width=0.6\textwidth]{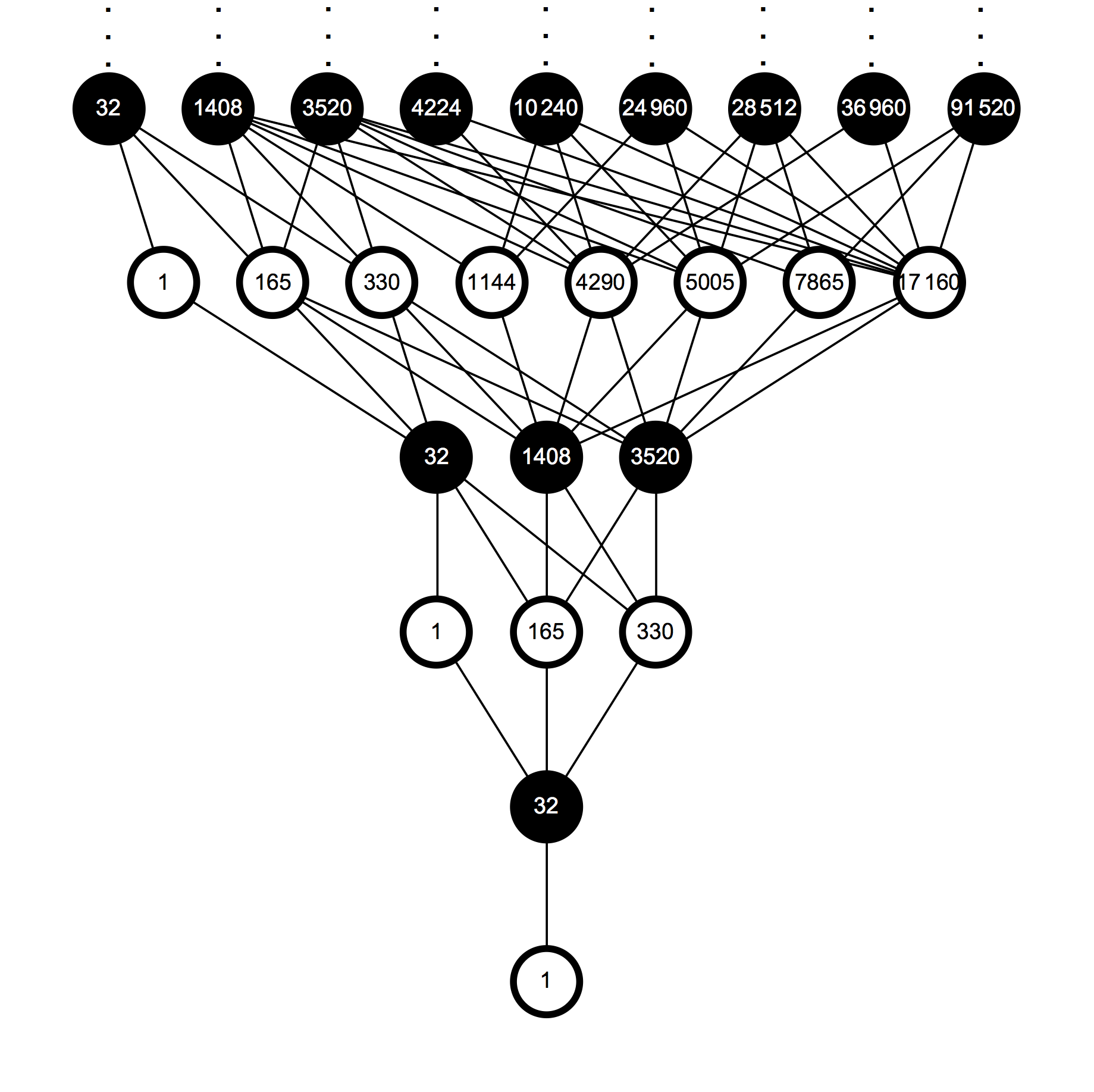}
\caption{Adinkra Diagram for 11D, $\mathcal{N} = 1$ (using dimensions)}
\label{Fig:11D}
\end{figure}

\begin{figure}[htp!]
\centering
\includegraphics[width=0.6\textwidth]{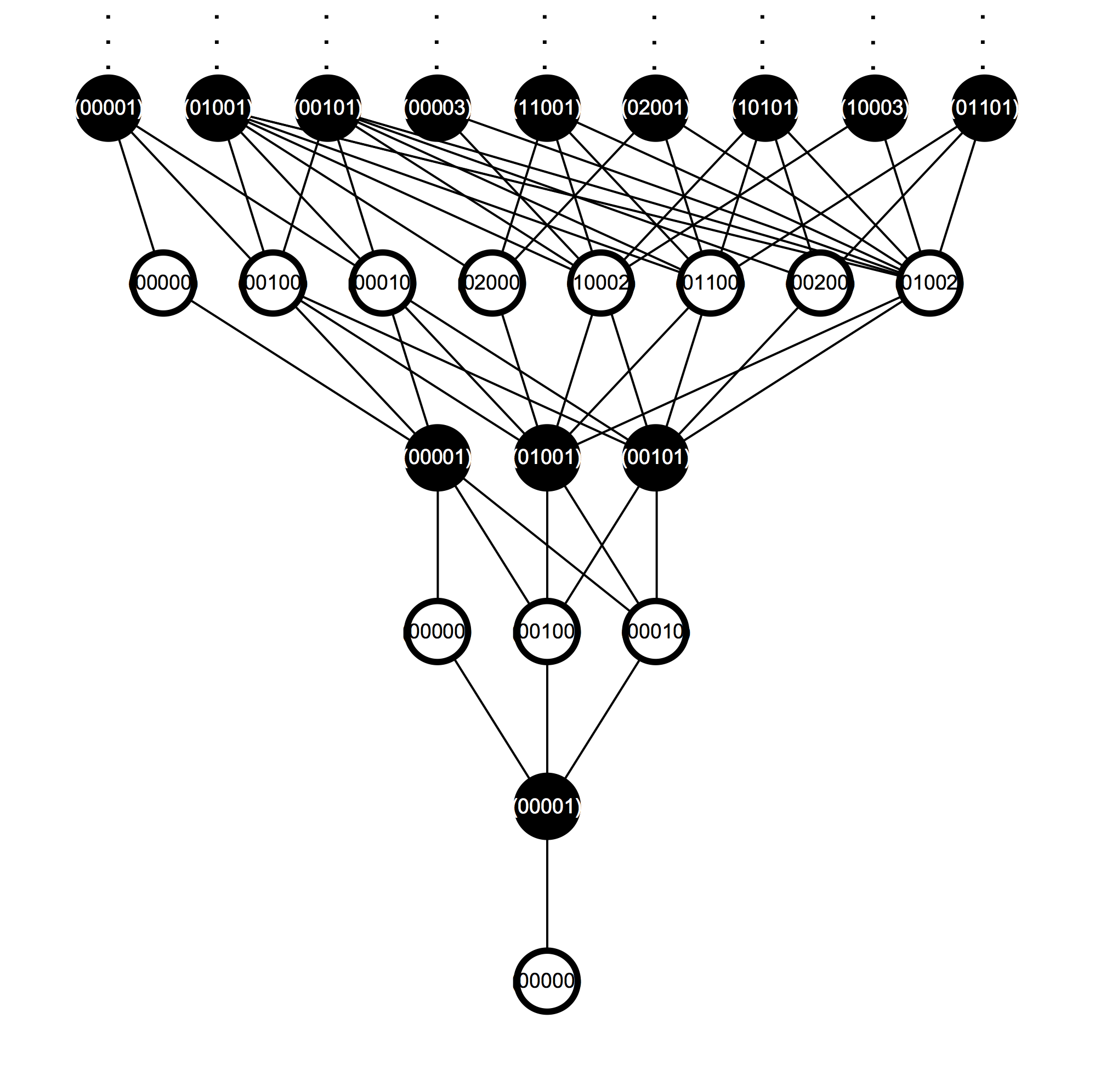}
\caption{Adinkra Diagram for 11D, $\mathcal{N} = 1$ (using Dynkin labels)}
\label{Fig:11D-Dynkin}
\end{figure}

The Adinkra diagram for 11D, $\mathcal{N} = 1$ up to level-5 can be represented using dimensions in Figure~\ref{Fig:11D} or Dynkin labels in Figure~\ref{Fig:11D-Dynkin}. 

$$~$$
\newpage
$$~$$
\newpage
\section{Conclusion}

As seen from Table \ref{Tab:graviton16}, the 11D, $\cal N$ = 1 scalar superfield is the simplest bosonic superfield that contains all the on-shell states of eleven dimensional supergravity and has the unique attribute of containing a single candidate for the graviton. This raises delightful possibilitities that we cast into the form of conjectures.

\begin{quotation}
    \noindent Conjecture \# 1: \vspace{0.5em}\newline
    {\it{Let $\cal V$ denote the scalar superfield in a Lorentz superspace of signature}} SO(1, 10){\it{, the facts that at the middle level of its adinkra both the conformal graviton and gauge 3-form (as well the conformal gravitino at one higher level) show up, imply $\cal V$ is a superfield limit of M-Theory, with $\cal V$ being a supergravity prepotential superfield or possibly a semi-prepotential superfield.}}
\end{quotation}
\begin{quotation}
    \noindent Conjecture \# 2: \vspace{0.5em}\newline
    {\it{Under the branchings of ${\mathfrak{su}(32)\supset\mathfrak{so}(10)}$ and ${\mathfrak{su}(32)\supset\mathfrak{so}(4)}$ respectively, the scalar superfield $\cal V$ describes the superfield limit of Type-IIA superstring theory and the prepotential or semi-prepotential
    superfield for $4D, \, {\cal N} = 8$ supergravity.}}
\end{quotation}

\vspace{.05in}

\noindent
To our knowledge, there exists no previous suggestions of these possibilities.

However, our calculations, discussions, and explorations also point to something else.

In the section entitled ``Traditional Path to Superfield Component Decompositions," we showed explicitly at low orders in the $\theta$-expansion the practical difficulty of using the conventional $\theta$-expansion to access the component field contents of high dimensional superfields.  This suggests the possible value of searching for expansions over quantities other than the Grassmann $\theta$-coordinates of the Salam-Strathdee superspace.

As discussed extensively in Appendix \ref{appen:handicraft} the approach of introducing two classes of Young Tableaux, one for bosonic representations and one for fermionic representations, is quite useful in both conceptual and calculational efficiency.  As Young Tableaux have a well understood definition of multiplication, one can build upon this fact.  Our bosonic Young Tableaux correspond to a set of tableaux that obey the usual multiplication rules of such objects.  On the other hand, only the totally antisymmetric products of the spinorial Young Tableaux are considered, i.e. single column Tableaux. Furthermore, as shown in this appendix, as single column spinorial Tableaux are also equivalent to a sum of bosonic Young Tableaux, they yield a more efficient method for representing supermultiplets consist of replacing $\theta$-expansions by products of elements taken from the two classes of Young Tableaux.

Thus, we are in position to define ``adinkra fields'' as an alternative to superfields.  These would take the form of conventional superfields, but with the differences that the $\theta$ variables raised to all possible powers would be replaced by products of the irreducible Tableaux and the Dynkin Labels play the role of the component fields. The Dynkin Labels implicitly carry the indices on the component field variable coefficients which saturate the indices represented by the boxes of the Tableaux.  There is currently
a puzzling feature of our results that requires more study.  Namely, at variously fixed levels there can
be seen to occur multiple numbers of the same irrep.  It is our suspicious this is related to the inequivalence of pathways to reach these multiple occurrences of the same irrep in the same level.  This 
is a topic for future study.

We wish to end on a note of historical observation.  Many years ago, the foundation for superfield 
supergravity in four dimensions occurred with two works \cite{OS1,OS2} where the prepotential for 
supergravity was proposed. This current work has now set in place, we hope, a similar period for 
superfield supergravity in eleven dimensions. Whether this step will be proven as successful as its 
precedent can only also be settled by research in the future.

\vspace{.05in}
 \begin{center}
\parbox{4in}{{\it ``True ignorance is not the absence of knowledge, \\ $~~$ 
but the refusal to acquire it.'' \\ ${~}$ 
 \\ ${~}$ 
\\ ${~}$ }\,\,-\,\, Karl Popper}
 \parbox{4in}{
 $~~$}  
 \end{center}
 \noindent
{\bf Acknowledgements}\\[.1in] \indent

We wish to acknowledge discussions with S.\ Kuzenko, W.\ Linch, W.\ Siegel, M.\ Ro\v cek, and E. Witten.
The research of S.\ J.\ G., Y.\ Hu, and S.-N.\ Mak is supported in part by the endowment of the Ford Foundation Professorship of Physics at Brown University and they gratefully acknowledge the support of the Brown Theoretical Physics Center. The authors would like to thank Renato Fonseca for discussing the SUSYno package and the Plethysm function.   Part of this research was conducted using computational resources and services at the Center for Computation and Visualization (CCV), Brown University. SNHM would like to thank research software engineers Paul Hall 
and Raj Shukla for their help in using the Brown CCV resources.

\newpage
\noindent
\textbf{Added Note In Proof}\\[.1in]

Recall the decomposition of the inverse frame and gravitino fields in 11D, yields
\begin{alignat}{4}
    &e_{\un a}{}^{\un m} ~=~ \{ &&h_{(\un{a}\un{b})}~+~ &&\eta_{\un{a}\un{b}}h ~+~ &&h_{[\un{a}\un{b}]} \}\, \eta^{\un{b}\un{m}} ~~~,\\
    &\{121\} && \CMTB{\{65\}} &&\CMTB{\{1\}} &&\CMTB{\{55\}} \nonumber
\end{alignat}
where $h_{(\un{a}\un{b})}$ is the conformal graviton, $h$ is the trace, and $h_{[\un{a}\un{b}]}$ is the two-form; and
\begin{alignat}{3}
    &\tilde{\psi}_{\un a}{}^{\a} ~=~  &&\psi_{\un a}{}^{\a}~-~\frac{1}{11}&& (\g_{\un a})^{\a\b}\psi_{\b} ~~~, \\
    &\{352\} && \CMTred{\{320\}} &&\CMTred{\{32\}} \nonumber
\end{alignat}
where $\psi_{\un a}{}^{\a}$ is the conformal gravitino and $\psi_{\b} \equiv (\g^{\un{a}})_{\a\b} \tilde{\psi}_{\un{a}}{}^{\a}$ is the $\gamma$-trace. Since on-shell 11D supergravity also contains the three-form with d.o.f. $\CMTB{\{165\}}$, the prepotential superfield must contain $\CMTB{\{1\}}$, $\CMTB{\{55\}}$, $\CMTB{\{65\}}$, and $\CMTB{\{165\}}$ at level-$n$ and contain $\CMTred{\{32\}}$ and $\CMTred{\{320\}}$ at level-$(n+1)$. As presented in Section \ref{sec:decomp_result}, the scalar superfield $\cal{V}$ contains 
the $\CMTB{\{1\}}$, $\CMTB{\{65\}}$, and $\CMTB{\{165\}}$ irreps at level-$16$ and contains the $\CMTred{\{32\}}$ and $\CMTred{\{320\}}$ irreps at level-$17$, but does not contain the $\CMTB{\{55\}}$ irrep at level-$16$, which suggests that $\cal{V}$ may be 
a semi-prepotential, i.e. some spinorial derivatives of the fundamental prepotential. 

Next, consider the spinor superfield $\Psi_{\a}$ satisfying 
\begin{equation}
\cal{V} ~=~ {\rm D}^{\a} {\rm \Psi}_{\a} ~~~.
\end{equation}
Since the D-operator acting on a superfield always lowers a component field by one level, in the spinor superfield the
$\CMTB{\{1\}}$, $\CMTB{\{65\}}$, and $\CMTB{\{165\}}$ irreps must appear at Level-17 and the $\CMTred{\{32\}}$ and 
$\CMTred{\{320\}}$ irreps must appear at Level-18. Table \ref{Tab:graviton_Psialpha} summarizes the occurrences of these important component fields we care about. Note that $\Psi_{\a}$ satisfies the criterion and we conject that $\cal{V}$ is a supergravity semi-prepotential superfield and $\Psi_{\a}$ is a supergravity prepotential superfield.

\begin{table}[h!]
\centering
\begin{tabular}{|c|c|c|c|c|c|c|} \hline
   Level & $b_{\CMTB{\{1\}}}$ & $b_{\CMTB{\{55\}}}$  &$b_{\CMTB{\{65\}}}$ & $b_{\CMTB{\{165\}}}$ & $b_{\CMTred{\{32\}}}$ & $b_{\CMTred{\{320\}}}$\\ \hline \hline
    17 & 2 &5 & 2 & 8 & 0 & 0   \\ \hline
    18 & 0 & 0 & 0 & 0 & 5 & 8  \\ \hline 
\end{tabular}
\caption{Summary of important component fields contained in the superfield $\Psi_{\a}$. \label{Tab:graviton_Psialpha}}
\end{table}

\newpage
\appendix
\section{SO(11) Irreducible Representations}
\label{appen:so11}

In this appendix, we list some of the SO(11) irreducible representations by Dynkin labels and dimensions and thus give a dictionary between the two methods for describing irreps \cite{yamatsu2015}.

\begin{table}[h!]
\centering
\begin{tabular}{|c|c|c|}
\hline
 Dynkin label & Dimension\\
 \hline
 $\CMTB{(10000)}$ & $\CMTB{11}$\\
 \hline
 $\CMTred{(00001)}$ & $\CMTred{32}$\\
 \hline
 $\CMTB{(01000)}$ & $\CMTB{55}$\\
 \hline
 $\CMTB{(20000)}$ & $\CMTB{65}$\\
 \hline
 $\CMTB{(00100)}$ & $\CMTB{165}$\\
 \hline
 $\CMTB{(30000)}$ & $\CMTB{275}$\\
 \hline
 $\CMTred{(10001)}$ & $\CMTred{320}$\\
 \hline
 $\CMTB{(00010)}$ & $\CMTB{330}$\\
 \hline
 $\CMTB{(11000)}$ & $\CMTB{429}$\\
 \hline
 $\CMTB{(00002)}$ & $\CMTB{462}$\\
 \hline
 $\CMTB{(40000)}$ & $\CMTB{935}$\\
 \hline
 $\CMTB{(02000)}$ & $\CMTB{1,144}$\\
 \hline
 $\CMTred{(01001)}$ & $\CMTred{1,408}$\\
 \hline
 $\CMTB{(10100)}$ & $\CMTB{1,430}$\\
 \hline
 $\CMTred{(20001)}$ & $\CMTred{1,760}$\\
 \hline
 $\CMTB{(21000)}$ & $\CMTB{2,025}$\\
 \hline
 $\CMTB{(50000)}$ & $\CMTB{2,717}$\\
 \hline
 $\CMTB{(10010)}$ & $\CMTB{3,003}$\\
 \hline
 $\CMTred{(00101)}$ & $\CMTred{3,520}$\\
 \hline
 $\CMTred{(00003)}$ & $\CMTred{4,224}$\\
 \hline
 $\CMTB{(10002)}$ & $\CMTB{4,290}$\\
 \hline
 $\CMTB{(01100)}$ & $\CMTB{5,005}$\\
 \hline
 $\CMTred{(00011)}$ & $\CMTred{5,280}$\\
 \hline
 $\CMTB{(60000)}$ & $\CMTB{7,007}$\\
 \hline
 $\CMTred{(30001)}$ & $\CMTred{7,040}$\\
 \hline
 $\CMTB{(20100)}$ & $\CMTB{7,128}$\\
 \hline
 $\CMTB{(12000)}$ & $\CMTB{7,150}$\\
 \hline
 $\CMTB{(31000)}$ & $\CMTB{7,293}$\\
 \hline
 $\CMTB{(00200)}$ & $\CMTB{7,865}$\\
 \hline
 $\CMTred{(11001)}$ & $\CMTred{10,240}$\\
 \hline
 $\CMTB{(01010)}$ & $\CMTB{11,583}$\\
 \hline
 $\CMTB{(03000)}$ & $\CMTB{13,650}$\\
 \hline
 $\CMTB{(20010)}$ & $\CMTB{15,400}$\\
 \hline
\end{tabular}
\end{table}

\newpage
\begin{table}[h!]
\centering
\begin{tabular}{|c|c|c|}
\hline
 Dynkin label & Dimension\\
 \hline
 $\CMTB{(70000)}$ & $\CMTB{16,445}$\\
 \hline
 $\CMTB{(01002)}$ & $\CMTB{17,160}$\\
 \hline
 $\CMTB{(41000)}$ & $\CMTB{21,945}$\\
 \hline
 $\CMTB{(20002)}$ & $\CMTB{22,275}$\\
 \hline
 $\CMTred{(40001)}$ & $\CMTred{22,880}$\\
 \hline
 $\CMTB{(00110)}$ & $\CMTB{23,595}$\\
 \hline
 $\CMTB{(00020)}$ & $\CMTB{23,595'}$\\
 \hline
 $\CMTred{(02001)}$ & $\CMTred{24,960}$\\
 \hline
 $\CMTB{(30100)}$ & $\CMTB{26,520}$\\
 \hline
 $\CMTB{(00004)}$ & $\CMTB{28,314}$\\
 \hline
 $\CMTred{(10101)}$ & $\CMTred{28,512}$\\
 \hline
 $\CMTB{(22000)}$ & $\CMTB{28,798}$\\
 \hline
 $\CMTB{(11100)}$ & $\CMTB{33,033}$\\
 \hline
 $\CMTB{(80000)}$ & $\CMTB{35,750}$\\
 \hline
 $\CMTred{(10003)}$ & $\CMTred{36,960}$\\
 \hline
 $\CMTB{(00102)}$ & $\CMTB{37,752}$\\
 \hline
 $\CMTred{(10011)}$ & $\CMTred{45,056}$\\
 \hline
 $\CMTred{(21001)}$ & $\CMTred{45,760}$\\
 \hline
 $\CMTB{(00012)}$ & $\CMTB{47,190}$\\
 \hline
 \end{tabular}
\caption{SO(11) irreducible representations \protect\cite{yamatsu2015}}
\end{table}

\newpage
\section{11D Gamma Matrix Multiplication Table\label{appen:gamma_matrix_mul}}

In the work of \cite{CoDeX}, we have given previously
the definitions we use for the 11D $\gamma$-matrices.
For the convenience of our readers, we review a number of
these below and as well present some new results.

\subsection{Identities with unique expressions}
\label{sec:IUE}

$$
 \g^{\un{a}} \g_{\un{b}} ~=~ \g^{\un{a}}{}_{\un{b}} + \d_{\un{b}}{}^{\un{a}} \mathbb{I} ~~~,
 \eqno(\text{B}.1.1)
$$
$$
 \g^{\un{a}} \g_{\un{b}\un{c}} ~=~ \g^{\un{a}}{}_{\un{b}\un{c}} + \d_{[\un{b}}{}^{\un{a}} \g_{\un{c}]}~~~,
 \eqno(\text{B}.1.2)
$$
$$
 \g^{\un{a}} \g_{\un{b}\un{c}\un{d}} ~=~ \g^{\un{a}}{}_{\un{b}\un{c}\un{d}} + \fracm12 \d_{[\un{b}}{}^{\un{a}} \g_{\un{c}\un{d}]}~~~,
 \eqno(\text{B}.1.3)
$$
$$
 \g^{\un{a}} \g_{\un{b}\un{c}\un{d}\un{e}} ~=~ \g^{\un{a}}{}_{\un{b}\un{c}\un{d}\un{e}} + \fracm1{3!} \d_{[\un{b}}{}^{\un{a}} \g_{\un{c}\un{d}\un{e}]}~~~,
 \eqno(\text{B}.1.4)
$$
$$
 \g^{\un{a}} \g_{\un{b}\un{c}\un{d}\un{e}\un{f}} ~=~ \fracm1{5!} \e^{\un{a}}{}_{\un{b}\un{c}\un{d}\un{e}\un{f}}{}^{[5]} \g_{[5]} + \fracm1{4!} \d_{[\un{b}}{}^{\un{a}} \g_{\un{c}\un{d}\un{e}\un{f}]}~~~.
 \eqno(\text{B}.1.5)
$$

$$
 \g^{\un{a}\un{b}} \g_{\un{c}} ~=~ \g^{\un{a}\un{b}}{}_{\un{c}} - \d_{\un{c}}{}^{[\un{a}} \g^{\un{b}]}~~~,
 \eqno(\text{B}.2.1)
$$
$$
 \g^{\un{a}\un{b}} \g_{\un{c}\un{d}} ~=~ \g^{\un{a}\un{b}}{}_{\un{c}\un{d}} + \d_{[\un{c}}{}^{[\un{a}} \g_{\un{d}]}{}^{\un{b}]} - \d_{\un{c}}{}^{[\un{a}} \d_{\un{d}}{}^{\un{b}]}~~~,
 \eqno(\text{B}.2.2)
$$
$$
 \g^{\un{a}\un{b}} \g_{\un{c}\un{d}\un{e}} ~=~ \g^{\un{a}\un{b}}{}_{\un{c}\un{d}\un{e}} - \fracm12 \d_{[\un{c}}{}^{[\un{a}} \g_{\un{d}\un{e}]}{}^{\un{b}]} - \d_{[\un{c}}{}^{\un{a}} \d_{\un{d}}{}^{\un{b}} \g_{\un{e}]}~~~,
 \eqno(\text{B}.2.3)
$$
$$
 \g^{\un{a}\un{b}} \g_{\un{c}\un{d}\un{e}\un{f}} ~=~ \fracm1{5!} \e^{\un{a}\un{b}}{}_{\un{c}\un{d}\un{e}\un{f}}{}^{[5]} \g_{[5]} + \fracm1{3!} \d_{[\un{c}}{}^{[\un{a}} \g_{\un{d}\un{e}\un{f}]}{}^{\un{b}]} - \fracm12 \d_{[\un{c}}{}^{\un{a}} \d_{\un{d}}{}^{\un{b}} \g_{\un{e}\un{f}]}~~~,
 \eqno(\text{B}.2.4)
$$
$$
 \g^{\un{a}\un{b}} \g_{\un{c}\un{d}\un{e}\un{f}\un{g}} ~=~ \fracm1{4!} \e^{\un{a}\un{b}}{}_{\un{c}\un{d}\un{e}\un{f}\un{g}}{}^{[4]} \g_{[4]} - \fracm1{4!} \d_{[\un{c}}{}^{[\un{a}} \g_{\un{d}\un{e}\un{f}\un{g}]}{}^{\un{b}]} - \fracm1{3!} \d_{[\un{c}}{}^{\un{a}} \d_{\un{d}}{}^{\un{b}} \g_{\un{e}\un{f}\un{g}]}~~~.
 \eqno(\text{B}.2.5)
$$

$$
 \g^{\un{a}\un{b}\un{c}} \g_{\un{d}} ~=~ \g^{\un{a}\un{b}\un{c}}{}_{\un{d}} +\fracm12 \d_{\un{d}}{}^{[\un{a}} \g^{\un{b}\un{c}]}~~~,
 \eqno(\text{B}.3.1)
$$
$$
 \g^{\un{a}\un{b}\un{c}} \g_{\un{d}\un{e}} ~=~ \g^{\un{a}\un{b}\un{c}}{}_{\un{d}\un{e}} + \fracm12 \d_{[\un{d}}{}^{[\un{a}} \g_{\un{e}]}{}^{\un{b}\un{c}]} - \d_{\un{d}}{}^{[\un{a}} \d_{\un{e}}{}^{\un{b}} \g^{\un{c}]}~~~,
 \eqno(\text{B}.3.2)
$$
$$
 \g^{\un{a}\un{b}\un{c}} \g_{\un{d}\un{e}\un{f}} ~=~ \fracm1{5!} \e^{\un{a}\un{b}\un{c}}{}_{\un{d}\un{e}\un{f}}{}^{[5]} \g_{[5]} + \fracm14 \d_{[\un{d}}{}^{[\un{a}} \g_{\un{e}\un{f}]}{}^{\un{b}\un{c}]} + \fracm12 \d_{[\un{d}}{}^{[\un{a}} \d_{\un{e}}{}^{\un{b}} \g_{\un{f}]}{}^{\un{c}]} - \d_{\un{d}}{}^{[\un{a}} \d_{\un{e}}{}^{\un{b}} \d_{\un{f}}{}^{\un{c}]}~~~,
 \eqno(\text{B}.3.3)
$$
$$
 \g^{\un{a}\un{b}\un{c}} \g_{\un{d}\un{e}\un{f}\un{g}} ~=~ \fracm1{4!} \e^{\un{a}\un{b}\un{c}}{}_{\un{d}\un{e}\un{f}\un{g}}{}^{[4]} \g_{[4]} + \fracm1{12} \d_{[\un{d}}{}^{[\un{a}} \g_{\un{e}\un{f}\un{g}]}{}^{\un{b}\un{c}]} - \fracm14 \d_{[\un{d}}{}^{[\un{a}} \d_{\un{e}}{}^{\un{b}} \g_{\un{f}\un{g}]}{}^{\un{c}]} - \d_{[\un{d}}{}^{\un{a}} \d_{\un{e}}{}^{\un{b}} \d_{\un{f}}{}^{\un{c}} \g_{\un{g}]}~~~.
 \eqno(\text{B}.3.4)
$$

$$
 \g^{\un{a}\un{b}\un{c}\un{d}} \g_{\un{e}} ~=~ \g^{\un{a}\un{b}\un{c}\un{d}}{}_{\un{e}} - \fracm1{3!} \d_{\un{e}}{}^{[\un{a}} \g^{\un{b}\un{c}\un{d}]}~~~,
 \eqno(\text{B}.4.1)
$$
$$
 \g^{\un{a}\un{b}\un{c}\un{d}} \g_{\un{e}\un{f}} ~=~ \fracm1{5!} \e^{\un{a}\un{b}\un{c}\un{d}}{}_{\un{e}\un{f}}{}^{[5]} \g_{[5]} + \fracm1{3!} \d_{[\un{e}}{}^{[\un{a}} \g_{\un{f}]}{}^{\un{b}\un{c}\un{d}]} - \fracm12 \d_{\un{e}}{}^{[\un{a}} \d_{\un{f}}{}^{\un{b}} \g^{\un{c}\un{d}]}~~~,
 \eqno(\text{B}.4.2)
$$
$$
 \g^{\un{a}\un{b}\un{c}\un{d}} \g_{\un{e}\un{f}\un{g}} ~=~ \fracm1{4!} \e^{\un{a}\un{b}\un{c}\un{d}}{}_{\un{e}\un{f}\un{g}}{}^{[4]} \g_{[4]} - \fracm1{12} \d_{[\un{e}}{}^{[\un{a}} \g_{\un{f}\un{g}]}{}^{\un{b}\un{c}\un{d}]} - \fracm14 \d_{[\un{e}}{}^{[\un{a}} \d_{\un{f}}{}^{\un{b}} \g_{\un{g}]}{}^{\un{c}\un{d}]} + \d_{\un{e}}{}^{[\un{a}} \d_{\un{f}}{}^{\un{b}} \d_{\un{g}}{}^{\un{c}} \g^{\un{d}]}~~~.
 \eqno(\text{B}.4.3)
$$

$$
 \g^{\un{a}\un{b}\un{c}\un{d}\un{e}} \g_{\un{f}} ~=~ \fracm1{5!} \e^{\un{a}\un{b}\un{c}\un{d}\un{e}}{}_{\un{f}}{}^{[5]} \g_{[5]} + \fracm1{4!} \d_{\un{f}}{}^{[\un{a}} \g^{\un{b}\un{c}\un{d}\un{e}]}~~~,
 \eqno(\text{B}.5.1)
$$
$$
 \g^{\un{a}\un{b}\un{c}\un{d}\un{e}} \g_{\un{f}\un{g}} ~=~ \fracm1{4!} \e^{\un{a}\un{b}\un{c}\un{d}\un{e}}{}_{\un{f}\un{g}}{}^{[4]} \g_{[4]} + \fracm1{4!} \d_{[\un{f}}{}^{[\un{a}} \g_{\un{g}]}{}^{\un{b}\un{c}\un{d}\un{e}]} - \fracm1{3!} \d_{\un{f}}{}^{[\un{a}} \d_{\un{g}}{}^{\un{b}} \g^{\un{c}\un{d}\un{e}]}~~~.
 \eqno(\text{B}.5.2)
$$

\subsection{Identities with multiple expressions}
\label{sec:IME}

Sometimes, multiplication of our general 11D $\gamma$ matrices
can yield multiple equivalent expressions.  The following cases of this phenomenon are relevant to know about in the discussion of irreducible monomials.

For $\g^{[3]}\g_{[5]}$, the $\g_{[5]}$-term has multiple expressions.
$$ \eqalign{ {~~~~~~~~~~~~~}
 &~ \g^{\un{a}\un{b}\un{c}} \g_{\un{d}\un{e}\un{f}\un{g}\un{h}} \cr
 ~=&~  \textcolor{blue}{\fracm1{5!4!2!} \d_{[\un{d}}{}^{[\un{a}} \e_{\un{e}\un{f}\un{g}\un{h}]}{}^{\un{b}c]}{}^{[5]} \g_{[5]} } - \fracm1{3!} \e^{\un{a}\un{b}\un{c}}{}_{\un{d}\un{e}\un{f}\un{g}\un{h}}{}^{[3]} \g_{[3]} + \fracm1{12} \d_{[\un{d}}{}^{[\un{a}} \d_{\un{e}}{}^{\un{b}} \g_{\un{f}\un{g}\un{h}]}{}^{\un{c}]} - \fracm12 \d_{[\un{d}}{}^{\un{a}} \d_{\un{e}}{}^{\un{b}} \d_{\un{f}}{}^{\un{c}} \g_{\un{g}\un{h}]}  \cr
 ~=&~  \textcolor{blue}{\fracm1{4!2!} \e^{[4]}{}_{\un{d}\un{e}\un{f}\un{g}\un{h}}{}^{[\un{a}\un{b}} \g^{\un{c}]}{}_{[4]} } - \fracm1{3!} \e^{\un{a}\un{b}\un{c}}{}_{\un{d}\un{e}\un{f}\un{g}\un{h}}{}^{[3]} \g_{[3]} + \fracm1{12} \d_{[\un{d}}{}^{[\un{a}} \d_{\un{e}}{}^{\un{b}} \g_{\un{f}\un{g}\un{h}]}{}^{\un{c}]} - \fracm12 \d_{[\un{d}}{}^{\un{a}} \d_{\un{e}}{}^{\un{b}} \d_{\un{f}}{}^{\un{c}} \g_{\un{g}\un{h}]}  \cr
 ~=&~ \textcolor{blue}{\fracm1{4!4!} \e^{[4]\un{a}\un{b}\un{c}}{}_{[\un{d}\un{e}\un{f}\un{g}} \g_{\un{h}] [4]} } - \fracm1{3!} \e^{\un{a}\un{b}\un{c}}{}_{\un{d}\un{e}\un{f}\un{g}\un{h}}{}^{[3]} \g_{[3]} + \fracm1{12} \d_{[\un{d}}{}^{[\un{a}} \d_{\un{e}}{}^{\un{b}} \g_{\un{f}\un{g}\un{h}]}{}^{\un{c}]} - \fracm12 \d_{[\un{d}}{}^{\un{a}} \d_{\un{e}}{}^{\un{b}} \d_{\un{f}}{}^{\un{c}} \g_{\un{g}\un{h}]}
 ~~~.}
 \eqno(\text{B}.3.5)
$$
For $\g^{[4]}\g_{[\bar{4}]}$, the $\g_{[5]}$-term has multiple expressions.
$$ \eqalign{
 &~ \g^{\un{a}\un{b}\un{c}\un{d}} \g_{\un{e}\un{f}\un{g}\un{h}} \cr
 ~=&~  \textcolor{blue}{ \fracm1{5!3!3!} \d_{[\un{e}}{}^{[\un{a}} \e_{\un{f}\un{g}\un{h}]}{}^{\un{b}\un{c}\un{d}]}{}^{[5]} \g_{[5]} } - \fracm1{3!} \e^{\un{a}\un{b}\un{c}\un{d}}{}_{\un{e}\un{f}\un{g}\un{h}}{}^{[3]} \g_{[3]} - \fracm18 \d_{[\un{e}}{}^{[\un{a}} \d_{\un{f}}{}^{\un{b}} \g_{\un{g}\un{h}]}{}^{\un{c}\un{d}]} - \fracm1{3!} \d_{[\un{e}}{}^{[\un{a}} \d_{\un{f}}{}^{\un{b}} \d_{\un{g}}{}^{\un{c}} \g_{\un{h}]}{}^{\un{d}]} + \d_{\un{e}}{}^{[\un{a}} \d_{\un{f}}{}^{\un{b}} \d_{\un{g}}{}^{\un{c}} \d_{\un{h}}{}^{\un{d}]}  \cr
 ~=&~  \textcolor{blue}{ \fracm1{4!3!} \e^{[4]\un{a}\un{b}\un{c}\un{d}}{}_{[\un{e}\un{f}\un{g}} \g_{\un{h}] [4]} } - \fracm1{3!} \e^{\un{a}\un{b}\un{c}\un{d}}{}_{\un{e}\un{f}\un{g}\un{h}}{}^{[3]} \g_{[3]} - \fracm18 \d_{[\un{e}}{}^{[\un{a}} \d_{\un{f}}{}^{\un{b}} \g_{\un{g}\un{h}]}{}^{\un{c}\un{d}]} - \fracm1{3!} \d_{[\un{e}}{}^{[\un{a}} \d_{\un{f}}{}^{\un{b}} \d_{\un{g}}{}^{\un{c}} \g_{\un{h}]}{}^{\un{d}]} + \d_{\un{e}}{}^{[\un{a}} \d_{\un{f}}{}^{\un{b}} \d_{\un{g}}{}^{\un{c}} \d_{\un{h}}{}^{\un{d}]}  \cr
 ~=&~  \textcolor{blue}{ - \fracm1{4!3!} \e^{[4]}{}_{\un{e}\un{f}\un{g}\un{h}}{}^{[\un{a}\un{b}\un{c}} \g^{\un{d}]}{}_{[4]} } - \fracm1{3!} \e^{\un{a}\un{b}\un{c}\un{d}}{}_{\un{e}\un{f}\un{g}\un{h}}{}^{[3]} \g_{[3]} - \fracm18 \d_{[\un{e}}{}^{[\un{a}} \d_{\un{f}}{}^{\un{b}} \g_{\un{g}\un{h}]}{}^{\un{c}\un{d}]} - \fracm1{3!} \d_{[\un{e}}{}^{[\un{a}} \d_{\un{f}}{}^{\un{b}} \d_{\un{g}}{}^{\un{c}} \g_{\un{h}]}{}^{\un{d}]} + \d_{\un{e}}{}^{[\un{a}} \d_{\un{f}}{}^{\un{b}} \d_{\un{g}}{}^{\un{c}} \d_{\un{h}}{}^{\un{d}]}
 ~~~.}
 \eqno(\text{B}.4.4)
$$
For $\g^{[4]}\g_{[5]}$, the $\g_{[4]}$-term has multiple expressions.
$$ \eqalign{
 &~ \g^{\un{a}\un{b}\un{c}\un{d}} \g_{\un{e}\un{f}\un{g}\un{h}\un{i}} \cr 
 ~=&~ \textcolor{blue}{ - \fracm1{4!4!3!} \d_{[\un{e}}{}^{[\un{a}} \e_{\un{f}\un{g}\un{h}\un{i}]}{}^{\un{b}\un{c}\un{d}] [4]} \g_{[4]} } - \fracm1{2} \e^{\un{a}\un{b}\un{c}\un{d}}{}_{\un{e}\un{f}\un{g}\un{h}\un{i}}{}^{[2]} \g_{[2]} - \fracm1{4!} \d_{[\un{e}}{}^{[\un{a}} \d_{\un{f}}{}^{\un{b}} \g_{\un{g}\un{h}\un{i}]}{}^{\un{c}\un{d}]} + \fracm1{12} \d_{[\un{e}}{}^{[\un{a}} \d_{\un{f}}{}^{\un{b}} \d_{\un{g}}{}^{\un{c}} \g_{\un{h}\un{i}]}{}^{\un{d}]} + \d_{[\un{e}}{}^{\un{a}} \d_{\un{f}}{}^{\un{b}} \d_{\un{g}}{}^{\un{c}} \d_{\un{h}}{}^{\un{d}} \g_{\un{i}]}  \cr
 ~=&~ \textcolor{blue}{ - \fracm1{3!3!} \e^{[3]}{}_{\un{e}\un{f}\un{g}\un{h}\un{i}}{}^{[\un{a}\un{b}\un{c}} \g^{\un{d}]}{}_{[3]} } - \fracm1{2} \e^{\un{a}\un{b}\un{c}\un{d}}{}_{\un{e}\un{f}\un{g}\un{h}\un{i}}{}^{[2]} \g_{[2]} - \fracm1{4!} \d_{[\un{e}}{}^{[\un{a}} \d_{\un{f}}{}^{\un{b}} \g_{\un{g}\un{h}\un{i}]}{}^{\un{c}\un{d}]} + \fracm1{12} \d_{[\un{e}}{}^{[\un{a}} \d_{\un{f}}{}^{\un{b}} \d_{\un{g}}{}^{\un{c}} \g_{\un{h}\un{i}]}{}^{\un{d}]} + \d_{[\un{e}}{}^{\un{a}} \d_{\un{f}}{}^{\un{b}} \d_{\un{g}}{}^{\un{c}} \d_{\un{h}}{}^{\un{d}} \g_{\un{i}]}  \cr
 ~=&~ \textcolor{blue}{ \fracm1{4!3!} \e^{[3]\un{a}\un{b}\un{c}\un{d}}{}_{[\un{e}\un{f}\un{g}\un{h}} \g_{\un{i}] [3]} } - \fracm1{2} \e^{\un{a}\un{b}\un{c}\un{d}}{}_{\un{e}\un{f}\un{g}\un{h}\un{i}}{}^{[2]} \g_{[2]} - \fracm1{4!} \d_{[\un{e}}{}^{[\un{a}} \d_{\un{f}}{}^{\un{b}} \g_{\un{g}\un{h}\un{i}]}{}^{\un{c}\un{d}]} + \fracm1{12} \d_{[\un{e}}{}^{[\un{a}} \d_{\un{f}}{}^{\un{b}} \d_{\un{g}}{}^{\un{c}} \g_{\un{h}\un{i}]}{}^{\un{d}]} + \d_{[\un{e}}{}^{\un{a}} \d_{\un{f}}{}^{\un{b}} \d_{\un{g}}{}^{\un{c}} \d_{\un{h}}{}^{\un{d}} \g_{\un{i}]} ~~~.
 }
 \eqno(\text{B}.4.5)
$$
For $\g^{[5]}\g_{[3]}$, the $\g_{[5]}$-term has multiple expressions.
$$ \eqalign{
 \g^{\un{a}\un{b}\un{c}\un{d}\un{e}} \g_{\un{f}\un{g}\un{h}} =& \textcolor{blue}{ \textcolor{blue}{\fracm1{5!4!2!}} \d_{[\un{f}}{}^{[\un{a}} \e_{\un{g}\un{h}]}{}^{\un{b}\un{c}\un{d}\un{e}] [5]} \g_{[5]} } - \fracm1{3!} \e^{\un{a}\un{b}\un{c}\un{d}\un{e}}{}_{\un{f}\un{g}\un{h}}{}^{[3]} \g_{[3]} + \fracm1{12} \d_{[\un{f}}{}^{[\un{a}} \d_{\un{g}}{}^{\un{b}} \g_{\un{h}]}{}^{\un{c}\un{d}\un{e}]} - \fracm12 \d_{\un{f}}{}^{[\un{a}} \d_{\un{g}}{}^{\un{b}} \d_{\un{h}}{}^{\un{c}} \g^{\un{d}\un{e}]}  \cr 
 =& \textcolor{blue}{\fracm1{4!2!} \e^{[4]\un{a}\un{b}\un{c}\un{d}\un{e}}{}_{[\un{f}\un{g}} \g_{\un{h}] [4]} } - \fracm1{3!} \e^{\un{a}\un{b}\un{c}\un{d}\un{e}}{}_{\un{f}\un{g}\un{h}}{}^{[3]} \g_{[3]} + \fracm1{12} \d_{[\un{f}}{}^{[\un{a}} \d_{\un{g}}{}^{\un{b}} \g_{\un{h}]}{}^{\un{c}\un{d}\un{e}]} - \fracm12 \d_{\un{f}}{}^{[\un{a}} \d_{\un{g}}{}^{\un{b}} \d_{\un{h}}{}^{\un{c}} \g^{\un{d}\un{e}]} \cr
 =& \textcolor{blue}{\fracm1{4!4!} \e^{[4]}{}_{\un{f}\un{g}\un{h}}{}^{[\un{a}\un{b}\un{c}\un{d}} \g^{\un{e}]}{}_{[4]} } - \fracm1{3!} \e^{\un{a}\un{b}\un{c}\un{d}\un{e}}{}_{\un{f}\un{g}\un{h}}{}^{[3]} \g_{[3]} + \fracm1{12} \d_{[\un{f}}{}^{[\un{a}} \d_{\un{g}}{}^{\un{b}} \g_{\un{h}]}{}^{\un{c}\un{d}\un{e}]} - \fracm12 \d_{\un{f}}{}^{[\un{a}} \d_{\un{g}}{}^{\un{b}} \d_{\un{h}}{}^{\un{c}} \g^{\un{d}\un{e}]}~~~.
 }
 \eqno(\text{B}.5.3)
$$
For $\g^{[5]}\g_{[4]}$, the $\g_{[4]}$-term has multiple expressions.
$$ \eqalign{
 &~ \g^{\un{a}\un{b}\un{c}\un{d}\un{e}} \g_{\un{f}\un{g}\un{h}\un{i}} \cr 
 ~=&~  \textcolor{blue}{ \fracm1{4!4!3!} \d_{[\un{f}}{}^{[\un{a}} \e_{\un{g}\un{h}\un{i}]}{}^{\un{b}\un{c}\un{d}\un{e}] [4]} \g_{[4]} } - \fracm1{2} \e^{\un{a}\un{b}\un{c}\un{d}\un{e}}{}_{\un{f}\un{g}\un{h}\un{i}}{}^{[2]} \g_{[2]} - \fracm1{4!} \d_{[\un{f}}{}^{[\un{a}} \d_{\un{g}}{}^{\un{b}} \g_{\un{h}\un{i}]}{}^{\un{c}\un{d}\un{e}]} - \fracm1{12} \d_{[\un{f}}{}^{[\un{a}} \d_{\un{g}}{}^{\un{b}} \d_{\un{h}}{}^{\un{c}} \g_{\un{i}]}{}^{\un{d}\un{e}]} + \d_{\un{f}}{}^{[\un{a}} \d_{\un{g}}{}^{\un{b}} \d_{\un{h}}{}^{\un{c}} \d_{\un{i}}{}^{\un{d}} \g^{\un{e}]}  \cr
 ~=&~  \textcolor{blue}{ \fracm1{3!3!} \e^{[3]\un{a}\un{b}\un{c}\un{d}\un{e}}{}_{[\un{f}\un{g}\un{h}} \g_{\un{i}] [3]} } - \fracm1{2} \e^{\un{a}\un{b}\un{c}\un{d}\un{e}}{}_{\un{f}\un{g}\un{h}\un{i}}{}^{[2]} \g_{[2]} - \fracm1{4!} \d_{[\un{f}}{}^{[\un{a}} \d_{\un{g}}{}^{\un{b}} \g_{\un{h}\un{i}]}{}^{\un{c}\un{d}\un{e}]} - \fracm1{12} \d_{[\un{f}}{}^{[\un{a}} \d_{\un{g}}{}^{\un{b}} \d_{\un{h}}{}^{\un{c}} \g_{\un{i}]}{}^{\un{d}\un{e}]} + \d_{\un{f}}{}^{[\un{a}} \d_{\un{g}}{}^{\un{b}} \d_{\un{h}}{}^{\un{c}} \d_{\un{i}}{}^{\un{d}} \g^{\un{e}]}  \cr
 ~=&~  \textcolor{blue}{ - \fracm1{4!3!} \e^{[3]}{}_{\un{f}\un{g}\un{h}\un{i}}{}^{[\un{a}\un{b}\un{c}\un{d}} \g^{\un{e}]}{}_{[3]} } - \fracm1{2} \e^{\un{a}\un{b}\un{c}\un{d}\un{e}}{}_{\un{f}\un{g}\un{h}\un{i}}{}^{[2]} \g_{[2]} - \fracm1{4!} \d_{[\un{f}}{}^{[\un{a}} \d_{\un{g}}{}^{\un{b}} \g_{\un{h}\un{i}]}{}^{\un{c}\un{d}\un{e}]} - \fracm1{12} \d_{[\un{f}}{}^{[\un{a}} \d_{\un{g}}{}^{\un{b}} \d_{\un{h}}{}^{\un{c}} \g_{\un{i}]}{}^{\un{d}\un{e}]} + \d_{\un{f}}{}^{[\un{a}} \d_{\un{g}}{}^{\un{b}} \d_{\un{h}}{}^{\un{c}} \d_{\un{i}}{}^{\un{d}} \g^{\un{e}]}  ~~~.
 }
 \eqno(\text{B}.5.4)
$$

\newpage

\noindent
For $\g^{[5]}\g_{[\bar{5}]}$, both $\g_{[5]}$-term and $\g_{[3]}$-term have multiple expressions. The different $\g_{[5]}$-term expressions are:
\begin{table}[h!]
\centering
\begin{tabular}{c|c} 
 $\g_{[5]}$-term & coefficient \\ \hline
 $\d_{[\un{f}}{}^{[\un{a}} \d_{\un{g}}{}^{\un{b}} \e_{\un{h}\un{i}\un{j}]}{}^{\un{c}\un{d}\un{e}][5]} \g_{[5]}$ & $\fracm1{5!3!3!2!}$ \\
 $\d_{[\un{f}}{}^{[\un{a}} \e_{\un{g}\un{h}\un{i}}{}^{\un{b}\un{c}\un{d}\un{e}][4]} \g_{\un{j}] [4]}$ & $\fracm1{4!4!3!2!}$ \\
 $\d_{[\un{f}}{}^{[\un{a}} \e_{\un{g}\un{h}\un{i}\un{j}]}{}^{\un{b}\un{c}\un{d}}{}_{[4]} \g^{\un{e}] [4]}$ & $-\fracm1{4!4!3!2!}$  \\
 $\e^{[3]\un{a}\un{b}\un{c}\un{d}\un{e}}{}_{[\un{f}\un{g}\un{h}} \g_{\un{i}\un{j}] [3]}$ & $-\fracm1{3!3!2!}$ \\
 $\e^{[3]}{}_{\un{f}\un{g}\un{h}\un{i}\un{j}}{}^{[\un{a}\un{b}\un{c}} \g^{\un{d}\un{e}]}{}_{[3]}$ & $\fracm1{3!3!2!}$ \\ 
\end{tabular}
\end{table}

\noindent
The different $\g_{[3]}$-term expressions are:
\begin{table}[h!]
\centering
\begin{tabular}{c|c}
 $\g_{[3]}$-term & coefficient \\ \hline
 $\d_{[\un{f}}{}^{[\un{a}} \e_{\un{g}\un{h}\un{i}\un{j}]}{}^{\un{b}\un{c}\un{d}\un{e}][3]} \g_{[3]}$ & $-\fracm1{4!4!3!}$ \\
 $\e^{[2]\un{a}\un{b}\un{c}\un{d}\un{e}}{}_{[\un{f}\un{g}\un{h}\un{i}} \g_{\un{j}] [2]}$ & $-\fracm1{4!2!}$ \\
 $\e^{[2]}{}_{\un{f}\un{g}\un{h}\un{i}\un{j}}{}^{[\un{a}\un{b}\un{c}\un{d}} \g^{\un{e}]}{}_{[2]}$ & $-\fracm1{4!2!}$ 
\end{tabular}
\end{table}

\noindent
Therefore, we have $5 \times 3 = 15$ different expressions of $\g^{\un{a}\un{b}\un{c}\un{d}\un{e}}\g_{\un{f}\un{g}\un{h}\un{i}\un{j}}$ with right symmetries and good coefficients. One example would be
$$ \eqalign{
 \g^{\un{a}\un{b}\un{c}\un{d}\un{e}} \g_{\un{f}\un{g}\un{h}\un{i}\un{j}} ~=&~ \textcolor{blue}{\fracm1{5!3!3!2!} \d_{[\un{f}}{}^{[\un{a}} \d_{\un{g}}{}^{\un{b}} \e_{\un{h}\un{i}\un{j}]}{}^{\un{c}\un{d}\un{e}][5]} \g_{[5]}} \textcolor{blue}{ - \fracm1{4!4!3!} \d_{[\un{f}}{}^{[\un{a}} \e_{\un{g}\un{h}\un{i}\un{j}]}{}^{\un{b}\un{c}\un{d}\un{e}][3]} \g_{[3]}} \cr
 & + \e^{\un{a}\un{b}\un{c}\un{d}\un{e}}{}_{\un{f}\un{g}\un{h}\un{i}\un{j}}{}^{[1]} \g_{[1]} - \fracm1{4!} \d_{[\un{f}}{}^{[\un{a}} \d_{\un{g}}{}^{\un{b}} \d_{\un{h}}{}^{\un{c}} \g_{\un{i}\un{j}]}{}^{\un{d}\un{e}]} - \fracm1{4!} \d_{[\un{f}}{}^{[\un{a}} \d_{\un{g}}{}^{\un{b}} \d_{\un{h}}{}^{\un{c}} \d_{\un{i}}{}^{\un{d}} \g_{\un{j}]}{}^{\un{e}]} + \d_{\un{f}}{}^{[\un{a}} \d_{\un{g}}{}^{\un{b}} \d_{\un{h}}{}^{\un{c}} \d_{\un{i}}{}^{\un{d}} \d_{\un{j}}{}^{\un{e}]}~~~.
 }
 \eqno(\text{B}.5.5)
$$

\newpage
\section{Additional Useful Identities for 11D Gamma Matrices}
Over and above previous results \cite{CoDeX}, the list of identities below are useful for any reader who wishes to reproduce the results given in Chapter \ref{sec:analytical}
particularly with regards to the discussion
on deriving irreducible $\theta$ monomials.
\begin{align}
    \g_{[1]}\g^{[1]} ~&=~ 11~~~,\\
    \g_{[2]}\g^{[2]} ~&=~ -110~~~,\\
    \g_{[3]}\g^{[3]} ~&=~ -990~~~,\\
    \g_{[4]}\g^{[4]} ~&=~ 7920~~~,\\
    \g_{[5]}\g^{[5]} ~&=~ 55440~~~,\\
    \g^{\un{a} [2]}\g_{[2]} ~&=~ -90 \g^{\un{a}} ~~~,\\
    \g^{\un{a} [2]} \g_{\un{d}\un{e}\un{f}} \g_{[2]} ~&=~ -6 \g^{\un{a}}{}_{\un{d}\un{e}\un{f}} -13 \d_{[\un{d}}{}^{\un{a}}\g_{\un{e}\un{f}]} ~~~,\\
    \g^{\un{a} [2]} \g_{\un{d}\un{e}\un{f}\un{g}}\g_{[2]} ~&=~ 6\g^{\un{a}}{}_{\un{d}\un{e}\un{f}\un{g}} -\d_{[\un{d}}{}^{\un{a}} \g_{\un{e}\un{f}\un{g}]} ~~~,\\
    \g_{[3]}\g^{[3]\un{a}} ~&=~ -720\g^{\un{a}}~~~,\\
    \g^{\un{a}[3]}\g_{[3]} ~&=~ -720\g^{\un{a}}~~~,\\
    \g_{[3]}\g_{\un{e}\un{f}\un{g}}\g^{[3]\un{a}} ~&=~ 48\g^{\un{a}}{}_{\un{e}\un{f}\un{g}} + 24\d_{[\un{e}}{}^{\un{a}} \g_{\un{f}\un{g}]} ~~~,\\
    \g^{\un{a}[3]}\g_{\un{e}\un{f}\un{g}}\g_{[3]} ~&=~ -48\g^{\un{a}}{}_{\un{e}\un{f}\un{g}} + 24\d_{[\un{e}}{}^{\un{a}} \g_{\un{f}\un{g}]} ~~~,\\
    \g_{[3]}\g_{\un{e}\un{f}\un{g}\un{h}}\g^{[3]\un{a}} ~&=~ 48\g^{\un{a}}{}_{\un{e}\un{f}\un{g}\un{h}} - 8\d_{[\un{e}}{}^{\un{a}} \g_{\un{f}\un{g}\un{h}]}~~~,\\
    \g^{\un{a}[3]}\g_{\un{e}\un{f}\un{g}\un{h}}\g_{[3]} ~&=~ 48\g^{\un{a}}{}_{\un{e}\un{f}\un{g}\un{h}} + 8\d_{[\un{e}}{}^{\un{a}} \g_{\un{f}\un{g}\un{h}]}~~~,\\
    \g_{[4]}\g^{[4]\un{a}} ~&=~ 5040\g^{\un{a}}~~~,\\
    \g_{[4]}\g_{\un{f}\un{g}\un{h}}\g^{[4]\un{a}} ~&=~ 336\g^{\un{a}}{}_{\un{f}\un{g}\un{h}} -168 \d_{[\un{f}}{}^{\un{a}} \g_{\un{g}\un{h}]} ~~~,\\
    \g_{[4]}\g_{\un{f}\un{g}\un{h}\un{i}}\g^{[4]\un{a}} ~&=~ 48\g^{\un{a}}{}_{\un{f}\un{g}\un{h}\un{i}} + 56\d_{[\un{f}}{}^{\un{a}} \g_{\un{g}\un{h}\un{i}]}~~~,\\
    \g^{[\un{a}} \g_{\un{h}\un{i}\un{j}} \g^{\un{b}]} ~&=~ -2 \g^{\un{a}\un{b}}{}_{\un{h}\un{i}\un{j}} - 2 \d_{[\un{h}}{}^{\un{a}} \d_{\un{i}}{}^{\un{b}} \g_{\un{j}]} ~~~,\\
    \g^{\un{a}\un{b}[2]} \g_{\un{h}\un{i}\un{j}}\g_{[2]} ~&=~ 8\d_{[\un{h}}{}^{[\un{a}} \g^{\un{b}]}{}_{\un{i}\un{j}]} + 40 \d_{[\un{h}}{}^{\un{a}} \d_{\un{i}}{}^{\un{b}} \g_{\un{j}]} ~~~,\\
    \g^{[\un{a}} \g_{\un{i}\un{j}\un{l}\un{m}} \g^{\un{b}]} ~&=~ \frac{2}{5!} \e^{\un{a}\un{b}}{}_{\un{i}\un{j}\un{l}\un{m}[5]} \g^{[5]} + \d_{[\un{i}}{}^{\un{a}} \d_{\un{j}}{}^{\un{b}} \g_{\un{l}\un{m}]} ~~~,\\
    \g^{\un{a}\un{b}[2]} \g_{\un{i}\un{j}\un{l}\un{m}} \g_{[2]} ~&=~ \frac{1}{15} \e^{\un{a}\un{b}}{}_{\un{i}\un{j}\un{l}\un{m}[5]} \g^{[5]} + 8 \d_{\un{i}}{}^{[\un{a}} \d_{\un{j}}{}^{\un{b}} \g_{\un{l}\un{m}]} ~~~,\\
    \g^{[3][\un{a}} \g^{\un{b}]}{}_{[3]} ~&=~ -1008 \g^{\un{a}\un{b}} ~~~,\\
    \g^{[3][\un{a}} \g_{\un{f}\un{g}\un{h}} \g^{\un{b}]}{}_{[3]} ~&=~ 96 \g^{\un{a}\un{b}}{}_{\un{f}\un{g}\un{h}} - 336 \d_{[\un{f}}{}^{\un{a}} \d_{\un{g}}{}^{\un{b}} \g_{\un{h}]} ~~~,\\
    \g^{[3][\un{a}} \g_{\un{f}\un{g}\un{h}\un{i}} \g^{\un{b}]}{}_{[3]} ~&=~ \frac{2}{5} \e^{\un{a}\un{b}}{}_{\un{f}\un{g}\un{h}\un{i}[5]} \g^{[5]} ~~~,\\
    \g^{\un{a}\un{b}[1]} \g_{\un{h}\un{i}\un{j}} \g_{[1]} ~&=~ -3\g^{\un{a}\un{b}}{}_{\un{h}\un{i}\un{j}} + \frac{5}{2} \d_{[\un{h}}{}^{[\un{a}} \g^{\un{b}]}{}_{\un{i}\un{j}]} + 7 \d_{[\un{h}}{}^{\un{a}} \d_{\un{i}}{}^{\un{b}} \g_{\un{j}]} ~~~,\\
    \g^{\un{a}\un{b}[1]}\g_{\un{i}\un{j}\un{l}\un{m}}\g_{[1]} ~&=~ \frac{1}{5!} \e^{\un{a}\un{b}}{}_{\un{i}\un{j}\un{l}\un{m}[5]} \g^{[5]}  +\frac{1}{2} \d_{[\un{i}}{}^{[\un{a}} \g_{\un{j}\un{l}\un{m}]}{}^{\un{b}]} -\frac{5}{2} \d_{[\un{i}}{}^{\un{a}} \d_{\un{j}}{}^{\un{b}} \g_{\un{l}\un{m}]} ~~~,\\
    \g^{[2][\un{a}} \g^{\un{b}]}{}_{[2]} ~&=~ -144 \g^{\un{a}\un{b}} ~~~,\\
    \g^{[2][\un{a}} \g_{\un{h}\un{i}\un{j}} \g^{\un{b}]}{}_{[2]} ~&=~ 80 \d_{[\un{h}}{}^{\un{a}} \d_{\un{i}}{}^{\un{b}} \g_{\un{j}]} ~~~,\\
    \g^{[2][\un{a}} \g_{\un{i}\un{j}\un{l}\un{m}} \g^{\un{b}]}{}_{[2]} ~&=~ \frac{2}{15}\e^{\un{a}\un{b}}{}_{\un{i}\un{j}\un{l}\un{m}[5]} \g^{[5]} - 16 \d_{[\un{i}}{}^{\un{a}} \d_{\un{j}}{}^{\un{b}} \g_{\un{l}\un{m}]} ~~~,\\
    \g^{[3]} \g^{\un{a}\un{b}}{}_{[3]} ~&=~ -504 \g^{\un{a}\un{b}} ~~~,\\
    \g^{[3]} \g_{\un{i}\un{j}\un{k}} \g^{\un{a}\un{b}}{}_{[3]} ~&=~ -48 \g^{\un{a}\un{b}}{}_{\un{i}\un{j}\un{k}} - 168 \d_{[\un{i}}{}^{\un{a}} \d_{\un{j}}{}^{\un{b}} \g_{\un{k}]} ~~~,\\
    \g^{[3]} \g_{\un{j}\un{k}\un{l}\un{m}} \g^{\un{a}\un{b}}{}_{[3]} ~&=~ \frac{1}{5} \e^{\un{a}\un{b}}{}_{\un{j}\un{k}\un{l}\un{m}[5]}\g^{[5]}  + 8 \d_{[j}{}^{[\un{a}} \g^{\un{b}]}{}_{\un{k}\un{l}\un{m}]} ~~~.
\end{align}

\newpage
\section{Fierz Identities for Analytical Expressions of Cubic Monomials}
\label{appen:cubicfierz}

In the derivation of the cubic irreducible monomials, we 
encountered a number of Fierz identities.  In this appendix
we list the ones relevant to derive our results.

\subsection{For $\CMTred{\{32\}}$ $\theta$-monomials}
\label{sec:Tmon1}

\begin{align}
    C_{[\d\e}\, C_{\b]\a} ~=&~ \frac{1}{32} \Big\{        C_{[\e\b}\, C_{\d]\a} + \frac{1}{3!}(\g{}^{[3]})_{[\e\b} \,  (\g_{[3]}){}_{\d]\a} - \frac{1}{4!} (\g{}^{[4]})_{[\e\b} \, (\g_{[4]}){}_{\d]\a}  \Big\}  ~~~,\\
    (\g{}^{ [3] })_{[\d\e}\,  (\g_{[3]}){}_{\b]\a} ~=&~ \frac{1}{32} \Big\{ 990 C_{[\e\b}\, C_{\d]\a} - 5 (\g{}^{[3]})_{[\e\b} \,  (\g_{[3]}){}_{\d]\a} - \frac{11}{4} (\g{}^{[4]})_{[\e\b} \,  (\g_{[4]}){}_{\d]\a} \Big\}  ~~~,\\
    (\g{}^{ [4] })_{[\d\e}\,  (\g_{[4]}){}_{\b]\a} ~=&~ - \frac{495}{2} C_{[\e\b}\, C_{\d]\a}  - \frac{11}{4} (\g{}^{[3]})_{[\e\b}\,  (\g_{[3]}){}_{\d]\a} + \frac{3}{16} (\g{}^{[4]})_{[\e\b} \, (\g_{[4]}){}_{\d]\a}~~~.
\end{align}

\subsection{For $\CMTred{\{320\}}$ $\theta$-monomials}
\label{sec:Tmon2}

\begin{align}
\begin{split}
    C_{[\d\e}(\g^{\un{a}})_{\b]\a} ~=&~ \frac{1}{32} \Big\{  - C_{[\e\b}(\g^{\un{a}})_{\d]\a}  - \frac{1}{3!}  (\g^{[3]})_{[\e\b}(\g^{\un a}_{\ [3]})_{\d]\a} + \frac{1}{2} (\g^{\un{a}[2]})_{[\e\b} (\g_{[2]})_{\d]\a}  \\
    & \qquad - \frac{1}{4!} (\g^{[4]})_{[\e\b} (\g^{\un a}_{\ [4]})_{\d]\a}+ \frac{1}{3!} (\g^{\un{a}[3]})_{[\e\b} (\g_{[3]})_{\d]\a} \Big\} ~~~,
\end{split}  \\
\begin{split}
    (\g^{\un{a}[2] })_{[\d\e}\,  (\g_{[2]}){}_{\b]\a} ~=&~ \frac{1}{32} \Big\{ 90 C_{[\e\b}(\g^{\un{a}})_{\d]\a} - (\g^{[3]})_{[\e\b}(\g^{\un a}_{\ [3]})_{\d]\a} - 13 (\g^{\un{a}[2]})_{[\e\b}(\g_{[2]})_{\d]\a} \\ 
    & \qquad - \frac{1}{4} (\g^{[4]})_{[\e\b}(\g^{\un a}_{\ [4]})_{\d]\a} + (\g^{\un{a}[3]})_{[\e\b}(\g_{[3]})_{\d]\a}  \Big\} ~~~,
\end{split}  \\
\begin{split}
    (\g^{[3]})_{[\d\e}\, ( \g^{ \un{a} }{}_{[3]})_{\b]\a} ~=&~  - \frac{45}{2} C_{[\e\b}(\g^{\un{a}})_{\d]\a}  - \frac{1}{4} (\g^{[3]})_{[\e\b}(\g^{\un a}_{\ [3]})_{\d]\a} - \frac{3}{4} (\g^{\un{a}[2]})_{[\e\b}(\g_{[2]})_{\d]\a}  \\
    & \qquad + \frac{1}{16} (\g^{[4]})_{[\e\b}(\g^{\un a}_{\ [4]})_{\d]\a} - \frac{1}{4} (\g^{\un{a}[3]})_{[\e\b} (\g_{[3]})_{\d]\a} ~~~,
\end{split} \\
\begin{split}
    (\g^{\un{a}[3]})_{[\d\e} \,  (\g_{[3]}){}_{\b]\a} ~=&~  \frac{45}{2} C_{[\e\b}(\g^{\un{a}})_{\d]\a} - \frac{1}{4} (\g^{[3]})_{[\e\b}(\g^{\un a}_{\ [3]})_{\d]\a} + \frac{3}{4} (\g^{\un{a}[2]})_{[\e\b}(\g_{[2]})_{\d]\a} \\
    & \qquad - \frac{1}{16} (\g^{[4]})_{[\e\b}(\g^{\un a}_{\ [4]})_{\d]\a} - \frac{1}{4} (\g^{\un{a}[3]})_{[\e\b} (\g_{[3]})_{\d]\a} ~~~,
\end{split}   \\
\begin{split}
    (\g^{[4]})_{[\d\e} \, ( \g^{\un{a} } {}_{[4]})_{\b]\a}  ~=&~         -\frac{315}{2} C_{[\e\b}(\g^{\un{a}})_{\d]\a} + \frac{7}{4} (\g^{[3]})_{[\e\b}(\g^{\un a}_{\ [3]})_{\d]\a} - \frac{21}{4} (\g^{\un{a}[2]})_{[\e\b}(\g_{[2]})_{\d]\a}  \\
    & \qquad - \frac{1}{16} (\g^{[4]})_{[\e\b}(\g^{\un a}_{\ [4]})_{\d]\a} - \frac{7}{4} (\g^{\un{a}[3]})_{[\e\b} (\g_{[3]})_{\d]\a} ~~~.
\end{split} 
\end{align}

\subsection{For $\CMTred{\{1,408\}}$ $\theta$-monomials}
\label{sec:Tmon3}

\begin{align}
\begin{split}
    (\g^{\un{a}\un{b}[2] })_{[\d\e}\,  (\g_{[2]})_{\b]\a} ~=&~ \frac{9}{4} C_{[\e\b}(\g^{\un{a}\un{b}})_{\d]\a} + \frac{1}{4} (\g^{[2][\un a})_{[\e\b}(\g^{\un b]}{}_{[2]})_{\d]\a} + \frac{5}{4} (\g^{\un{a}\un{b}[1]})_{[\e\b} (\g_{[1]})_{\d]\a}  \\  
    & - \frac{1}{4\times4!5!}\e^{\un{a}\un{b}}{}_{[4][5]} (\g^{[4]})_{[\e\b}(\g^{[5]})_{\d]\a} - \frac{1}{4} (\g^{\un{a}\un{b}[2]})_{[\e\b} (\g_{[2]})_{\d]\a} ~~~,
\end{split}  \\
\begin{split}
    (\g{}^{[3] [\un{a}})_{[\d\e} \, ( \g^{ \un{b} ]} {}_{[3]})_{\b]\a} ~=&~ \frac{63}{2} C_{[\e\b}(\g^{\un{a}\un{b}})_{\d]\a} + \frac{1}{2} (\g^{[3]})_{[\e\b}(\g^{\un{a}\un{b}}{}_{ [3]})_{\d]\a} - \frac{21}{2} (\g^{\un{a}\un{b}[1]})_{[\e\b}(\g_{[1]})_{\d]\a} \\
    &  - \frac{1}{16\times 5!} \e^{\un{a}\un{b}}{}_{[4][5]} (\g^{[4]})_{[\e\b}(\g^{[5]})_{\d]\a} ~~~,
\end{split}  \\
\begin{split}
    (\g^{\un{a}\un{b}[1] })_{[\d\e}\,  (\g_{[1]})_{\b]\a} ~=&~ -\frac{9}{32} C_{[\e\b}(\g^{\un{a}\un{b}})_{\d]\a} - \frac{1}{64} (\g^{[3]})_{[\e\b}(\g^{\un{a}\un{b}}{}_{[3]})_{\d]\a} + \frac{5}{64} (\g^{[2][\un a})_{[\e\b}(\g^{\un b]}{}_{[2]})_{\d]\a} \\
    & + \frac{7}{32} (\g^{\un{a}\un{b}[1]})_{[\e\b}(\g_{[1]})_{\d]\a} - \frac{1}{64} (\g^{[3][\un a})_{[\e\b}(\g^{\un b]}{}_{[3]})_{\d]\a} + \frac{5}{64} (\g^{\un{a}\un{b}[2]})_{[\e\b}(\g_{[2]})_{\d]\a} \\
    & - \frac{1}{32\times 4!5!} \e^{\un{a}\un{b}}{}_{[4][5]} (\g^{[4]})_{[\e\b}(\g^{[5]})_{\d]\a} ~~~,
\end{split}  \\
\begin{split}
    (\g^{ [2] [ \un{a} })_{[\d\e}\,  ( \g^{ \un{b} ]}{}_{[2]})_{\b]\a} ~=&~ \frac{9}{2} C_{[\e\b}(\g^{\un{a}\un{b}})_{\d]\a} + \frac{5}{2}(\g^{\un{a}\un{b}[1]})_{[\e\b}(\g_{[1]})_{\d]\a} + \frac{1}{2} (\g^{\un{a}\un{b}[2]})_{[\e\b} (\g_{[2]})_{\d]\a} \\
    & - \frac{1}{2\times4!5!} \e^{\un{a}\un{b}}{}_{[4][5]} (\g^{[4]})_{[\e\b}(\g^{[5]})_{\d]\a} ~~~,
\end{split}  \\
\begin{split}
    (\g^{[3]})_{[\d\e}\,  (\g^{\un{a}\un{b}}{}_{[3]})_{\b]\a} ~=&~ \frac{63}{4} C_{[\e\b}(\g^{\un{a}\un{b}})_{\d]\a} - \frac{1}{4} (\g^{[3]})_{[\e\b} (\g^{\un{a}\un{b}}{}_{ [3]})_{\d]\a} - \frac{21}{4} (\g^{\un{a}\un{b}[1]})_{[\e\b}(\g_{[1]})_{\d]\a}  \\
    & + \frac{1}{4} (\g^{[3][\un a})_{[\e\b} (\g^{\un b]}{}_{ [3]})_{\d]\a} - \frac{1}{32\times5!} \e^{\un{a}\un{b}}{}_{[4][5]} (\g^{[4]})_{[\e\b}(\g^{[5]})_{\d]\a} ~~~,
\end{split}  \\
\begin{split}
    C_{[\d\e}(\g^{\un{a}\un{b}})_{\b]\a } ~=&~ \frac{1}{32} \Big\{ - C_{[\e\b}(\g^{\un{a}\un{b}})_{\d]\a} + \frac{1}{3!} (\g^{[3]})_{[\e\b}(\g^{\un{a}\un{b}}{}_{ [3]})_{\d]\a} + \frac{1}{2} (\g^{[2][\un a})_{[\e\b}(\g^{\un b]}{}_{ [2]})_{\d]\a}  \\
    & \qquad - (\g^{\un{a}\un{b}[1]})_{[\e\b} (\g_{[1]})_{\d]\a} + \frac{1}{3!} (\g^{[3][\un a})_{[\e\b}(\g^{\un b]}{}_{ [3]})_{\d]\a} + \frac{1}{2} (\g^{\un{a}\un{b}[2]})_{[\e\b}(\g_{[2]})_{\d]\a}  \\
    & \qquad - \frac{1}{4!5!}\e^{\un{a}\un{b}}{}_{[4][5]} (\g^{[4]})_{[\e\b}(\g^{[5]})_{\d]\a} \Big\} ~~~.
\end{split}
\end{align}

\subsection{For $\CMTred{\{3,520\}}$ $\theta$-monomials}
\label{sec:Tmon4}

\begin{align}
\begin{split}
    (\g^{\un{a}\un{b}\un{c}})_{[\d\e}C_{\b]\a} ~=&~ \frac{1}{32}\,\Big\{ 
    ~-~ C_{[\e\b}(\g^{\un{a}\un{b}\un{c}})_{\d]\a}
    ~+~ \frac{1}{5!3!}\,\e^{\un{a}\un{b}\un{c}[3][5]}\,(\g_{[3]})_{[\e\b}(\g_{[5]})_{\d]\a} 
    ~+~ \frac{1}{4}\, (\g^{[2][\un{a}})_{[\e\b}(\g^{\un{b}\un{c}]}_{\ \ \ [2]})_{\d]\a}  \\
    & ~-~ \frac{1}{2}\, (\g^{[1][\un{a}\un{b}})_{[\e\b}(\g^{\un{c}]}_{\ \ [1]})_{\d]\a} 
    ~-~ (\g^{\un{a}\un{b}\un{c}})_{[\e\b}C_{\d]\a} ~-~ \frac{1}{4!4!}\, \e^{\un{a}\un{b}\un{c}[4][\bar{4}]}\, (\g_{[4]})_{[\e\b}(\g_{[\bar{4}]})_{\d]\a}  \\
    & ~+~ \frac{1}{12}\, (\g^{[3][\un{a}})_{[\e\b}(\g^{\un{b}\un{c}]}_{\ \ \ [3]})_{\d]\a} 
    ~+~ \frac{1}{4}\, (\g^{[2][\un{a}\un{b}})_{[\e\b}(\g^{\un{c}]}_{\ \ [2]})_{\d]\a} 
    ~+~ (\g^{\un{a}\un{b}\un{c}[1]})_{[\e\b}(\g_{[1]})_{\d]\a} \Big\} ~~~,
\end{split}  \\
\begin{split}
    C_{[\d\e}(\g^{\un{a}\un{b}\un{c}})_{\b]\a} ~=&~ \frac{1}{32}\,\Big\{ ~-~ C_{[\e\b}(\g^{\un{a}\un{b}\un{c}})_{\d]\a} 
    ~-~ \frac{1}{5!3!}\,\e^{\un{a}\un{b}\un{c}[3][5]}\,(\g_{[3]})_{[\e\b}(\g_{[5]})_{\d]\a} 
    ~+~ \frac{1}{4}\, (\g^{[2][\un{a}})_{[\e\b}(\g^{\un{b}\un{c}]}_{\ \ \ [2]})_{\d]\a}  \\
    & ~+~ \frac{1}{2}\, (\g^{[1][\un{a}\un{b}})_{[\e\b}(\g^{\un{c}]}_{\ \ [1]})_{\d]\a} 
    ~-~ (\g^{\un{a}\un{b}\un{c}})_{[\e\b}C_{\d]\a} ~-~ \frac{1}{4!4!}\, \e^{\un{a}\un{b}\un{c}[4][\bar{4}]}\, (\g_{[4]})_{[\e\b}(\g_{[\bar{4}]})_{\d]\a}  \\
    & ~-~ \frac{1}{12}\, (\g^{[3][\un{a}})_{[\e\b}(\g^{\un{b}\un{c}]}_{\ \ \ [3]})_{\d]\a} 
    ~+~ \frac{1}{4}\, (\g^{[2][\un{a}\un{b}})_{[\e\b}(\g^{\un{c}]}_{\ \ [2]})_{\d]\a} 
    ~-~ (\g^{\un{a}\un{b}\un{c}[1]})_{[\e\b}(\g_{[1]})_{\d]\a} \Big\} ~~~,
\end{split}  \\
\begin{split}
    (\g^{\un{a}\un{b}\un{c}[1]})_{[\d\e}(\g_{[1]})_{\b]\a} ~=&~ \frac{1}{32}\,\Big\{ 
    ~-~ 8 C_{[\e\b}(\g^{\un{a}\un{b}\un{c}})_{\d]\a} 
    ~-~ \frac{2}{5!3!}\,\e^{\un{a}\un{b}\un{c}[3][5]}\,(\g_{[3]})_{[\e\b}(\g_{[5]})_{\d]\a} 
    ~-~  (\g^{[2][\un{a}})_{[\e\b}(\g^{\un{b}\un{c}]}_{\ \ \ [2]})_{\d]\a}  \\
    & ~+~ 3\, (\g^{[1][\un{a}\un{b}})_{[\e\b}(\g^{\un{c}]}_{\ \ [1]})_{\d]\a} 
    ~+~ 8\, (\g^{\un{a}\un{b}\un{c}})_{[\e\b}C_{\d]\a} 
    ~+~ \frac{1}{6}\, (\g^{[3][\un{a}})_{[\e\b}(\g^{\un{b}\un{c}]}_{\ \ \ [3]})_{\d]\a}  \\
    & ~+~  (\g^{[2][\un{a}\un{b}})_{[\e\b}(\g^{\un{c}]}_{\ \ [2]})_{\d]\a} 
    ~+~ 6\, (\g^{\un{a}\un{b}\un{c}[1]})_{[\e\b}(\g_{[1]})_{\d]\a} \Big\} ~~~,
\end{split}  \\
\begin{split}
    (\g^{[1][\un{a}\un{b}})_{[\d\e}(\g^{\un{c}]}_{\ \ [1]})_{\b]\a} ~=&~ \frac{1}{32}\,\Big\{ ~ 
    48\, C_{[\e\b}(\g^{\un{a}\un{b}\un{c}})_{\d]\a} 
    ~-~ \frac{12}{5!3!}\,\e^{\un{a}\un{b}\un{c}[3][5]}\,(\g_{[3]})_{[\e\b}(\g_{[5]})_{\d]\a} 
    ~-~ 2\, (\g^{[2][\un{a}})_{[\e\b}(\g^{\un{b}\un{c}]}_{\ \ \ [2]})_{\d]\a}\\ 
    & ~-~ 6\, (\g^{[1][\un{a}\un{b}})_{[\e\b}(\g^{\un{c}]}_{\ \ [1]})_{\d]\a} 
    ~-~ 48\, (\g^{\un{a}\un{b}\un{c}})_{[\e\b}C_{\d]\a}
    ~-~ \frac{1}{3}\, (\g^{[3][\un{a}})_{[\e\b}(\g^{\un{b}\un{c}]}_{\ \ \ [3]})_{\d]\a} \\
    & ~+~ 2\, (\g^{[2][\un{a}\un{b}})_{[\e\b}(\g^{\un{c}]}_{\ \ [2]})_{\d]\a} 
    ~+~ 36\, (\g^{\un{a}\un{b}\un{c}[1]})_{[\e\b}(\g_{[1]})_{\d]\a} \Big\} ~~~,
\end{split}  \\
\begin{split}
    (\g^{[2][\un{a}\un{b}})_{[\d\e}(\g^{\un{c}]}_{\ \ [2]})_{\b]\a} ~=&~ \frac{1}{32}\,\Big\{ ~ 
    336\, C_{[\e\b}(\g^{\un{a}\un{b}\un{c}})_{\d]\a} 
    ~-~ \frac{4!}{5!3!}\,\e^{\un{a}\un{b}\un{c}[3][5]}\,(\g_{[3]})_{[\e\b}(\g_{[5]})_{\d]\a}
    ~+~ 4\, (\g^{[2][\un{a}})_{[\e\b}(\g^{\un{b}\un{c}]}_{\ \ \ [2]})_{\d]\a}  \\
    & ~+~ 28\, (\g^{[1][\un{a}\un{b}})_{[\e\b}(\g^{\un{c}]}_{\ \ [1]})_{\d]\a}
    ~+~ 336\, (\g^{\un{a}\un{b}\un{c}})_{[\e\b}C_{\d]\a} 
    ~-~ \frac{48}{4!4!}\, \e^{\un{a}\un{b}\un{c}[4][\bar{4}]}\, (\g_{[4]})_{[\e\b}(\g_{[\bar{4}]})_{\d]\a} \\
    & ~+~ \frac{2}{3}\, (\g^{[3][\un{a}})_{[\e\b}(\g^{\un{b}\un{c}]}_{\ \ \ [3]})_{\d]\a} 
    ~+~ 4\, (\g^{[2][\un{a}\un{b}})_{[\e\b}(\g^{\un{c}]}_{\ \ [2]})_{\d]\a} 
    ~+~ 168\, (\g^{\un{a}\un{b}\un{c}[1]})_{[\e\b}(\g_{[1]})_{\d]\a} \Big\} ~~~,
\end{split}  \\
\begin{split}
    (\g^{[2][\un{a}})_{[\d\e}(\g^{\un{b}\un{c}]}_{\ \ \ [2]})_{\b]\a} ~=&~ \frac{1}{32}\,\Big\{ ~ 
    336\, C_{[\e\b}(\g^{\un{a}\un{b}\un{c}})_{\d]\a} 
    ~+~ \frac{4!}{5!3!}\,\e^{\un{a}\un{b}\un{c}[3][5]}\,(\g_{[3]})_{[\e\b}(\g_{[5]})_{\d]\a} 
    ~+~ 4\, (\g^{[2][\un{a}})_{[\e\b}(\g^{\un{b}\un{c}]}_{\ \ \ [2]})_{\d]\a}  \\
    & ~-~ 28\, (\g^{[1][\un{a}\un{b}})_{[\e\b}(\g^{\un{c}]}_{\ \ [1]})_{\d]\a} 
    ~+~ 336\, (\g^{\un{a}\un{b}\un{c}})_{[\e\b}C_{\d]\a} 
    ~-~ \frac{48}{4!4!}\, \e^{\un{a}\un{b}\un{c}[4][\bar{4}]}\, (\g_{[4]})_{[\e\b}(\g_{[\bar{4}]})_{\d]\a}  \\
    &~-~ \frac{2}{3}\, (\g^{[3][\un{a}})_{[\e\b}(\g^{\un{b}\un{c}]}_{\ \ \ [3]})_{\d]\a}
    ~+~ 4\, (\g^{[2][\un{a}\un{b}})_{[\e\b}(\g^{\un{c}]}_{\ \ [2]})_{\d]\a}
    ~-~ 168\, (\g^{\un{a}\un{b}\un{c}[1]})_{[\e\b}(\g_{[1]})_{\d]\a} \Big\} ~~~.
\end{split}
\end{align}

\newpage
\section{Handicraft Approach to Scalar Superfield Decomposition in 11D $\mathcal{N}=1$}\label{appen:handicraft}

In this appendix, we will apply the ``Handicraft" approach in
a manner similar as was done in 10D \cite{counting10d} but now to derive the Lorentz descriptions of the component fields that
occur in the eleven dimensional scalar superfield. 

First, we introduce the spinorial Young Tableau as an extension of the normal Young Tableau which is a useful tool in group theory. In order to distinguish the bosonic Young tableaux and spinorial Young tableaux, we apply different colors to the boxes: Young Tableaux with blue boxes are bosonic and the ones with red boxes are spinorial. Namely, when calculating the dimension of a representation associated with any Young Tableau, we put ``11" into the box at the uppermost left corner of the tableau if it is bosonic and ``32" if it is spinorial in 11D. We also color the irreps: blue if it's bosonic
and red if it's spinorial.

At every level of the $\theta-$expansion, the d.o.f. (degree of
freedom) of each component field or $\theta-$monomial corresponds to one irreducible representation of $\mathfrak{so}(11)$. The zeroth level is $\CMTB{\{1\}}$ which is the trivial representation of $\mathfrak{so}(11)$. The first level is also trivially irreducible, since $\CMTred{\{32\}}$ is already an irreducible representation corresponding to the spinor representation. 
However, in the higher levels, the story is a nontrivial one.
In the following subsections, we will present the step-by-step calculations in quadratic, cubic, and quartic level. We will also show the results of the quintic level.  This method will only give us unique solutions up to the quintic level. 

\subsection{Quadratic Level}
\label{sec:QL1}

Starting with the quadratic level first, We can still use Young Tableaux to denote \emph{reducible} representations of so(11). The rules of tensor product of two Young tableaux are still valid. Thus, we have
\begin{equation}
    \CMTred {\yng(1)} ~\otimes~ \CMTred {\yng(1)} ~=~ \CMTred {\yng(1,1)} ~\oplus~ \CMTred {\yng(2)}~~~,
\end{equation}
where the entries in $\CMTred {\yng(1,1)}$ are completely anti-symmetric spinor indices and the entries in $\CMTred {\yng(2)}$ are completely symmetric spinor indices. Therefore the dimensions of these two reducible representations are $496$ and $528$ respectively. 
Moreover, $\CMTred {\yng(1,1)}$ and $\CMTred {\yng(2)}$ are all Young tableaux that contain two boxes. By using the Mathematica application ÒLieARTÓ (Lie Algebras and Representation Theory) provided by \cite{LieART}, the following result for the tensor product decomposition in SO(11) is seen:
\begin{equation}
      \CMTred {\{32\}} ~\otimes~ \CMTred {\{32\}} ~=~ \CMTB {\{1\}} \oplus \CMTB {\{11\}} \oplus \CMTB {\{55\}} \oplus \CMTB {\{165\}} \oplus \CMTB {\{330\}} \oplus \CMTB {\{462\}}   ~~~.
      \label{equ:32times32}
    \end{equation}
Now we know what are the decompositions of $\CMTred {\yng(1,1)}$ and $\CMTred {\yng(2)}$~:

\begin{align}
    \CMTred {\yng(1,1)} ~&=~ \CMTB {\{1\}} ~\oplus~ \CMTB {\{165\}} ~\oplus~ \CMTB {\{330\}} ~~~,\label{equ:quartic_anti}\\
    \CMTred {\yng(2)} ~&=~ \CMTB {\{11\}} ~\oplus~ \CMTB {\{55\}} ~\oplus~ \CMTB {\{462\}} ~~~,
\end{align}
since there is only one way to pick numbers that add up in the r.h.s. of equation (\ref{equ:32times32}) such that their sum is $496$ (or $528$). We have a Python code to do this type of searching. The Python code is attached in the end of this appendix \ref{appen:code} and a brief instruction is also included. 

This lowest order example shows us something of interest.  
If we were to impose as a definition the rule that $\CMTred{\yng(1)}$ is equivalent to the Grassmann coordinate $\theta^{\a}$, such that  {\it {only}} the totally antisymmetric product is meaningful,
then we would immediate retain only the single column Tableau
and its $\CMTB {\{1\}}$, $\CMTB {\{165\}}$, and $\CMTB {\{330\}}$ representations.

Note that in Sec.~\ref{sec:analytical} we discuss the quadratic level from the analytical aspect and all possible quadratic $\theta$-monomials are $\CMTB{\{1\}}$, $\CMTB{\{165\}}$, and $\CMTB{\{330\}}$, which are consistent with equation (\ref{equ:quartic_anti}). 

\subsection{Cubic Level}
\label{sec:Cmon1}

Using similar logic as in the quadratic level, we construct the following tensor product decomposition first:

\begin{equation}
\begin{split}
\label{equ:cubic_YT}
    \CMTred {\yng(1)} ~\otimes~ \CMTred {\yng(1)}~\otimes~ \CMTred {\yng(1)} ~=&~ \CMTred {\yng(1,1)}~\otimes~ \CMTred {\yng(1)} ~\oplus~ \CMTred {\yng(2)}~\otimes~ \CMTred {\yng(1)}\\
    ~=&~ \CMTred {\yng(3)}  ~\oplus~ \CMTred {\yng(2,1)}  ~\oplus~ \CMTred {\yng(2,1)} ~\oplus~\CMTred {\yng(1,1,1)}\\
    ~=&~ \Big[~\CMTB {\{1\}} ~\oplus~ \CMTB {\{165\}} ~\oplus~ \CMTB {\{330\}}~\Big]~\otimes~ \CMTred {\{32\}}\\ 
    &~\oplus~ \Big[~\CMTB {\{11\}} ~\oplus~ \CMTB {\{55\}} ~\oplus~ \CMTB {\{462\}}~\Big]~\otimes~ \CMTred {\{32\}}\\
    ~=&~ \Big[~ (3)\CMTred {\{32\}}~\oplus~(3)\CMTred {\{320\}}~\oplus~ (2)\CMTred {\{1,408\}}~\oplus~\CMTred {\{3,520\}}~\oplus~\CMTred {\{4,224\}}~\oplus~\CMTred {\{5,280\}} ~\Big] \\
    &~\oplus~ \Big[~ (3)\CMTred {\{32\}}~\oplus~(2)\CMTred {\{320\}}~\oplus~ (2)\CMTred {\{1,408\}}~\oplus~(2)\CMTred {\{3,520\}}~\oplus~\CMTred {\{5,280\}}~\Big] \\
    ~=&~ (6) \CMTred {\{32\}}~\oplus~ (5)\CMTred {\{320\}}~\oplus~(4)\CMTred {\{1,408\}}~\oplus~(3)\CMTred {\{3,520\}}~\oplus~\CMTred {\{4,224\}} \\
    ~~&~~\oplus~(2)\CMTred {\{5,280\}} ~~~.
\end{split}
\end{equation}

From this process we can obtain some important pieces of information: 
\begin{align}
\label{equ:3x3-1}
    \CMTred {\yng(3)}  ~\oplus~ \CMTred {\yng(2,1)} ~=&~ (3)\CMTred {\{32\}}~\oplus~(3)\CMTred {\{320\}}~\oplus~ (2)\CMTred {\{1,408\}}~\oplus~\CMTred {\{3,520\}}\nonumber\\
    &~\oplus~\CMTred {\{4,224\}}~\oplus~\CMTred {\{5,280\}} ~~~,\\
    \CMTred {\yng(2,1)} ~\oplus~\CMTred {\yng(1,1,1)} ~=&~(3)\CMTred {\{32\}}~\oplus~(2)\CMTred {\{320\}}~\oplus~ (2)\CMTred {\{1,408\}}~\oplus~(2)\CMTred {\{3,520\}}
    ~\oplus~\CMTred {\{5,280\}}\label{equ:3x3-2} ~~~.
\end{align}

Since we know $32~\times~31~\times~30/3!~=~4,960$, we can use the program discussed in Appendix~\ref{appen:code} with two assumptions and obtain the unique solution for the decomposition of the completely antisymmetric part:
\begin{equation}\label{equ:cubic_anti}
    \CMTred {\yng(1,1,1)} ~=~ \CMTred {\{32\}} \oplus \CMTred {\{1,408\}}\oplus \CMTred {\{3,520\}} ~~~.
\end{equation}
The two assumptions are: \newline \indent
(a.) assume the cubic level must include the linear level, i.e. $\CMTred{\{32\}}$ must show up in the solution,
\newline \indent
(b.) assume each irrep only appears once. 
\newline \noindent
Note in Sec.~\ref{sec:analytical} we discuss the cubic level from analytical aspect and all possible cubic $\theta$-monomials are $\CMTred{\{32\}}$, $\CMTred{\{1,408\}}$, and $\CMTred{\{3,520\}}$, which are consistent with equation (\ref{equ:cubic_anti}). 

Next we can solve the linear equations (\ref{equ:3x3-1}), (\ref{equ:3x3-2}) and obtain the decompositions of all possible spinorial Young Tableaux (SYT) at the cubic level as following:
\begin{equation}
\begin{split}
\label{equ:all_cubic_YT}
    \CMTred {\yng(1,1,1)} ~=&~ \CMTred {\{32\}} \oplus \CMTred {\{1,408\}}\oplus \CMTred {\{3,520\}}  ~~~,\\
    \CMTred {\yng(3)}~=&~ \CMTred {\{32\}} \oplus \CMTred {\{320\}} \oplus\CMTred {\{1,408\}}\oplus \CMTred {\{4,224\}} ~~~,\\
    \CMTred {\yng(2,1)}~=&~ (2)\CMTred {\{32\}} \oplus (2)\CMTred {\{320\}} \oplus\CMTred {\{1,408\}}\oplus \CMTred {\{3,520\}}\oplus \CMTred {\{5,280\}} ~~~,
\end{split}
\end{equation}
and of course, this is without applying the rule that only the
single column SYT contributes at the end.  It is often convenient to not impose this condition until the end of
a calculation at a given level.

From equation~(\ref{equ:all_cubic_YT}) we see that $\CMTred{\{320\}}$ only shows up in the decompositions of $\CMTred {\yng(3)}$ and $\CMTred {\yng(2,1)}$ which have more than one columns. $\CMTred{\{5,280\}}$ only shows up in the decomposition of $\CMTred {\yng(2,1)}$. Based on the symmetry properties, we can state that the $\CMTred{\{320\}}$ and $\CMTred{\{5,280\}}$ cubic monomials should be identically zero, which is consistent with our analytical discussion in Sec.~\ref{sec:analytical}. 

If one checks the dimension, it will be found the dimensions calculated by YT rules in the l.h.s of the equation (\ref{equ:all_cubic_YT}) are exactly the sums of numbers on the r.h.s.. Note that, the following equations are the only two independent equations we can find that all SYT in the r.h.s. have three boxes. However, there are three kinds of SYT containing three boxes. So we have two equations with three undetermined variables. If we want to know all of the irreducible decompositions of these three SYT, we have to introduce extra information. This situation is general, as 
it will be shown that at the quartic level we find the same problem. 

\begin{align}
    \CMTred {\yng(2)}~\otimes~ \CMTred {\yng(1)} ~=&~ \CMTred {\yng(3)}  ~\oplus~ \CMTred {\yng(2,1)}     ~~~,\\
    \CMTred {\yng(1,1)}~\otimes~ \CMTred {\yng(1)} ~=&~ \CMTred {\yng(2,1)} ~\oplus~\CMTred {\yng(1,1,1)} ~~~.
\end{align}

\subsection{Quartic Level}
\label{sec:QLz}

As before, we construct the following tensor product decomposition first:
\begin{equation}
    \begin{split}
    \label{equ:quartic_product}
        \CMTred {\{32\}} ~\otimes~ \CMTred {\{32\}} ~\otimes~ \CMTred {\{32\}}  ~\otimes~ \CMTred {\{32\}} ~=&~ (6)\CMTB {\{1\}}~\oplus~(11)\CMTB {\{11\}}~\oplus~(15)\CMTB {\{55\}}\\
        & ~\oplus~(5)\CMTB {\{65\}}~\oplus~(18)\CMTB {\{165\}}~\oplus~(20)\CMTB {\{330\}}~\oplus~(9)\CMTB {\{429\}}\\
        &~\oplus~(21)\CMTB {\{462\}}~\oplus~(4)\CMTB {\{1,144\}}~\oplus~(12)\CMTB {\{1,430\}}\\
        &~\oplus~(14)\CMTB {\{3,003\}}~\oplus~(15)\CMTB {\{4,290\}}~\oplus~(7)\CMTB {\{5,005\}}\\
        &~\oplus~(3)\CMTB {\{7,865\}}~\oplus~(9)\CMTB {\{11,583\}}~\oplus~(10)\CMTB {\{17,160\}}\\
        &~\oplus~(5)\CMTB {\{23,595\}}~\oplus~(2)\CMTB {\{23,595^{'}\}}~\oplus~\CMTB {\{28,314\}}\\
        &~\oplus~(6)\CMTB {\{37,752\}}~\oplus~(3)\CMTB {\{47,190\}} ~~~.
    \end{split}
    \end{equation}

From Equation (\ref{equ:quartic_product}), we want to find which irreducible representations appear anti-symmetrically. Since we know $32\times31\times30\times29/4!=35,960$, using the program discussed in Appendix~\ref{appen:code} with two assumptions, we find two possible solutions: 
\begin{equation}\label{equ:proposal1}
    \begin{split}
       [\CMTred {\{32\}} ~\otimes~ \CMTred {\{32\}}~\otimes~ \CMTred {\{32\}}\otimes~ \CMTred {\{32\}}]_A ~=&~ \CMTB {\{1\}} ~\oplus~ \CMTB {\{165\}} ~\oplus~ \CMTB {\{330\}} ~\oplus~ \CMTB {\{1,144\}} ~\oplus~ \CMTB {\{1,430\}} \\
        &~\oplus~ \CMTB {\{4,290\}} ~\oplus~ \CMTB {\{5,005\}} ~\oplus~ \CMTB {\{23,595\}}  ~~~,
    \end{split}
\end{equation}
and 
\begin{equation}\label{equ:proposal2}
    \begin{split}
       [\CMTred {\{32\}} ~\otimes~ \CMTred {\{32\}}~\otimes~ \CMTred {\{32\}}\otimes~ \CMTred {\{32\}}]_A ~=&~ \CMTB {\{1\}} ~\oplus~ \CMTB {\{165\}} ~\oplus~ \CMTB {\{330\}} ~\oplus~ \CMTB {\{1,144\}} ~\oplus~ \CMTB {\{4,290\}} \\
        &~\oplus~ \CMTB {\{5,005\}} ~\oplus~ \CMTB {\{7,865\}} ~\oplus~ \CMTB {\{17,160\}}  ~~~.
    \end{split}
\end{equation}

The two assumptions we made are: (a.) assume  $\CMTB {\{1\}}$, $ \CMTB {\{165\}}$, and $ \CMTB {\{330\}}$ must show up in the solution; (b.) assume each irrep only appears once. 

Then, our task is to find which solution is correct, and we can apply the SYT analysis:
\begin{equation}
    \begin{split}
    \label{equ:quartic_YT}
        \CMTred {\yng(1)} ~\otimes~ \CMTred {\yng(1)}~\otimes~ \CMTred {\yng(1)}~\otimes~ \CMTred {\yng(1)} ~=&~ \Big[~ \CMTred {\yng(3)}  ~\oplus~ (2) \CMTred {\yng(2,1)}  ~\oplus~\CMTred {\yng(1,1,1)} ~\Big] ~\otimes~ \CMTred {\yng(1)} \\
        ~=&~ \Big[~ \CMTred {\yng(4)}~\oplus~\CMTred {\yng(3,1)}  ~\Big] \\
        &~ \oplus~ (2)\Big[~ \CMTred {\yng(3,1)}~\oplus~\CMTred {\yng(2,2)}~\oplus~\CMTred {\yng(2,1,1)}~\Big] \\
         &~ \oplus~ \Big[~ \CMTred {\yng(1,1,1,1)} ~\oplus~ \CMTred {\yng(2,1,1)}~\Big]  ~~~.
    \end{split}
\end{equation}

Since we are only interested in the decomposition of $\CMTred {\yng(1,1,1,1)}$, calculate the last line in equation (\ref{equ:quartic_YT}) first:
\begin{equation}
    \begin{split}
    \label{equ:YT111t1_result}
        \CMTred {\yng(1,1,1)} ~\otimes~ \CMTred {\yng(1)}  ~=&~ \CMTred {\yng(1,1,1,1)} ~\oplus~ \CMTred {\yng(2,1,1)}\\
        ~=&~ \Big[ \CMTred {\{32\}} \oplus \CMTred {\{1,408\}}\oplus \CMTred {\{3,520\}}\Big] ~\otimes~ \CMTred {\{32\}}\\
        ~=&~ \CMTB {\{1\}} ~\oplus~ \CMTB {\{11\}} ~\oplus~ (2)\CMTB {\{55\}} ~\oplus~ (3)\CMTB {\{165\}}~\oplus~ (3)\CMTB {\{330\}}\\
        &~\oplus~ \CMTB {\{429\}}~\oplus~ (3)\CMTB {\{462\}}~\oplus~ \CMTB {\{1,144\}}~\oplus~ (2)\CMTB {\{1,430\}}~\oplus~ (2)\CMTB {\{3,003\}}\\
        &~\oplus~ (2)\CMTB {\{4,290\}}~\oplus~ (2)\CMTB {\{5,005\}}~\oplus~ \CMTB {\{7,865\}}~\oplus~ (2)\CMTB {\{11,583\}}\\
        &~\oplus~ (2)\CMTB {\{17,160\}}~\oplus~ \CMTB {\{23,595\}}~\oplus~ \CMTB {\{37,752\}} ~~~.
    \end{split}
\end{equation}

However, at this stage there is no clue how to divide the final result in equation (\ref{equ:YT111t1_result}) into two sets which are corresponding to $\CMTred {\yng(1,1,1,1)}$ and $\CMTred {\yng(2,1,1)}$ respectively. 

Since we understand the quadratic and cubic levels very well, we can derive the following set of linear equations. 
\begin{align}
    \CMTred {\yng(3)} ~\otimes~ \CMTred {\yng(1)}~=&~ \CMTred {\yng(4)} ~\oplus~ \CMTred {\yng(3,1)}\label{equ:YT3t1} ~~~,\\
    \CMTred {\yng(2,1)} ~\otimes~ \CMTred {\yng(1)}~=&~ \CMTred {\yng(3,1)} ~\oplus~ \CMTred {\yng(2,2)} ~\oplus~\CMTred {\yng(2,1,1)} \label{equ:YT21t1} ~~~,\\
    \CMTred {\yng(1,1,1)} ~\otimes~ \CMTred {\yng(1)}~=&~ \CMTred {\yng(1,1,1,1)} ~\oplus~ \CMTred {\yng(2,1,1)}\label{equ:YT111t1}  ~~~,\\
    \CMTred {\yng(1,1)} ~\otimes~ \CMTred {\yng(1,1)}~=&~ \CMTred {\yng(2,2)}~\oplus~\CMTred {\yng(2,1,1)}~\oplus~\CMTred {\yng(1,1,1,1)} \label{equ:YT11t11} ~~~,\\
    \CMTred {\yng(2)} ~\otimes~ \CMTred {\yng(2)}~=&~ \CMTred {\yng(4)} ~\oplus~ \CMTred {\yng(2,2)}~\oplus~\CMTred {\yng(3,1)}\label{equ:YT2t2} ~~~,\\
    \CMTred {\yng(1,1)} ~\otimes~ \CMTred {\yng(2)}~=&~\CMTred {\yng(3,1)} ~\oplus~ \CMTred {\yng(2,1,1)}\label{equ:YT11t2} ~~~.
\end{align}

The l.h.s. in Equations (\ref{equ:YT3t1}) - (\ref{equ:YT11t2}) are known.
\begin{align}
    \CMTred {\yng(3)} ~\otimes~ \CMTred {\yng(1)}~=&~ 
    \CMTB {\{1\}} 
    ~\oplus~(2)\CMTB {\{11\}} 
    ~\oplus~(3)\CMTB {\{55\}} \nonumber\\
    &~\oplus~\CMTB {\{65\}}
    ~\oplus~(3)\CMTB {\{165\}}
    ~\oplus~(3)\CMTB {\{330\}}\nonumber\\
    &~\oplus~ (2)\CMTB {\{429\}}
    ~\oplus~ (4)\CMTB {\{462\}}
    ~\oplus~\CMTB {\{1,144\}} \nonumber\\
    &~\oplus~(2)\CMTB {\{1,430\}} 
    ~\oplus~(2)\CMTB {\{3,003\}}
    ~\oplus~(3)\CMTB {\{4,290\}} \nonumber\\
    &~\oplus~\CMTB {\{5,005\}}
    ~\oplus~\CMTB {\{11,583\}}
    ~\oplus~(2)\CMTB {\{17,160\}}\nonumber\\
     & ~\oplus~\CMTB {\{28,314\}}
     ~\oplus~\CMTB {\{37,752\}} 
    ~\oplus~\CMTB {\{47,190\}} ~~~, \label{equ:YT3t1_result}\\
    \CMTred {\yng(2,1)} ~\otimes~ \CMTred {\yng(1)}~=&~ 
    (2)\CMTB {\{1\}} 
    ~\oplus~(4)\CMTB {\{11\}} 
    ~\oplus~(5)\CMTB {\{55\}} \nonumber\\
    &~\oplus~(2)\CMTB {\{65\}}
    ~\oplus~(6)\CMTB {\{165\}}
    ~\oplus~(7)\CMTB {\{330\}}\nonumber\\
    &~\oplus~ (3)\CMTB {\{429\}}
    ~\oplus~ (7)\CMTB {\{462\}}
    ~\oplus~\CMTB {\{1,144\}} \nonumber\\
    &~\oplus~(4)\CMTB {\{1,430\}} 
    ~\oplus~(5)\CMTB {\{3,003\}}
    ~\oplus~(5)\CMTB {\{4,290\}} \nonumber\\
    &~\oplus~(2)\CMTB {\{5,005\}}
    ~\oplus~\CMTB {\{7,865\}}
    ~\oplus~(3)\CMTB {\{11,583\}}\nonumber\\
    &~\oplus~(3)\CMTB {\{17,160\}}
    ~\oplus~(2)\CMTB {\{23,595\}}
    ~\oplus~\CMTB {\{23,595^{'}\}} \nonumber\\
    & ~\oplus~(2)\CMTB {\{37,752\}} 
   ~\oplus~(2)\CMTB {\{47,190\}}  ~~~,\\
    \CMTred {\yng(1,1,1)} ~\otimes~ \CMTred {\yng(1)}~=&~ 
    \CMTB {\{1\}} 
    ~\oplus~\CMTB {\{11\}} 
    ~\oplus~(2)\CMTB {\{55\}} \nonumber\\
    &~\oplus~(3)\CMTB {\{165\}}
    ~\oplus~(3)\CMTB {\{330\}}
    ~\oplus~ \CMTB {\{429\}}\nonumber\\
    &~\oplus~ (3)\CMTB {\{462\}}
    ~\oplus~\CMTB {\{1,144\}} 
    ~\oplus~(2)\CMTB {\{1,430\}} \nonumber\\
    &~\oplus~(2)\CMTB {\{3,003\}}
    ~\oplus~(2)\CMTB {\{4,290\}} 
    ~\oplus~(2)\CMTB {\{5,005\}}\nonumber\\
    &~\oplus~\CMTB {\{7,865\}}
    ~\oplus~(2)\CMTB {\{11,583\}}
    ~\oplus~(2)\CMTB {\{17,160\}}\nonumber\\
   & ~\oplus~\CMTB {\{23,595\}}
     ~\oplus~\CMTB {\{37,752\}}  ~~~,\\
    \CMTred {\yng(1,1)} ~\otimes~ \CMTred {\yng(1,1)}~=&~ 
  (3)\CMTB {\{1\}} 
    ~\oplus~(2)\CMTB {\{11\}} 
    ~\oplus~(2)\CMTB {\{55\}} \nonumber\\
    &~\oplus~(2)\CMTB {\{65\}}
    ~\oplus~(5)\CMTB {\{165\}}
    ~\oplus~(6)\CMTB {\{330\}}\nonumber\\
    &~\oplus~ (2)\CMTB {\{429\}}
    ~\oplus~ (4)\CMTB {\{462\}}
    ~\oplus~(2)\CMTB {\{1,144\}} \nonumber\\
    &~\oplus~(2)\CMTB {\{1,430\}} 
    ~\oplus~(3)\CMTB {\{3,003\}}
    ~\oplus~(4)\CMTB {\{4,290\}} \nonumber\\
    &~\oplus~(2)\CMTB {\{5,005\}}
    ~\oplus~(2)\CMTB {\{7,865\}}
    ~\oplus~(2)\CMTB {\{11,583\}}\nonumber\\
    &~\oplus~(3)\CMTB {\{17,160\}}
    ~\oplus~(2)\CMTB {\{23,595\}}
    ~\oplus~\CMTB {\{23,595^{'}\}} \nonumber\\
    & ~\oplus~\CMTB {\{37,752\}}  ~~~,\\
    \CMTred {\yng(2)} ~\otimes~ \CMTred {\yng(2)}~=&~ 
   (3)\CMTB {\{1\}} 
    ~\oplus~(3)\CMTB {\{11\}} 
    ~\oplus~(3)\CMTB {\{55\}}\nonumber\\
    &~\oplus~(3)\CMTB {\{65\}}
    ~\oplus~(5)\CMTB {\{165\}}
    ~\oplus~(6)\CMTB {\{330\}}\nonumber\\
    &~\oplus~ (3)\CMTB {\{429\}}
    ~\oplus~ (5)\CMTB {\{462\}}
    ~\oplus~(2)\CMTB {\{1,144\}} \nonumber\\
    &~\oplus~(2)\CMTB {\{1,430\}} 
    ~\oplus~(3)\CMTB {\{3,003\}}
    ~\oplus~(5)\CMTB {\{4,290\}} \nonumber\\
    &~\oplus~\CMTB {\{5,005\}}
    ~\oplus~\CMTB {\{7,865\}}
    ~\oplus~\CMTB {\{11,583\}}\nonumber\\
    &~\oplus~(3)\CMTB {\{17,160\}}
    ~\oplus~\CMTB {\{23,595\}}
    ~\oplus~\CMTB {\{23,595^{'}\}}\nonumber\\
    &~\oplus~\CMTB {\{28,314\}}
     ~\oplus~\CMTB {\{37,752\}} 
     ~\oplus~\CMTB {\{47,190\}}  ~~~,\\
    \CMTred {\yng(1,1)} ~\otimes~ \CMTred {\yng(2)}~=&~~
    ~(3)\CMTB {\{11\}} 
    ~\oplus~(5)\CMTB {\{55\}}
    ~\oplus~(4)\CMTB {\{165\}}\nonumber\\
    &~\oplus~(4)\CMTB {\{330\}}
    ~\oplus~ (2)\CMTB {\{429\}}
    ~\oplus~ (6)\CMTB {\{462\}}\nonumber\\
    &~\oplus~(4)\CMTB {\{1,430\}} 
    ~\oplus~(4)\CMTB {\{3,003\}}
    ~\oplus~(3)\CMTB {\{4,290\}} \nonumber\\
    &~\oplus~(2)\CMTB {\{5,005\}}
    ~\oplus~(3)\CMTB {\{11,583\}}
    ~\oplus~(2)\CMTB {\{17,160\}}\nonumber\\
    &~\oplus~\CMTB {\{23,595\}}
     ~\oplus~(2)\CMTB {\{37,752\}} 
      ~\oplus~\CMTB {\{47,190\}} ~~~. \label{equ:YT11t2_result}
\end{align}

Notice that there are only four independent equations and two identities:
\begin{align}
    \CMTred {\yng(1,1,1,1)} ~=&~ \CMTred {\yng(1,1,1)} ~\otimes~ \CMTred {\yng(1)}  ~-~ \CMTred {\yng(2,1,1)}  ~~~,\\
    \CMTred {\yng(2,2)} ~=&~  \CMTred {\yng(1,1)} ~\otimes~ \CMTred {\yng(1,1)} ~-~ \CMTred {\yng(1,1,1)} ~\otimes~ \CMTred {\yng(1)}   ~~~,\\
    \CMTred {\yng(3,1)} ~=&~ \CMTred {\yng(1,1)} ~\otimes~ \CMTred {\yng(2)} ~-~ \CMTred {\yng(2,1,1)}  ~~~,\\
    \CMTred {\yng(4)} ~=&~ \CMTred {\yng(3)} ~\otimes~ \CMTred {\yng(1)}  ~-~\CMTred {\yng(1,1)} ~\otimes~ \CMTred {\yng(2)}~\oplus~ \CMTred {\yng(2,1,1)}  ~~~,\\
    \CMTred {\yng(2,1)} ~\otimes~ \CMTred {\yng(1)} ~=&~  \CMTred {\yng(1,1)} ~\otimes~ \CMTred {\yng(1,1)} ~-~ \CMTred {\yng(1,1,1)} ~\otimes~ \CMTred {\yng(1)} ~\oplus~ \CMTred {\yng(1,1)} ~\otimes~ \CMTred {\yng(2)} \label{equ:quartic_identity1} ~~~,\\
    \CMTred {\yng(2)} ~\otimes~ \CMTred {\yng(2)}~=&~ \CMTred {\yng(1,1)} ~\otimes~ \CMTred {\yng(1,1)} ~-~ \CMTred {\yng(1,1,1)} ~\otimes~ \CMTred {\yng(1)} ~\oplus~ \CMTred {\yng(3)} ~\otimes~ \CMTred {\yng(1)}\label{equ:quartic_identity2} ~~~.
\end{align}

Clearly these two identities (\ref{equ:quartic_identity1}) and (\ref{equ:quartic_identity2}) are correct based on equations (\ref{equ:YT3t1_result}) - (\ref{equ:YT11t2_result}). Now we are facing the same problem as in the cubic level that we have more undetermined variables than independent linear equations. Thus, we have to impose extra informations: equation (\ref{equ:proposal1}) and equation (\ref{equ:proposal2}). 
Observe that there is no $\CMTB{\{7,865\}}$ in the decomposition of $\CMTred {\yng(1,1)} ~\otimes~ \CMTred {\yng(2)}$ , which implies that there must be no $\CMTB{\{7,865\}}$ in the decomposition of $\CMTred {\yng(2,1,1)}$. However, there is $\CMTB{\{7,865\}}$ showing up in the r.h.s. of Equation (\ref{equ:YT111t1_result}). Thus, $\CMTB{\{7,865\}}$ must come from $\CMTred {\yng(1,1,1,1)}$ , which means the second solution is correct.

Collecting all of the information we have so far, we derive the decompositions of all SYT with four boxes:
\begin{equation}
    \begin{split}
    \CMTred {\yng(1,1,1,1)} ~=&~
        \CMTB {\{1\}} 
    ~\oplus~\CMTB {\{165\}}
    ~\oplus~\CMTB {\{330\}}
    ~\oplus~\CMTB {\{1,144\}} \\
   & ~\oplus~\CMTB {\{4,290\}} 
    ~\oplus~\CMTB {\{5,005\}}
    ~\oplus~\CMTB {\{7,865\}}
    ~\oplus~\CMTB {\{17,160\}}
     ~~~,
    \end{split}
\end{equation}

\begin{equation}
    \begin{split}
    \CMTred {\yng(2,1,1)} ~=&~
    \CMTB {\{11\}} 
    ~\oplus~(2)\CMTB {\{55\}}
    ~\oplus~(2)\CMTB {\{165\}}
    ~\oplus~(2)\CMTB {\{330\}}\\
    &~\oplus~ \CMTB {\{429\}}
    ~\oplus~ (3)\CMTB {\{462\}}
    ~\oplus~(2)\CMTB {\{1,430\}} 
    ~\oplus~(2)\CMTB {\{3,003\}}\\
    &~\oplus~\CMTB {\{4,290\}} 
    ~\oplus~\CMTB {\{5,005\}}
    ~\oplus~(2)\CMTB {\{11,583\}}
    ~\oplus~\CMTB {\{17,160\}}\\
    &~\oplus~\CMTB {\{23,595\}}
     ~\oplus~\CMTB {\{37,752\}} 
      ~~~,
    \end{split}
\end{equation}

\begin{equation}
    \begin{split}
     \CMTred {\yng(2,2)} ~=&~
        (2)\CMTB {\{1\}} 
    ~\oplus~\CMTB {\{11\}} 
    ~\oplus~(2)\CMTB {\{65\}}
    ~\oplus~(2)\CMTB {\{165\}} \\
    &~\oplus~(3)\CMTB {\{330\}}
    ~\oplus~ \CMTB {\{429\}}
    ~\oplus~ \CMTB {\{462\}}
    ~\oplus~ \CMTB {\{1,144\}} \\
    &~\oplus~ \CMTB {\{3,003\}}
    ~\oplus~(2)\CMTB {\{4,290\}} 
    ~\oplus~\CMTB {\{7,865\}}
    ~\oplus~\CMTB {\{17,160\}}\\
    &~\oplus~\CMTB {\{23,595\}}
    ~\oplus~\CMTB {\{23,595^{'}\}}
    ~~~,
    \end{split}
\end{equation}

\begin{equation}
    \begin{split}
    \CMTred {\yng(3,1)} ~=&~ 
   (2)\CMTB {\{11\}} 
    ~\oplus(3)\CMTB {\{55\}}
    ~\oplus~(2)\CMTB {\{165\}}
    ~\oplus~(2)\CMTB {\{330\}}\\
    &~\oplus~ \CMTB {\{429\}}
    ~\oplus~ (3)\CMTB {\{462\}}
    ~\oplus~(2)\CMTB {\{1,430\}} 
    ~\oplus~(2)\CMTB {\{3,003\}}\\
    &~\oplus~(2)\CMTB {\{4,290\}} 
    ~\oplus~\CMTB {\{5,005\}}
    ~\oplus~\CMTB {\{11,583\}}
    ~\oplus~\CMTB {\{17,160\}}\\
    & ~\oplus~\CMTB {\{37,752\}} 
     ~\oplus~\CMTB {\{47,190\}} 
      ~~~,
    \end{split}
\end{equation}

\begin{equation}
    \begin{split}
    \CMTred {\yng(4)} ~=&~ 
        \CMTB {\{1\}} 
    ~\oplus~\CMTB {\{65\}}
    ~\oplus~\CMTB {\{165\}}
    ~\oplus~\CMTB {\{330\}}\\
    &~\oplus~ \CMTB {\{429\}}
    ~\oplus~\CMTB {\{462\}}
    ~\oplus~\CMTB {\{1,144\}} 
    ~\oplus~\CMTB {\{4,290\}} \\
    &~\oplus~\CMTB {\{17,160\}}
    ~\oplus~\CMTB {\{28,314\}}
     ~~~.
    \end{split}
\end{equation}

\subsection{Quintic Level}
\label{sec:QLy}

Using similar methods as described above, we can solve for all the spinorial Young tableaux in the quintic level. For the totally antisymmetric SYT, there are two assumptions: (a.) the quintic level one includes the cubic level one, i.e. it must include $\CMTred{ \{32\} }$, $\CMTred{ \{1,408\} }$ and $\CMTred{ \{3,520\} }$; (b.) each irrep only appears once. Then a set of unique solutions comes to the fore.

\begin{equation}
\begin{split}
\label{equ:all_quintic_YT}
    \CMTred {\yng(5)} ~=&~ \CMTred {\{32\}} \oplus \CMTred {\{320\}} \oplus \CMTred {\{1,408\}} \oplus \CMTred {\{1,760\}} \oplus \CMTred {\{3,520\}} \oplus \CMTred {\{4,224\}} \oplus \CMTred {\{5,280\}} \\
    & \oplus \CMTred {\{10,240\}} \oplus \CMTred {\{24,960\}} \oplus \CMTred {\{36,960\}} \oplus \CMTred {\{137,280\}} \oplus \CMTred {\{151,008\}}   ~~~,
    \end{split}
\end{equation}
\begin{equation}
\begin{split}
    \CMTred {\yng(4,1)} ~=&~ (3) \CMTred {\{32\}} \oplus (5) \CMTred {\{320\}} \oplus (6) \CMTred {\{1,408\}} \oplus (2) \CMTred {\{1,760\}} \oplus (4) \CMTred {\{3,520\}} \oplus (3) \CMTred {\{4,224\}} \\
    & \oplus (3) \CMTred {\{5,280\}} \oplus (3) \CMTred {\{10,240\}} \oplus \CMTred {\{24,960\}} \oplus (2) \CMTred {\{28,512\}} \oplus (2) \CMTred {\{36,960\}} \oplus (2) \CMTred {\{45,056\}} \\
    & \oplus \CMTred {\{91,520\}} \oplus \CMTred {\{137,280\}} \oplus \CMTred {\{160,160\}} \oplus \CMTred {\{274,560\}} \oplus \CMTred {\{302,016\}}   ~~~,
    \end{split}
\end{equation}
\begin{equation}
\begin{split}
    \CMTred {\yng(3,2)} ~=&~ (5) \CMTred {\{32\}} \oplus (7) \CMTred {\{320\}} \oplus (6) \CMTred {\{1,408\}} \oplus (3) \CMTred {\{1,760\}} \oplus (6) \CMTred {\{3,520\}} \oplus (2) \CMTred {\{4,224\}} \\
    & \oplus (5) \CMTred {\{5,280\}} \oplus (3) \CMTred {\{10,240\}} \oplus \CMTred {\{24,960\}} \oplus (3) \CMTred {\{28,512\}} \oplus (2) \CMTred {\{36,960\}} \oplus (3) \CMTred {\{45,056\}} \\
    & \oplus \CMTred {\{91,520\}} \oplus \CMTred {\{128,128\}} \oplus \CMTred {\{137,280\}} \oplus \CMTred {\{160,160\}} \oplus \CMTred {\{251,680\}} \oplus \CMTred {\{292,864\}}   ~~~,
    \end{split}
\end{equation}
\begin{equation}
\begin{split}
    \CMTred {\yng(3,1,1)} ~=&~ (4) \CMTred {\{32\}} \oplus (7) \CMTred {\{320\}} \oplus (8) \CMTred {\{1,408\}} \oplus (2) \CMTred {\{1,760\}} \oplus (7) \CMTred {\{3,520\}} \oplus (3) \CMTred {\{4,224\}} \\
    & \oplus (5) \CMTred {\{5,280\}} \oplus (4) \CMTred {\{10,240\}} \oplus \CMTred {\{24,960\}} \oplus (5) \CMTred {\{28,512\}} \oplus \CMTred {\{36,960\}} \oplus (3) \CMTred {\{45,056\}} \\
    & \oplus (2) \CMTred {\{91,520\}} \oplus \CMTred {\{137,280\}} \oplus (2) \CMTred {\{160,160\}} \oplus \CMTred {\{274,560\}} \oplus \CMTred {\{292,864\}} ~~~,
    \end{split}
\end{equation}
\begin{equation}
\begin{split}
    \CMTred {\yng(2,2,1)} ~=&~ (4) \CMTred {\{32\}} \oplus (6) \CMTred {\{320\}} \oplus (6) \CMTred {\{1,408\}} \oplus (3) \CMTred {\{1,760\}} \oplus (6) \CMTred {\{3,520\}} \oplus (2) \CMTred {\{4,224\}} \\
    & \oplus (5) \CMTred {\{5,280\}} \oplus (3) \CMTred {\{10,240\}} \oplus \CMTred {\{24,960\}} \oplus (3) \CMTred {\{28,512\}} \oplus \CMTred {\{36,960\}} \oplus (3) \CMTred {\{45,056\}} \\
    & \oplus (2) \CMTred {\{91,520\}} \oplus \CMTred {\{128,128\}} \oplus (2) \CMTred {\{160,160\}} \oplus \CMTred {\{292,864\}}   ~~~,
\end{split}
\end{equation}
\begin{equation}
\begin{split}
    \CMTred {\yng(2,1,1,1)} ~=&~ (2) \CMTred {\{32\}} \oplus (3) \CMTred {\{320\}} \oplus (5) \CMTred {\{1,408\}} \oplus \CMTred {\{1,760\}} \oplus (5) \CMTred {\{3,520\}} \oplus \CMTred {\{4,224\}} \\
    & \oplus (3) \CMTred {\{5,280\}} \oplus (3) \CMTred {\{10,240\}} \oplus (2) \CMTred {\{24,960\}} \oplus (3) \CMTred {\{28,512\}} \oplus \CMTred {\{36,960\}} \oplus (2) \CMTred {\{45,056\}} \\
    & \oplus (2) \CMTred {\{91,520\}} \oplus \CMTred {\{128,128\}} \oplus \CMTred {\{137,280\}} \oplus \CMTred {\{160,160\}}  ~~~, 
\end{split}
\end{equation}
\begin{equation}
\begin{split}
    \CMTred {\yng(1,1,1,1,1)} ~=&~ \CMTred {\{32\}} \oplus \CMTred {\{1,408\}} \oplus \CMTred {\{3,520\}} \oplus \CMTred {\{4,224\}} \oplus \CMTred {\{10,240\}} \oplus \CMTred {\{24,960\}} \oplus \CMTred {\{28,512\}} \\
    & \oplus \CMTred {\{36,960\}} \oplus \CMTred {\{91,520\}} ~~~.
\end{split}
\end{equation}

We will close this appendix here.  However, there is one matter for future study that is raised 
by the results presented here.  In equations (\ref{equ:Exp2}) - (\ref{equ:Exp5}), there was given
a step-by-step recursive argument given for expanding a superfield.  It is possible that
adapting that $\theta$-coordinate based argument into the language of Young Tableaux
might provide extra information in the context of the handicraft approach.
\newpage

\subsection{Original script for the program in the Python language}\label{appen:code}
\label{sec:PyS1}

In this section, the original script for the program we used to do the searching is attached. The function of this program is explained as following.

First we have a set of numbers called ``candidates'' in the code. For example, in the code attached below, this set of numbers is $\{1, 11, 55, 165, 330, 462\}$, which contains the numbers showing up in the r.h.s. of equation (\ref{equ:32times32}). Then we have a target sum, and in this case our target sum is $496$. This program basically solves the equation
\begin{equation}
    b_1 ~+~ b_2 11~+~ b_3 55 ~+~ b_4 165 ~+~ b_5 330 ~+~ b_6 462 ~=~ 496 ~~~,
\end{equation}
where $b_1$ to $b_6$ are coefficients to be solved and they can be either $0$ or $1$. The output of this code is $[1, 165, 330]$ and then we have equation (\ref{equ:quartic_anti}). 

\perlscript{searching}{Original script for the program in the Python language\label{script}}
$~$

\newpage
\section{Lorentz Decomposition Results of the 11D, $\mathcal{N} = 1$ Scalar Superfield by Dynkin Labels\label{appen:dynkin}}

In Section \ref{sec:decomp_result}, we listed the component field content of the 11D, $\mathcal{N}=1$ scalar superfield by the dimensions of the SO(1,10) representations. Here we list them in terms of Dynkin labels.

\begin{itemize}
\sloppy
    \item Level-0: $\CMTB{(00000)}$
    \item Level-1: $\CMTred{(00001)}$
    \item Level-2: $\CMTB{(00000)} \oplus \CMTB{(00100)} \oplus \CMTB{(00010)}$
    \item Level-3: $\CMTred{(00001)} \oplus \CMTred{(01001)} \oplus \CMTred{(00101)}$
    \item Level-4: $\CMTB{(00000)} \oplus \CMTB{(00100)} \oplus \CMTB{(00010)} \oplus \CMTB{(02000)} \oplus \CMTB{(10002)} \oplus \CMTB{(01100)} \oplus \CMTB{(00200)} \oplus \CMTB{(01002)}$
    \item Level-5: $\CMTred{(00001)} \oplus \CMTred{(01001)} \oplus \CMTred{(00101)} \oplus \CMTred{(00003)} \oplus \CMTred{(11001)} \oplus \CMTred{(02001)} \oplus \CMTred{(10101)} \oplus \CMTred{(10003)} \oplus \CMTred{(01101)}$
    \item Level-6: $\CMTB{(00000)} \oplus \CMTB{(00100)} \oplus \CMTB{(00010)} \oplus \CMTB{(02000)} \oplus \CMTB{(10002)} \oplus \CMTB{(01100)} \oplus \CMTB{(20100)} \oplus \CMTB{(00200)} \oplus \CMTB{(20010)} \oplus (2) \CMTB{(01002)} \oplus \CMTB{(00004)} \oplus \CMTB{(11100)} \oplus \CMTB{(00102)} \oplus \CMTB{(02100)} \oplus \CMTB{(11010)} \oplus \CMTB{(11002)} \oplus \CMTB{(02010)} \oplus \CMTB{(10102)}$
    \item Level-7: $\CMTred{(00001)} \oplus \CMTred{(01001)} \oplus \CMTred{(00101)} \oplus \CMTred{(00003)} \oplus \CMTred{(30001)} \oplus \CMTred{(11001)} \oplus \CMTred{(02001)} \oplus (2) \CMTred{(10101)} \oplus \CMTred{(10003)} \oplus \CMTred{(10011)} \oplus \CMTred{(21001)} \oplus (2) \CMTred{(01101)} \oplus \CMTred{(20101)} \oplus \CMTred{(01003)} \oplus \CMTred{(12001)} \oplus \CMTred{(01011)} \oplus \CMTred{(20011)} \oplus \CMTred{(03001)} \oplus \CMTred{(00103)} \oplus \CMTred{(11101)} \oplus \CMTred{(11011)}$
    \item Level-8: $\CMTB{(00000)} \oplus \CMTB{(00100)} \oplus \CMTB{(00010)} \oplus \CMTB{(40000)} \oplus \CMTB{(02000)} \oplus \CMTB{(10002)} \oplus \CMTB{(01100)} \oplus \CMTB{(20100)} \oplus \CMTB{(31000)} \oplus (2) \CMTB{(00200)} \oplus (2) \CMTB{(20010)} \oplus (2) \CMTB{(01002)} \oplus \CMTB{(20002)} \oplus \CMTB{(00110)} \oplus \CMTB{(00020)} \oplus \CMTB{(00004)} \oplus \CMTB{(22000)} \oplus \CMTB{(11100)} \oplus \CMTB{(00102)} \oplus \CMTB{(10200)} \oplus \CMTB{(30010)} \oplus \CMTB{(02100)} \oplus \CMTB{(13000)} \oplus (2) \CMTB{(11010)} \oplus \CMTB{(30002)} \oplus \CMTB{(04000)} \oplus (2) \CMTB{(11002)} \oplus (2) \CMTB{(02010)} \oplus \CMTB{(10110)} \oplus \CMTB{(10020)} \oplus \CMTB{(20200)} \oplus \CMTB{(02002)} \oplus (2) \CMTB{(10102)} \oplus \CMTB{(21010)} \oplus \CMTB{(10012)} \oplus \CMTB{(21002)} \oplus \CMTB{(20110)} \oplus \CMTB{(01102)} \oplus \CMTB{(20020)} \oplus \CMTB{(12010)} \oplus \CMTB{(01012)} \oplus \CMTB{(12002)}$
    \item Level-9: $\CMTred{(00001)} \oplus \CMTred{(01001)} \oplus \CMTred{(00101)} \oplus \CMTred{(00003)} \oplus (2) \CMTred{(30001)} \oplus \CMTred{(11001)} \oplus \CMTred{(40001)} \oplus \CMTred{(02001)} \oplus (3) \CMTred{(10101)} \oplus \CMTred{(10003)} \oplus (2) \CMTred{(10011)} \oplus (2) \CMTred{(21001)} \oplus (3) \CMTred{(01101)} \oplus \CMTred{(00201)} \oplus (2) \CMTred{(20101)} \oplus \CMTred{(01003)} \oplus (2) \CMTred{(12001)} \oplus \CMTred{(31001)} \oplus (2) \CMTred{(01011)} \oplus \CMTred{(20003)} \oplus (3) \CMTred{(20011)} \oplus \CMTred{(00021)} \oplus (2) \CMTred{(03001)} \oplus \CMTred{(00103)} \oplus \CMTred{(00111)} \oplus \CMTred{(30101)} \oplus \CMTred{(22001)} \oplus (2) \CMTred{(11101)} \oplus \CMTred{(30011)} \oplus \CMTred{(10201)} \oplus \CMTred{(11003)} \oplus (3) \CMTred{(11011)} \oplus \CMTred{(13001)} \oplus \CMTred{(10021)} \oplus \CMTred{(02003)} \oplus \CMTred{(10111)} \oplus \CMTred{(02011)} \oplus \CMTred{(21101)} \oplus \CMTred{(21011)}$
    \item Level-10: $\CMTB{(00000)} \oplus \CMTB{(00100)} \oplus \CMTB{(00010)} \oplus \CMTB{(40000)} \oplus \CMTB{(02000)} \oplus \CMTB{(10002)} \oplus \CMTB{(01100)} \oplus (2) \CMTB{(20100)} \oplus \CMTB{(31000)} \oplus (2) \CMTB{(00200)} \oplus (3) \CMTB{(20010)} \oplus (3) \CMTB{(01002)} \oplus \CMTB{(20002)} \oplus \CMTB{(00110)} \oplus \CMTB{(00020)} \oplus \CMTB{(30100)} \oplus \CMTB{(00004)} \oplus \CMTB{(22000)} \oplus (2) \CMTB{(11100)} \oplus (2) \CMTB{(00102)} \oplus \CMTB{(10200)} \oplus (2) \CMTB{(30010)} \oplus (2) \CMTB{(02100)} \oplus \CMTB{(13000)} \oplus (3) \CMTB{(11010)} \oplus \CMTB{(40100)} \oplus (2) \CMTB{(30002)} \oplus \CMTB{(04000)} \oplus (3) \CMTB{(11002)} \oplus \CMTB{(21100)} \oplus (3) \CMTB{(02010)} \oplus (2) \CMTB{(10110)} \oplus \CMTB{(40010)} \oplus \CMTB{(00300)} \oplus (2) \CMTB{(10020)} \oplus \CMTB{(20200)} \oplus \CMTB{(02002)} \oplus (4) \CMTB{(10102)} \oplus (2) \CMTB{(21010)} \oplus (2) \CMTB{(10012)} \oplus \CMTB{(12100)} \oplus \CMTB{(31100)} \oplus (3) \CMTB{(21002)} \oplus \CMTB{(01110)} \oplus \CMTB{(01020)} \oplus \CMTB{(00210)} \oplus \CMTB{(00030)} \oplus (2) \CMTB{(20110)} \oplus (2) \CMTB{(01102)} \oplus (2) \CMTB{(20020)} \oplus (2) \CMTB{(12010)} \oplus \CMTB{(00120)} \oplus \CMTB{(31010)} \oplus (2) \CMTB{(01012)} \oplus (2) \CMTB{(20102)} \oplus \CMTB{(22100)} \oplus (2) \CMTB{(12002)} \oplus \CMTB{(31002)} \oplus \CMTB{(20012)} \oplus \CMTB{(03002)} \oplus \CMTB{(11110)} \oplus \CMTB{(22010)} \oplus \CMTB{(11020)} \oplus \CMTB{(30102)} \oplus \CMTB{(11102)} \oplus \CMTB{(11012)}$
    \item Level-11: $\CMTred{(00001)} \oplus \CMTred{(01001)} \oplus \CMTred{(00101)} \oplus (2) \CMTred{(00003)} \oplus (2) \CMTred{(30001)} \oplus (2) \CMTred{(11001)} \oplus \CMTred{(40001)} \oplus (2) \CMTred{(02001)} \oplus (4) \CMTred{(10101)} \oplus (2) \CMTred{(10003)} \oplus (2) \CMTred{(10011)} \oplus (3) \CMTred{(21001)} \oplus (4) \CMTred{(01101)} \oplus \CMTred{(00201)} \oplus (4) \CMTred{(20101)} \oplus (2) \CMTred{(01003)} \oplus (3) \CMTred{(12001)} \oplus (2) \CMTred{(31001)} \oplus (3) \CMTred{(01011)} \oplus (2) \CMTred{(20003)} \oplus (4) \CMTred{(20011)} \oplus \CMTred{(00021)} \oplus (2) \CMTred{(03001)} \oplus (2) \CMTred{(00103)} \oplus (2) \CMTred{(00111)} \oplus \CMTred{(41001)} \oplus (3) \CMTred{(30101)} \oplus (2) \CMTred{(22001)} \oplus (4) \CMTred{(11101)} \oplus \CMTred{(30003)} \oplus (2) \CMTred{(30011)} \oplus (2) \CMTred{(10201)} \oplus (2) \CMTred{(11003)} \oplus (5) \CMTred{(11011)} \oplus \CMTred{(02101)} \oplus \CMTred{(13001)} \oplus \CMTred{(40101)} \oplus \CMTred{(32001)} \oplus (2) \CMTred{(10021)} \oplus \CMTred{(02003)} \oplus \CMTred{(10103)} \oplus \CMTred{(40003)} \oplus (3) \CMTred{(10111)} \oplus (2) \CMTred{(02011)} \oplus (3) \CMTred{(21101)} \oplus \CMTred{(01201)} \oplus \CMTred{(21003)} \oplus (3) \CMTred{(21011)} \oplus \CMTred{(01111)} \oplus \CMTred{(01021)} \oplus \CMTred{(12101)} \oplus \CMTred{(31101)} \oplus \CMTred{(20103)} \oplus \CMTred{(20111)} \oplus \CMTred{(12011)}$
    \item Level-12: $\CMTB{(00000)} \oplus \CMTB{(00100)} \oplus \CMTB{(00010)} \oplus \CMTB{(40000)} \oplus (2) \CMTB{(02000)} \oplus (2) \CMTB{(10002)} \oplus (2) \CMTB{(01100)} \oplus (2) \CMTB{(20100)} \oplus \CMTB{(12000)} \oplus \CMTB{(31000)} \oplus (3) \CMTB{(00200)} \oplus (3) \CMTB{(20010)} \oplus (4) \CMTB{(01002)} \oplus (2) \CMTB{(20002)} \oplus \CMTB{(00110)} \oplus \CMTB{(00020)} \oplus \CMTB{(30100)} \oplus (2) \CMTB{(00004)} \oplus (2) \CMTB{(22000)} \oplus (3) \CMTB{(11100)} \oplus (2) \CMTB{(00102)} \oplus \CMTB{(00012)} \oplus (2) \CMTB{(10200)} \oplus (2) \CMTB{(30010)} \oplus (2) \CMTB{(02100)} \oplus \CMTB{(13000)} \oplus (4) \CMTB{(11010)} \oplus \CMTB{(40100)} \oplus (3) \CMTB{(30002)} \oplus \CMTB{(32000)} \oplus \CMTB{(04000)} \oplus (5) \CMTB{(11002)} \oplus (2) \CMTB{(21100)} \oplus (4) \CMTB{(02010)} \oplus (3) \CMTB{(10110)} \oplus \CMTB{(40010)} \oplus \CMTB{(00300)} \oplus (2) \CMTB{(10020)} \oplus \CMTB{(10004)} \oplus \CMTB{(42000)} \oplus (3) \CMTB{(20200)} \oplus \CMTB{(40002)} \oplus (2) \CMTB{(02002)} \oplus (5) \CMTB{(10102)} \oplus (3) \CMTB{(21010)} \oplus (3) \CMTB{(10012)} \oplus \CMTB{(12100)} \oplus (2) \CMTB{(31100)} \oplus (5) \CMTB{(21002)} \oplus (2) \CMTB{(01110)} \oplus \CMTB{(01020)} \oplus \CMTB{(00210)} \oplus \CMTB{(00030)} \oplus \CMTB{(50002)} \oplus (4) \CMTB{(20110)} \oplus \CMTB{(30200)} \oplus (4) \CMTB{(01102)} \oplus (3) \CMTB{(20020)} \oplus \CMTB{(11200)} \oplus (3) \CMTB{(12010)} \oplus \CMTB{(00120)} \oplus (2) \CMTB{(31010)} \oplus \CMTB{(00202)} \oplus \CMTB{(20004)} \oplus \CMTB{(41100)} \oplus (3) \CMTB{(01012)} \oplus (3) \CMTB{(20102)} \oplus \CMTB{(22100)} \oplus (3) \CMTB{(12002)} \oplus (3) \CMTB{(31002)} \oplus \CMTB{(02200)} \oplus (2) \CMTB{(20012)} \oplus \CMTB{(00112)} \oplus \CMTB{(40200)} \oplus \CMTB{(03002)} \oplus \CMTB{(30110)} \oplus (3) \CMTB{(11110)} \oplus (2) \CMTB{(22010)} \oplus (2) \CMTB{(11020)} \oplus \CMTB{(30004)} \oplus (2) \CMTB{(30102)} \oplus \CMTB{(41002)} \oplus (3) \CMTB{(11102)} \oplus \CMTB{(22002)} \oplus \CMTB{(02110)} \oplus \CMTB{(30012)} \oplus \CMTB{(02020)} \oplus \CMTB{(10202)} \oplus (2) \CMTB{(11012)} \oplus \CMTB{(21110)} \oplus \CMTB{(10112)} \oplus \CMTB{(21102)}$
    \item Level-13: $\CMTred{(00001)} \oplus (2) \CMTred{(01001)} \oplus (2) \CMTred{(00101)} \oplus (2) \CMTred{(00003)} \oplus (2) \CMTred{(30001)} \oplus (3) \CMTred{(11001)} \oplus \CMTred{(40001)} \oplus (3) \CMTred{(02001)} \oplus (5) \CMTred{(10101)} \oplus (3) \CMTred{(10003)} \oplus (3) \CMTred{(10011)} \oplus (4) \CMTred{(21001)} \oplus (5) \CMTred{(01101)} \oplus (2) \CMTred{(00201)} \oplus (5) \CMTred{(20101)} \oplus (3) \CMTred{(01003)} \oplus (4) \CMTred{(12001)} \oplus (3) \CMTred{(31001)} \oplus (4) \CMTred{(01011)} \oplus (3) \CMTred{(20003)} \oplus (5) \CMTred{(20011)} \oplus \CMTred{(00021)} \oplus (2) \CMTred{(03001)} \oplus (3) \CMTred{(00103)} \oplus (2) \CMTred{(00111)} \oplus \CMTred{(00013)} \oplus (2) \CMTred{(41001)} \oplus (4) \CMTred{(30101)} \oplus (3) \CMTred{(22001)} \oplus (6) \CMTred{(11101)} \oplus (2) \CMTred{(30003)} \oplus (3) \CMTred{(30011)} \oplus (4) \CMTred{(10201)} \oplus (3) \CMTred{(11003)} \oplus (7) \CMTred{(11011)} \oplus (2) \CMTred{(02101)} \oplus \CMTred{(51001)} \oplus \CMTred{(13001)} \oplus (2) \CMTred{(40101)} \oplus (2) \CMTred{(32001)} \oplus (2) \CMTred{(10021)} \oplus (2) \CMTred{(02003)} \oplus (2) \CMTred{(10103)} \oplus (2) \CMTred{(40003)} \oplus (4) \CMTred{(10111)} \oplus (3) \CMTred{(02011)} \oplus (5) \CMTred{(21101)} \oplus \CMTred{(10013)} \oplus (2) \CMTred{(01201)} \oplus \CMTred{(40011)} \oplus (2) \CMTred{(21003)} \oplus \CMTred{(50101)} \oplus (2) \CMTred{(20201)} \oplus (5) \CMTred{(21011)} \oplus \CMTred{(01103)} \oplus (2) \CMTred{(01111)} \oplus \CMTred{(01021)} \oplus (2) \CMTred{(12101)} \oplus \CMTred{(00203)} \oplus (2) \CMTred{(31101)} \oplus (2) \CMTred{(20103)} \oplus (2) \CMTred{(20111)} \oplus \CMTred{(20013)} \oplus (2) \CMTred{(12011)} \oplus \CMTred{(31003)} \oplus \CMTred{(30201)} \oplus \CMTred{(11201)} \oplus \CMTred{(31011)} \oplus \CMTred{(11111)}$
    \item Level-14: $\CMTB{(00000)} \oplus (2) \CMTB{(00100)} \oplus (2) \CMTB{(00010)} \oplus \CMTB{(40000)} \oplus (2) \CMTB{(02000)} \oplus \CMTB{(10100)} \oplus \CMTB{(10010)} \oplus (2) \CMTB{(10002)} \oplus (2) \CMTB{(01100)} \oplus (3) \CMTB{(20100)} \oplus \CMTB{(12000)} \oplus \CMTB{(31000)} \oplus (3) \CMTB{(00200)} \oplus \CMTB{(01010)} \oplus (4) \CMTB{(20010)} \oplus (5) \CMTB{(01002)} \oplus (2) \CMTB{(20002)} \oplus (2) \CMTB{(00110)} \oplus \CMTB{(00020)} \oplus (2) \CMTB{(30100)} \oplus (2) \CMTB{(00004)} \oplus (2) \CMTB{(22000)} \oplus (3) \CMTB{(11100)} \oplus (3) \CMTB{(00102)} \oplus \CMTB{(00012)} \oplus (2) \CMTB{(10200)} \oplus (3) \CMTB{(30010)} \oplus (3) \CMTB{(02100)} \oplus \CMTB{(13000)} \oplus (5) \CMTB{(11010)} \oplus (2) \CMTB{(40100)} \oplus (3) \CMTB{(30002)} \oplus \CMTB{(32000)} \oplus \CMTB{(04000)} \oplus (6) \CMTB{(11002)} \oplus (2) \CMTB{(21100)} \oplus \CMTB{(01200)} \oplus (5) \CMTB{(02010)} \oplus (4) \CMTB{(10110)} \oplus (2) \CMTB{(40010)} \oplus (2) \CMTB{(00300)} \oplus (2) \CMTB{(10020)} \oplus \CMTB{(50100)} \oplus \CMTB{(10004)} \oplus \CMTB{(42000)} \oplus (3) \CMTB{(20200)} \oplus \CMTB{(40002)} \oplus (2) \CMTB{(02002)} \oplus (7) \CMTB{(10102)} \oplus (4) \CMTB{(21010)} \oplus (4) \CMTB{(10012)} \oplus (2) \CMTB{(12100)} \oplus (2) \CMTB{(31100)} \oplus \CMTB{(50010)} \oplus (6) \CMTB{(21002)} \oplus (3) \CMTB{(01110)} \oplus \CMTB{(60100)} \oplus \CMTB{(01020)} \oplus (2) \CMTB{(00210)} \oplus \CMTB{(00030)} \oplus \CMTB{(50002)} \oplus (5) \CMTB{(20110)} \oplus \CMTB{(01004)} \oplus \CMTB{(30200)} \oplus (5) \CMTB{(01102)} \oplus (3) \CMTB{(20020)} \oplus (2) \CMTB{(11200)} \oplus (4) \CMTB{(12010)} \oplus \CMTB{(00120)} \oplus (3) \CMTB{(31010)} \oplus \CMTB{(00202)} \oplus \CMTB{(20004)} \oplus \CMTB{(60010)} \oplus \CMTB{(10300)} \oplus \CMTB{(41100)} \oplus (4) \CMTB{(01012)} \oplus (5) \CMTB{(20102)} \oplus (2) \CMTB{(22100)} \oplus (4) \CMTB{(12002)} \oplus \CMTB{(00104)} \oplus (4) \CMTB{(31002)} \oplus \CMTB{(02200)} \oplus (3) \CMTB{(20012)} \oplus \CMTB{(00112)} \oplus \CMTB{(40200)} \oplus (2) \CMTB{(03002)} \oplus (2) \CMTB{(30110)} \oplus \CMTB{(41010)} \oplus (4) \CMTB{(11110)} \oplus \CMTB{(21200)} \oplus (3) \CMTB{(22010)} \oplus (2) \CMTB{(11020)} \oplus \CMTB{(30004)} \oplus \CMTB{(10210)} \oplus (4) \CMTB{(30102)} \oplus (2) \CMTB{(41002)} \oplus \CMTB{(20300)} \oplus (5) \CMTB{(11102)} \oplus \CMTB{(10030)} \oplus \CMTB{(22002)} \oplus \CMTB{(02110)} \oplus (2) \CMTB{(30012)} \oplus \CMTB{(02020)} \oplus (2) \CMTB{(10202)} \oplus \CMTB{(40110)} \oplus (3) \CMTB{(11012)} \oplus \CMTB{(02102)} \oplus \CMTB{(31200)} \oplus (2) \CMTB{(21110)} \oplus \CMTB{(40102)} \oplus \CMTB{(10112)} \oplus \CMTB{(20210)} \oplus \CMTB{(01202)} \oplus (2) \CMTB{(21102)} \oplus \CMTB{(21012)}$
    \item Level-15: $(2) \CMTred{(00001)} \oplus \CMTred{(10001)} \oplus (2) \CMTred{(01001)} \oplus \CMTred{(20001)} \oplus (3) \CMTred{(00101)} \oplus (2) \CMTred{(00003)} \oplus \CMTred{(00011)} \oplus (3) \CMTred{(30001)} \oplus (3) \CMTred{(11001)} \oplus (2) \CMTred{(40001)} \oplus (3) \CMTred{(02001)} \oplus (6) \CMTred{(10101)} \oplus (3) \CMTred{(10003)} \oplus (4) \CMTred{(10011)} \oplus (4) \CMTred{(21001)} \oplus \CMTred{(50001)} \oplus (6) \CMTred{(01101)} \oplus (3) \CMTred{(00201)} \oplus (6) \CMTred{(20101)} \oplus (3) \CMTred{(01003)} \oplus (4) \CMTred{(12001)} \oplus (3) \CMTred{(31001)} \oplus (5) \CMTred{(01011)} \oplus \CMTred{(60001)} \oplus (3) \CMTred{(20003)} \oplus (6) \CMTred{(20011)} \oplus \CMTred{(00021)} \oplus (3) \CMTred{(03001)} \oplus (3) \CMTred{(00103)} \oplus (3) \CMTred{(00111)} \oplus \CMTred{(00013)} \oplus \CMTred{(70001)} \oplus (2) \CMTred{(41001)} \oplus (5) \CMTred{(30101)} \oplus (3) \CMTred{(22001)} \oplus (7) \CMTred{(11101)} \oplus (2) \CMTred{(30003)} \oplus (4) \CMTred{(30011)} \oplus (5) \CMTred{(10201)} \oplus (4) \CMTred{(11003)} \oplus (8) \CMTred{(11011)} \oplus (3) \CMTred{(02101)} \oplus \CMTred{(51001)} \oplus \CMTred{(10005)} \oplus (2) \CMTred{(13001)} \oplus (3) \CMTred{(40101)} \oplus (2) \CMTred{(32001)} \oplus (2) \CMTred{(10021)} \oplus (3) \CMTred{(02003)} \oplus (3) \CMTred{(10103)} \oplus (2) \CMTred{(40003)} \oplus (5) \CMTred{(10111)} \oplus (3) \CMTred{(02011)} \oplus (6) \CMTred{(21101)} \oplus \CMTred{(10013)} \oplus (3) \CMTred{(01201)} \oplus (2) \CMTred{(40011)} \oplus \CMTred{(00301)} \oplus (3) \CMTred{(21003)} \oplus (2) \CMTred{(50101)} \oplus (3) \CMTred{(20201)} \oplus \CMTred{(01005)} \oplus (6) \CMTred{(21011)} \oplus (2) \CMTred{(01103)} \oplus (2) \CMTred{(01111)} \oplus \CMTred{(01021)} \oplus (3) \CMTred{(12101)} \oplus \CMTred{(50011)} \oplus \CMTred{(00203)} \oplus (3) \CMTred{(31101)} \oplus (3) \CMTred{(20103)} \oplus (3) \CMTred{(20111)} \oplus \CMTred{(12003)} \oplus \CMTred{(20013)} \oplus (2) \CMTred{(12011)} \oplus \CMTred{(31003)} \oplus (2) \CMTred{(30201)} \oplus (2) \CMTred{(11201)} \oplus (2) \CMTred{(31011)} \oplus \CMTred{(10301)} \oplus \CMTred{(30111)} \oplus \CMTred{(11103)} \oplus \CMTred{(11111)}$
    \item Level-16: $(2) \CMTB{(00000)} \oplus \CMTB{(10000)} \oplus \CMTB{(20000)} \oplus (2) \CMTB{(00100)} \oplus \CMTB{(30000)} \oplus (2) \CMTB{(00010)} \oplus \CMTB{(00002)} \oplus (2) \CMTB{(40000)} \oplus (2) \CMTB{(02000)} \oplus \CMTB{(10100)} \oplus \CMTB{(50000)} \oplus \CMTB{(10010)} \oplus (3) \CMTB{(10002)} \oplus (2) \CMTB{(01100)} \oplus \CMTB{(60000)} \oplus (3) \CMTB{(20100)} \oplus \CMTB{(12000)} \oplus \CMTB{(31000)} \oplus (4) \CMTB{(00200)} \oplus \CMTB{(01010)} \oplus (4) \CMTB{(20010)} \oplus \CMTB{(70000)} \oplus (5) \CMTB{(01002)} \oplus (3) \CMTB{(20002)} \oplus (3) \CMTB{(00110)} \oplus (2) \CMTB{(00020)} \oplus (2) \CMTB{(30100)} \oplus (2) \CMTB{(00004)} \oplus (2) \CMTB{(22000)} \oplus (3) \CMTB{(11100)} \oplus \CMTB{(80000)} \oplus (3) \CMTB{(00102)} \oplus \CMTB{(00012)} \oplus (3) \CMTB{(10200)} \oplus (3) \CMTB{(30010)} \oplus (3) \CMTB{(02100)} \oplus \CMTB{(13000)} \oplus (5) \CMTB{(11010)} \oplus (2) \CMTB{(40100)} \oplus (4) \CMTB{(30002)} \oplus \CMTB{(32000)} \oplus (2) \CMTB{(04000)} \oplus (6) \CMTB{(11002)} \oplus (2) \CMTB{(21100)} \oplus \CMTB{(01200)} \oplus (5) \CMTB{(02010)} \oplus (5) \CMTB{(10110)} \oplus (2) \CMTB{(40010)} \oplus (2) \CMTB{(00300)} \oplus (3) \CMTB{(10020)} \oplus \CMTB{(50100)} \oplus \CMTB{(10004)} \oplus \CMTB{(42000)} \oplus (4) \CMTB{(20200)} \oplus (2) \CMTB{(40002)} \oplus (3) \CMTB{(02002)} \oplus (7) \CMTB{(10102)} \oplus (4) \CMTB{(21010)} \oplus (4) \CMTB{(10012)} \oplus (2) \CMTB{(12100)} \oplus (2) \CMTB{(31100)} \oplus \CMTB{(50010)} \oplus (6) \CMTB{(21002)} \oplus (3) \CMTB{(01110)} \oplus \CMTB{(60100)} \oplus \CMTB{(01020)} \oplus \CMTB{(03100)} \oplus (2) \CMTB{(00210)} \oplus \CMTB{(00006)} \oplus \CMTB{(00030)} \oplus (2) \CMTB{(50002)} \oplus (6) \CMTB{(20110)} \oplus \CMTB{(01004)} \oplus (2) \CMTB{(30200)} \oplus (6) \CMTB{(01102)} \oplus (4) \CMTB{(20020)} \oplus (2) \CMTB{(11200)} \oplus (4) \CMTB{(12010)} \oplus \CMTB{(00120)} \oplus (3) \CMTB{(31010)} \oplus (2) \CMTB{(00202)} \oplus (2) \CMTB{(20004)} \oplus \CMTB{(60010)} \oplus \CMTB{(10300)} \oplus \CMTB{(41100)} \oplus (4) \CMTB{(01012)} \oplus (5) \CMTB{(20102)} \oplus (2) \CMTB{(22100)} \oplus (5) \CMTB{(12002)} \oplus \CMTB{(00104)} \oplus (4) \CMTB{(31002)} \oplus (2) \CMTB{(02200)} \oplus \CMTB{(60002)} \oplus (3) \CMTB{(20012)} \oplus \CMTB{(00112)} \oplus (2) \CMTB{(40200)} \oplus (2) \CMTB{(03002)} \oplus \CMTB{(00400)} \oplus (3) \CMTB{(30110)} \oplus \CMTB{(01300)} \oplus \CMTB{(30020)} \oplus \CMTB{(41010)} \oplus (4) \CMTB{(11110)} \oplus \CMTB{(21200)} \oplus (3) \CMTB{(22010)} \oplus (2) \CMTB{(11020)} \oplus \CMTB{(30004)} \oplus \CMTB{(10210)} \oplus (4) \CMTB{(30102)} \oplus (2) \CMTB{(41002)} \oplus \CMTB{(20300)} \oplus (6) \CMTB{(11102)} \oplus (2) \CMTB{(10030)} \oplus (2) \CMTB{(22002)} \oplus \CMTB{(02110)} \oplus (2) \CMTB{(30012)} \oplus \CMTB{(02020)} \oplus (3) \CMTB{(10202)} \oplus (2) \CMTB{(40110)} \oplus (3) \CMTB{(11012)} \oplus \CMTB{(40020)} \oplus \CMTB{(02102)} \oplus \CMTB{(02004)} \oplus \CMTB{(31200)} \oplus (2) \CMTB{(21110)} \oplus \CMTB{(40102)} \oplus \CMTB{(10112)} \oplus \CMTB{(20210)} \oplus \CMTB{(01202)} \oplus (3) \CMTB{(21102)} \oplus \CMTB{(20202)} \oplus \CMTB{(21012)}$
\end{itemize}

\newpage
$$~~$$


\begin{thebibliography}{99}
\small\frenchspacing\raggedright

\bibitem{SSP}
A. Salam, and J.A. Strathdee,
``Supersymmetry and Superfields,''
{\bf {Fortsch.Phys. 26}} (1978) 57,
DOI: 10.1002/prop.19780260202. 

\bibitem{Adnk1}
M.~Faux, S.~J.~Gates Jr., ``Adinkras: A Graphical Technology for Supersymmetric 
Representation Theory'', 
{\bf {Phys.Rev. D71}} (2005) 065002,
DOI: 10.1103/PhysRevD.71.065002
e-Print: hep-th/0408004.
 
\bibitem{GRana1}
S.\ J.\ Gates and L. Rana, Phys.\ Lett.\ {\bf {B352}} (1995) 50, 
DOI: 10.1016/0370-2693(95)00474-Y,
e-Print: hep-th/9504025.

\bibitem{GRana2}
S.\ J.\ Gates and L. Rana, Phys.\ Lett.\ {\bf {B369}} (1996) 262, DOI: 10.1016/0370-2693(95)01542-6,
e-Print: hep-th/9510151.

\bibitem{ENUF}
S.J. Gates, Jr., W.D. Linch, III,  and J. Phillips,
``When superspace is not enough,'' unpublished,  Nov 2002. 44 pp.,
UMDEPP-02-054, CALT-68-2387,
e-Print: hep-th/0211034. 

\bibitem{counting10d}
  S. J. Gates, Jr., Y. Hu and S.-N. H. Mak, ``Superfield Component Decompositions and the Scan for Prepotential Supermultiplets in 10D Superspaces,'' arXiv:1911.00807 (2019), to appear in {\bf {JHEP}}.

\bibitem{CoDeX}
 S.J.\ Gates, Jr., Y.\ Hu,, H.\ Jiang, and , S.-N. Hazel Mak, 
A codex on linearized Nordstr\" om supergravity in eleven and ten dimensional superspaces
{\bf {JHEP 1907}} (2019) 063,
DOI: 10.1007/JHEP07(2019)063,
e-Print: arXiv:1812.05097 [hep-th]. 

\bibitem{MTh}
E.\ Witten,
 "String theory dynamics in various dimensions". 
 {\bf {Nucl. Phys. B. 443}} (1): 85Ð126, DOI: 10.1016/0550-3213(95)00158-O, 
 arXiv:hep-th/9503124.

\bibitem{crD11a}
E.\ Cremmer, and S.\ Ferrara, ``Formulation of Eleven-Dimensional Supergravity 
in Superspace,''  Phys.\ Lett.\ {\bf {91B}} (1980) 61, DOI: 10.1016/0370-2693(80)90662-0.

\bibitem{crD11b}
L.\ Brink and P.\ S.\ Howe, ``Eleven-Dimensional Supergravity on the Mass-Shell in 
Superspace,'' Phys.\ Lett.\ {\bf {91B}} (1980) 384, DOI: 10.1016/0370-2693(80)91002-3.

\bibitem{PF1}
S.\ M.\ Chester, S.\ S.\ Pufu, and X.\ Yin,
``The M-Theory S-Matrix From ABJM: Beyond 11D Supergravity,''
{\bf {JHEP 1808}} (2018) 115,
DOI: 10.1007/JHEP08(2018)115,
e-Print: arXiv:1804.00949 [hep-th].

\bibitem{PF2}
D.\ J.\ Binder, S.\ M.\ Chester, and S.\ S.\ Pufu, 
``Absence of D${}^4$ R${}^4$ in M-Theory From ABJM,''
{\bf {JHEP 04}} (2020) 052, 
DOI: 10.1007/JHEP04(2020)052,
e-Print: arXiv:1808.10554 [hep-th].

\bibitem{PF3}
D.\ J.\ Binder, S.\ M.\ Chester, and S.\ S.\ Pufu, 
``AdS4/CFT3 from Weak to Strong String Coupling,''
{\bf {JHEP 2001}} (2020) 034, DOI: 10.1007/JHEP01(2020)034
e-Print: arXiv:1906.07195 [hep-th].

\bibitem{Howe:2003sa} 
  P.~S.~Howe, S.~F.~Kerstan, U.~Lindstrom and D.~Tsimpis,
  ``The Deformed M2-brane,''
  {\bf {JHEP 0309}}, 013 (2003)
  DOI:10.1088/1126-6708/2003/09/013,
  [hep-th/0307072].
  
\bibitem{Howe:2003cy} 
 P.~S.~Howe and D.~Tsimpis,
 ``On higher order corrections in M theory,''
 {\bf {JHEP 0309}}, 038 (2003)
 DOI:10.1088/1126-6708/2003/09/038, [hep-th/0305129].
 
\bibitem{GS}
M.\ B.\ Green, and S.\ Sethi,
ÒSupersymmetry Constraints on Type-IIB Supergravity,Ó
{\bf {Phys. Rev. D59}} (1999) 046006, DOI: 10.1103/PhysRevD.59.046006,
e-Print: hep-th/9808061 [hep-th].

\bibitem{MandL1}
S.\ Mandelstam,
``Ultra-Violet Finiteness of the N = 4 Model,"
{\bf {Stud.\ Nat\ .Sci.\ 20}} (1985) 167,
Contribution to: 20th Annual Orbis Scientiae: 
Dedicated to P.A.M. Dirac's 80th Year, 167-177,
DOI: https://doi.org/10.1007/978-1-4684-8848-7$_-$13. 

\bibitem{MandL2}
S.\ Mandelstam,
``Light Cone Superspace and the Ultraviolet Finiteness of the N=4 Model,"
{\bf {Nucl.\ Phys.\ B}} 213 (1983) 149,
DOI: https://doi.org/10.1016/0550-3213(83)90179-7.

\bibitem{MandL3}
S.\ Mandelstam,
``Lightcone superspace and the finiteness of the N=4 model,''
{\bf {Workshop on Problems in Unification and Supergravity, 
AIP Conf.\ Proc.\ 116}}, (1984) 99,
DOI: https://doi.org/10.1063/1.34597.

\bibitem{OK}
O.\ Ahl\' en, and A.\ Kleinschmidt,
ÒD${}^6$ R${}^4$ curvature corrections, modular graph functions and Poincar\' e series,Ó
JHEP 09 (2017) 155, DOI: 10.1007/JHEP05(2018)194,
e-Print: arXiv:1803.10250v1 [hep-th].

\bibitem{BB}
K.\ Becker, and D.\ Butter, 
``4D, $\cal N$ = 1 Kaluza-Klein superspace,''
arXiv: 2003.01790[hep-th].

\bibitem{FX}
S. J. Gates, Jr. and S. Vashakidze, "On D = 10, N = 1 Supersymmetry, 
Superspace Geometry and Superstring Effects (I)," {\bf {Nucl. Phys. B  291}} 
(1987) 172, https://doi.org/10.1016/0550-3213(87)90470-6
 
\bibitem{M2}
S.\ J.\ Gates, Jr., ``Superconformal symmetry in 11-D superspace and the M theory effective action,''
{\bf {Nucl.\ Phys.\ B616}} (2001) 85, DOI: 10.1016/S0550-3213(01)00421-7, e-Print: hep-th/0106150.

\bibitem{2MT}
S.\ J.\ Gates, Jr., and H.\ Nishino, 
``Deliberations on 11-D Superspace for the M-Theory Effective Action,''
{\bf {Phys.\ Lett.\ B508}} (2001) 115,
DOI: 10.1016/S0370-2693(01)00487-7,
e-Print: hep-th/0101037 [hep-th].

\bibitem{2MT0}
H.\ Nishino, and S.\ J.\ Gates, Jr., 
``Toward an Off - Shell 11D Supergravity Limit of M-Theory,''
{\bf {Phys.\ Lett.\ B388}} (1996) 504-511,
DOI: 10.1016/S0370-2693(96)01193-8,
e-Print: hep-th/9602011.


\bibitem{10DScLR}
E. Bergshoeff, and M. de Roo, 
``The Supercurrent in Ten Dimensions," 
 {\bf {Phys. Lett. 112B}} (1982) 53,
DOI: 10.1016/0370-2693(82)90904-2.

\bibitem{C0L0R}
S. J. Gates, Jr. and K. Stiffler, 
``Adinkra `Color' Confinement In Exemplary Off-Shell Constructions Of 4D, 
N = 2 Supersymmetry Representations,''
{\bf {JHEP 1407}} (2014) 051, DOI: 10.1007/JHEP07(2014)051
e-Print: arXiv:1405.0048 [hep-th] 

\bibitem{DKP1}
R.J. Duffin, ``On The Characteristic Matrices of Covariant Systems,''
{\bf {Phys. Rev. Lett., vol. 54}}, 1114 (1938), DOI:10.1103/PhysRev.54.1114.

\bibitem{DKP2}
N. Kemmer, ``The particle aspect of meson theory,'' {\bf {Proceedings of the Royal Society A, vol. 173}}, pp. 91Ð116 (1939), DOI:10.1098/rspa.1939.0131.

\bibitem{DKP3}
G. Petiau, University of Paris thesis (1936), published in {\bf {Acad. Roy. de Belg., A. Sci. Mem. Collect.vol. 16, N 2}}, 1 (1936)

\bibitem{yamatsu2015}
  N. Yamatsu, ``Finite-dimensional Lie algebras and their representations for unified model building,'' arXiv:1511.08771  (2015).
  
\bibitem{Plethysm}
Macdonald, Ian Grant., ``Symmetric functions and Hall polynomials," Oxford university press, 1998.
  
\bibitem{Susyno}
  R. M. Fonseca, ``Calculating the Renormalisation Group Equations of a SUSY Model with Susyno,'' {\bf {Comput. Phys. Commun. 183}} (2012) 2298Ð2306, 
  DOI: 10.1016/j.cpc.2012.05.017,
  arXiv:1106.5016 [hep-ph].
  
\bibitem{LiE}
Van Leeuwen, M. A. A., ``LiE, a software package for Lie group computations." Euromath Bull 1.2 (1994): 83-94.

\bibitem{LieART} 
  R.~Feger and T.~W.~Kephart,
  ``LieART--A Mathematica application for Lie algebras and representation theory,''
 {\bf {Comput.\ Phys.\ Commun.\  192}}, 166 (2015)
  DOI:10.1016/j.cpc.2014.12.023
  [arXiv:1206.6379 [math-ph]].
  
\bibitem{FT}
C.\ Vafa. ``Evidence for F-theory,"
{\bf {Nucl.\  Phys.\ B}}. 469 (3): 403Ð415, DOI: 10.1016/0550-3213(96)00172-1
arXiv:hep-th/9602022.

\bibitem{VaR}
W. Siegel, ``Supergravity Superfields Without a Supermetric,''
Nov 1977. 16 pp., Harvard Univ. Preprint, HUTP-77/A068, unpublished. 

\bibitem{0VRT}
K.S. Stelle, and P.C. West,
``Minimal Auxiliary Fields for Supergravity,'' 
{\bf {Phys.Lett.}} 74B (1978) 330-332,
DOI: 10.1016/0370-2693(78)90669-X.

\bibitem{SpRSp8c}
 S.J.\ Gates, Jr., M.T. Grisaru, M. Ro\v cek, and W. Siegel,
 {\it {Superspace Or One Thousand and One Lessons in Supersymmetry}},
{\bf {Front.Phys. 58}} (1983) 1-548,
e-Print: hep-th/0108200. 

\bibitem{OS1}
V. Ogievetsky, and E. Sokatchev,
"Supercurrent,"
{\bf {Sov.\ J.\ Nucl.\ Phys.\ 28}} (1978) 423, {\bf {Yad\ .Fiz.\ 28}} (1978) 825, 
{\bf {Yadern.\ Fiz.\ 28}} (1978) 825, 
preprint Dubna JINR - 11528 (78,REC.JUL) 25p

\bibitem{OS2}
V. Ogievetsky, and E. Sokatchev,
"On Vector Superfield Generated by Supercurrent,"
{\bf {Nucl.\ Phys.\ B 124}} (1977) 309, 
DOI: https://doi.org/10.1016/0550-3213(77)90318-2

\end{thebibliography}
\end{document}